\documentclass[traditabstract]{aa}

\usepackage{natbib}
\usepackage{graphicx}
\usepackage{txfonts}
\usepackage[bookmarks=false, colorlinks=true, citecolor=blue, linkcolor=blue]{hyperref}
\usepackage{subfig}
\usepackage{blindtext}

\newcommand{\Spitzer}{\textit{Spitzer}}
\newcommand{\Herschel}{\textit{Herschel}}
\newcommand{\IRAS}{IRAS}
\def\micron{\hbox{\,$\mu$m}}
\newcommand{\Lsun}{\hbox{$L_{\rm \odot}$}}
\newcommand{\Msun}{\hbox{$M_{\rm \odot}$}}
\newcommand{\LIR}{\hbox{$L_{\rm IR}$}}
\newcommand{\degree}{\ensuremath{^\circ}}
\newcommand\nodata{ ~$\cdots$~ }

\bibpunct{(}{)}{;}{a}{}{,}

\titlerunning{Molecular outflows in ULIRGs}
\authorrunning{Pereira-Santaella et al.}

\begin{document}

\title{Spatially resolved cold molecular outflows in ULIRGs}

\author{M.~Pereira-Santaella\inst{\ref{inst1}} \and L.~Colina\inst{\ref{inst2}} \and S.~Garc\'ia-Burillo\inst{\ref{inst3}}  \and F.~Combes\inst{\ref{inst4}} \and B.~Emonts\inst{\ref{inst5}} \and S.~Aalto\inst{\ref{inst6}} \and A.~Alonso-Herrero\inst{\ref{inst2}} \and S.~Arribas\inst{\ref{inst2}} \and C.~Henkel\inst{\ref{inst7},\ref{inst8}} \and A.~Labiano\inst{\ref{inst2}} \and S.~Muller\inst{\ref{inst6}} \and J.~Piqueras L\'opez\inst{\ref{inst2}} \and D.~Rigopoulou\inst{\ref{inst1}} \and P.~van der Werf\inst{\ref{inst9}}}

\institute{Department of Physics, University of Oxford, Keble Road, Oxford OX1 3RH, UK\\ \email{miguel.pereira@physics.ox.ac.uk}\label{inst1}
\and
Centro de Astrobiolog\'ia (CSIC/INTA), Ctra de Torrej\'on a Ajalvir, km 4, 28850, Torrej\'on de Ardoz, Madrid, Spain\label{inst2}
\and
Observatorio Astron\'omico Nacional (OAN-IGN)-Observatorio de Madrid, Alfonso XII, 3, 28014, Madrid, Spain\label{inst3}
\and
LERMA, Obs. de Paris, PSL Research Univ., Coll\'ege de France, CNRS, Sorbonne Univ., UPMC, Paris, France\label{inst4}
\and
National Radio Astronomy Observatory, 520 Edgemont Road, Charlottesville, VA 22903, US\label{inst5}
\and
Department of Space, Earth and Environment, Onsala Space Observatory, Chalmers University of Technology, 439 92 Onsala, Sweden\label{inst6}
\and
Max-Planck-Institut f\"ur Radioastronomie, Auf dem H\"ugel 69, D-53121 Bonn, Germany\label{inst7}
\and
Astron. Dept, King Abdulaziz University, PO Box 80203, Jeddah 21589, Saudi Arabia\label{inst8}
\and
Leiden Observatory, Leiden University, PO Box 9513, 2300, RA Leiden, The Netherlands\label{inst9}
}

\abstract{We present new CO(2--1) observations of three low-$z$ ($d\sim$350\,Mpc) ULIRG systems (6 nuclei) observed with ALMA at high-spatial resolution ($\sim$500\,pc). We detect massive cold molecular gas outflows in 5 out of 6 nuclei ($M_{\rm out}\sim(0.3-5)\times10^8$\,\Msun). These outflows are spatially resolved with deprojected { effective} radii between 250\,pc and 1\,kpc although high-velocity molecular gas is detected up to $R_{\rm max}\sim 0.5-1.8$\,kpc { ($1-6$\,kpc deprojected). The mass outflow rates are $12-400$\,\Msun\,yr$^{-1}$} and the inclination corrected average velocity of the outflowing gas $350-550$\,km\,s$^{-1}$ (v$_{\rm max}=500-900$\,km\,s$^{-1}$). The origin of these outflows can be explained by the strong nuclear starbursts although the contribution of an obscured AGN can not be completely ruled out.
The position angle (PA) of the outflowing gas along the kinematic minor axis of the nuclear molecular disk suggests that the outflow axis is perpendicular to the disk for three of these outflows. Only in one case, the outflow PA is clearly not along the kinematic minor axis and might indicate a different outflow geometry.
The outflow depletion times are { $15-80$\,Myr}. These are comparable to, although slightly shorter than the star-formation (SF) depletion times { ($30-80$\,Myr)}. 
However, we estimate that only $15-30$\% of the outflowing molecular gas will escape the gravitational potential of the nucleus. The majority of the outflowing gas will return to the disk after $5-10$\,Myr and become available to form new stars. Therefore, these outflows will not likely completely quench the nuclear starbursts.
These star-forming powered molecular outflows would be consistent with being driven by radiation pressure from young stars (i.e., momentum-driven) only if the coupling between radiation and dust increases with increasing SF rates. This can be achieved if the dust optical depth is higher in objects with higher SF. This is the case in, at least, one of the studied objects. Alternatively, if the outflows are mainly driven by supernovae (SNe), the coupling efficiency between the interstellar medium and SNe must increase with increasing SF levels. 
The relatively small sizes ($<$1\,kpc) and dynamical times ($<$3\,Myr) of the cold molecular outflows suggests that molecular gas cannot survive longer in the outflow environment or that it cannot form efficiently beyond these distances or times. 
In addition, the ionized and hot molecular phases have been detected for several of these outflows, so this suggests that outflowing gas can experience phase changes and indicates that the outflowing gas is intrinsically multiphase, likely sharing similar kinematics, but different mass and, therefore, energy and momentum contributions.
}

\date{Received 26 March 2018 / Accepted 9 May 2018}

\maketitle
 
\section{Introduction}\label{s:intro}

Negative feedback from active galactic nuclei (AGN) and starbursts plays a fundamental role in the evolution of galaxies according to theoretical models and numerical simulations (e.g., \citealt{Narayanan2011, Scannapieco2012, Hopkins2012, Nelson2015, Schaye2015}). This feedback occurs through the injection of material, energy, and momentum into the interstellar medium (ISM) and gives rise to massive gas outflows and regulates the growth of the stellar mass and black-hole accretion.

Such energetic and massive outflows have been detected in galaxies at low and high redshift. In particular, they have been detected in ultra-luminous infrared galaxies (ULIRGs; \LIR$>10^{12}$\Lsun) in their atomic ionized (e.g., \citealt{Westmoquette2012,Arribas2014}), atomic neutral (e.g., \citealt{Rupke2005, Cazzoli2016}), and cold molecular (e.g., \citealt{Fischer2010, Feruglio2010, Sturm2011, Cicone2014}) phases.
Local ULIRGs are major gas-rich mergers mainly powered by star-formation (SF), although AGN accounting for a significant fraction of the total IR luminosity ($10-60\%$) are usually detected too \citep{Farrah2003,Nardini2010}. 
Since local ULIRGs are the hosts of the most extreme starbursts in the local Universe with star-formation rates (SFR) greater than $\sim$150\,\Msun\,yr$^{-1}$, based on their IR luminosities \citep{Kennicutt2012}, they are adequate objects to study the negative feedback from both AGN and SF.

In this paper, we focus on the molecular phase of these outflows. This phase includes molecular gas with a wide range of temperatures. The hot ($T>1500$\,K) and the warm ($T>200$\,K) molecular phases can be observed using the near-IR ro-vibrational H$_2$ transitions (e.g., \citealt{Emonts2014, Dasyra2015, Emonts2017}) and the mid-IR rotational H$_2$ transitions (e.g., \citealt{Hill2014}). However, it is thought that the energy and mass of these outflows are dominated by the cold molecular phase { (e.g., \citealt{Feruglio2010, Cicone2014, Saito2018}}) although some observations and models seem to contradict this view (e.g., \citealt{Hopkins2012,Dasyra2016}). The cold molecular phase has been detected using multiple CO transitions (e.g., \citealt{Feruglio2010, Chung2011, Cicone2014, GarciaBurillo2015, Pereira2016b}), HCN transitions { (e.g., \citealt{Aalto2012, Walter2017, BarcosMunoz2018})}, and far-IR OH absorption \citealt{Fischer2010, Sturm2011, Spoon2013, GonzalezAlfonso2017}). All these observations have revealed that cold molecular outflows are common in ULIRGs and that they can be massive enough to play a relevant role in the regulation of the SF in their host galaxies. 

Knowing the distribution of the outflowing gas is important to derive accurate outflow properties, like the outflow mass, energy, and momentum rates, which are key to determine the impact of these outflows onto their host galaxies. However, spatially resolved observations of outflows in ULIRGs are { still limited to few sources (e.g., \citealt{GarciaBurillo2015, Veilleux2017, Saito2018, BarcosMunoz2018})}. Here, we present new high-angular resolution ($\sim0\farcs3-0\farcs4$) ALMA observations of the CO(2--1) transition in three low-$z$ ULIRGs where the cold molecular outflow phase is spatially resolved on scales of $\sim$500\,pc. This provides a direct measurement of the outflow size, and, therefore, allows us to derive more accurately the outflow properties.

This paper is organized as follows: the sample and the ALMA observations are described in Section~\ref{s:data}. In Section ~\ref{s:ana}, we analyze the 248\,GHz continuum and CO(2--1) emissions and measure the outflow properties in these systems. 
The energy source of the outflows, as well as their impact, launching mechanism, and multi-phase structure are discussed in Section \ref{s:discu}. Finally, in Section \ref{s:conclusions}, we summarize the main results of the paper.

Throughout this article we assume the following cosmology: $H_{\rm 0} = 70$\,km\,s$^{-1}$\,Mpc$^{-1}$, $\Omega_{\rm m}=0.3$, and $\Omega_{\rm \Lambda}=0.7$.

\section{Observations and data reduction}\label{s:data}

\subsection{Sample of ULIRGs}\label{s:sample}

\begin{table*}[ht]
\caption{Sample of local ULIRGs}
\label{tbl_sample}
\centering
\begin{small}
\begin{tabular}{llccccccccccc}
\hline \hline
\\
{IRAS} Name & Component & R.A.\tablefootmark{a} & Dec.\tablefootmark{a} &  v$_{\rm sys}$\tablefootmark{b} & $z$\tablefootmark{~c} & $d_{\rm L}$\tablefootmark{d} & Scale\tablefootmark{~d} & $\log \frac{L({\rm AGN})}{ L_\odot}$\tablefootmark{~e} &$\log \frac{L_{\rm IR}}{L_\odot}$\tablefootmark{~f} \\
&  & (ICRS) & (ICRS) & (km\,s$^{-1}$) & & (Mpc) & (pc\,arcsec$^{-1}$)\\
\hline\\[-2ex]
12112$+$0305 & & & & & 0.0731 & 330 & 1390 & 11.4 & 12.19 \\
 & SW & 12h13m45.939s & 2d48m39.10s & 21167 &  &  & \\
 & NE & 12h13m46.056s & 2d48m41.53s & 21045 &  &  & \\
14348$-$1447 & & & & & 0.0825 & 375 & 1554 & 11.6 & 12.27 \\
 & SW & 14h37m38.280s & --15d00m24.24s & 23766 &  &  & \\
 & NE & 14h37m38.396s & --15d00m21.29s & 23676 &  &  & \\
22491$-$1808 & & & & & 0.0778 & 353 & 1469 & 11.5 & 12.03 \\
 & E & 22h51m49.348s & --17d52m24.12s & 22412 &  &  & \\
 & W\tablefootmark{*} & \nodata & \nodata & \nodata \\
\hline
\end{tabular}
\end{small}
\tablefoot{
\tablefoottext{a}{Coordinates of the 248\,GHz rest-frame continuum emission for each nucleus (see Section \ref{ss:continuum}). The astrometric uncertainty is $\sim$25\,mas (see Section~\ref{ss:nuclear_kin}).}
\tablefoottext{b}{CO(2--1) systemic velocity using the relativistic velocity definition in the kinematic local standard of rest (LSRK; see Section \ref{ss:nuclear_kin}). Typical uncertainties are $\lesssim$10\,km\,s$^{-1}$.}
\tablefoottext{c}{Redshift using the average systemic velocity of the system.}
\tablefoottext{d}{Luminosity distance and scale for the assumed cosmology (see Sect~\ref{s:intro}).}
\tablefoottext{e}{Luminosity of the AGN in the system estimated from mid-IR spectroscopy \citep{Veilleux2009}.}
\tablefoottext{f}{IR luminosity of the system based on the SED fitting of the {\it Spitzer} and {\it Herschel} { mid- and far-IR} photometry (see Section~\ref{ss:ir_lum}).}
\tablefoottext{*}{No 248\,GHz continuum is detected at the position of the near-IR W nucleus of \IRAS~22491$-$1808.}}
\end{table*}

In this paper, we study three low-$z$ ($d\sim350$\,Mpc) ULIRGs (six individual nuclei) with \hbox{$\log L_{\rm IR}\slash L_\odot=12.0-12.3$} (see Table~ \ref{tbl_sample}) { based on their mid- and far-IR spectral energy distribution modeling (Section~\ref{ss:ir_lum})}. These three ULIRG systems seem to be in a similar dynamical state. They were classified as type III by \citet{Veilleux2002}, which corresponds to a pre-merger stage characterized by two identifiable nuclei with well defined tidal tails and bridges. They also belong to the subclass of ``close binary'' (i.e., ``b'') as the projected separation of their nuclei is smaller than 10\,kpc.

Their nuclei are classified as low ionization nuclear emission-line regions (LINER; \IRAS~12112+0305 and \IRAS~14348$-$1447; e.g., \citealt{Colina2000, Evans2002}) or \ion{H}{ii} (\IRAS~22491$-$1808; e.g., \citealt{Veilleux1999}) and in all systems a weak AGN contribution ($10-15$\%) is detected in their mid-IR \Spitzer\ spectra \citep{Veilleux2009}. For \IRAS~14348$-$1447, high-angular resolution mid-IR imaging { suggests} that the AGN is located in the SW nucleus \citep{AlonsoHerrero2016}. For \IRAS~12112+0305 and \IRAS~22491$-$1808, we assume that the AGN is at the brightest nucleus in the radio/sub-mm continuum, i.e., \IRAS~12112+0305 NE and \IRAS~22491$-$1808 E (see below). In addition, vibrationally excited HCN $J=4-3$ emission is detected in \IRAS~12112+0305 NE and \IRAS~22491$-$1808 E which can be a signature of hot dust heated by an AGN \citep{Imanishi2016, Imanishi2018}.

In addition, these three ULIRGs belong to a representative sample of local ULIRGs studied by \citet{GarciaMarin2009, GarciaMarin2009Part1}, \citet{Arribas2014}, and \citet{Piqueras2012} using optical and near-IR integral field spectroscopy.

\subsection{ALMA data}

We obtained Band 6 ALMA CO(2--1) 230.5\,GHz and continuum observations for these three local ULIRGs (see Table~\ref{tbl_sample}) as part of the ALMA projects 2015.1.00263.S and 2016.1.00170.S (PI: Pereira-Santaella). The observations were carried out between June 2016 and May 2017. The total on-source integration times per source were $\sim30-40$\,min split into two scheduling blocks. The baseline lengths range between 15 and 1100\,m providing a synthesized beam full-width at half-maximum (FWHM) of $\sim0\farcs3-0\farcs4$ ($400-500$\,pc at the distance of these ULIRGs). Details on the observations for each source are listed in Table~\ref{tbl_alma_log}.

Two spectral windows of 1.875\,GHz bandwidth (0.976\,MHz\,$\equiv\sim$1.3\,km\slash s channels) were centered at the sky frequency of the $^{12}$CO(2--1) and CS(5--4) transitions (see Table~\ref{tbl_alma_log}). In addition, a continuum spectral window was set at $\sim$248\,GHz ($\sim$1.2\,mm). In this paper, we analyze the CO(2--1) and continuum spectral windows, the CS(5--4) data will be presented in a future paper.

The data were calibrated using the ALMA reduction software CASA (v4.7; \citealt{McMullin2007}). The amplitude calibrators for each scheduling block are listed in Table \ref{tbl_alma_log}. For the CO(2--1) spectral window, a constant continuum level was estimated using the line free channels and then subtracted in the $uv$ plane. For the image cleaning, we used the Briggs weighting with a robustness parameter of 0.5 \citep{Briggs1995PhDT}. The synthesized beam ($\sim0\farcs3-0\farcs4$) and maximum recoverable scale ($\sim4\arcsec$) are presented in Table \ref{tbl_alma_log} for each observation. 
{ To our knowledge, there are no single-dish CO(2--1) fluxes published for these ULIRGs so it is not straightforward to estimate if we filter part of the extended emission. However, the bulk of the CO(2--1)} emission of these systems is relatively compact (see Section~\ref{ss:molgas} { and Appendix~\ref{apx_channels}}), so we expect to recover most of the CO(2--1) emission with these array configurations. 
The final datacubes have 300$\times$300 spatial pixels of 0\farcs08 and 220 spectral channels of 7.81\,MHz ($\sim10$\,km\slash s). 
For the CO(2--1) { cubes}, the 1$\sigma$ sensitivity is $\sim310-450$\,$\mu$Jy\,beam$^{-1}$ per channel and $\sim30-45$\,$\mu$Jy\,beam$^{-1}$ for the continuum images. A primary beam correction (FWHM$\sim$20\arcsec) was applied to the data. 

\subsection{Near-IR \textit{HST} imaging}

We downloaded the near-IR \textit{HST}\slash NICMOS F160W ($\lambda_{\rm c}=$1.60\micron, FWHM=0.34\micron) and F222M ($\lambda_{\rm c}=$2.21\micron, FWHM=0.15\micron) reduced images from the Mikulski Archive for Space Telescopes (MAST). The angular resolutions of these images are 0\farcs14 and 0\farcs20 for the F160W and F222M filters, respectively, which is slightly better than the resolution of the ALMA data. The ALMA and \textit{HST} images were aligned using the positions of the nuclei in the 248\,GHz and F222M images. The F222M filter was used because it is less affected by dust obscuration than F160W.
{ If the 2.2\micron\ near-IR and 248\,GHz continua have similar spatial distributions in these ULIRGs, the uncertainty of the image alignment is about 0\farcs08 ($\sim$120\,pc) limited by the centroid accuracy in the {\it HST} data.
}

\begin{table*}[ht]
\caption{ALMA observation log}
\label{tbl_alma_log}
\centering
\begin{small}
\begin{tabular}{lcccccccccccc}
\hline \hline
\\
Object & Date & Observed & On-source & Maximum & Synthesized & \multicolumn{2}{c}{Amplitude calibrator} & Sensitivity\tablefootmark{c}\\
\cline{7-8}\\[-2.1ex]
& & frequency\tablefootmark{a} & time & recoverable scale & beam\tablefootmark{b} & Name & Flux\\
& & (GHz) & (min) &  &  & & (Jy) & ($\mu$Jy\,beam$^{-1}$) \\
\hline\\[-2ex]
12112+0305   & 2017-05-08 & 214.81 & 37 & 3\farcs9 & 0\farcs36$\times$0\farcs27, --82\degree & J1229+0203 & 7.01$\pm$0.27 & 310\slash 33 \\
	     & 2017-05-09 & & & & & & \\
14348$-$1447 & 2017-05-09 & 212.87 & 39 & 4\farcs0 & 0\farcs32$\times$0\farcs26, --78\degree & J1517-2422 & 1.79$\pm$0.11 & 340\slash 32 \\
	     & 2017-05-22 & & & & & & 2.15$\pm$0.15 \\
22491$-$1808 & 2016-06-21 & 213.92 & 29 & 4\farcs0 & 0\farcs48$\times$0\farcs34, --84\degree & Pallas & Butler\tablefootmark{d} & 450\slash 46 \\
	     & 2016-07-21 & & & & & & \\
\hline
\end{tabular}
\end{small}
\tablefoot{
\tablefoottext{a}{Central observed frequency of the CO(2--1) spectral window.}
\tablefoottext{b}{FWHM and position angle of the synthesized beam using Briggs weighting with a robustness parameter of 0.5.}
\tablefoottext{c}{1$\sigma$ line\slash continuum sensitivities after combining the two scheduling blocks for each object. For the line sensitivity, we use the 7.8\,MHz ($\sim$10\,km\,s$^{-1}$) channels of the final data cube.}
\tablefoottext{d}{Flux estimated using the Butler-JPL-Horizons 2012 models and ephemeris information (see ALMA Memo \#594).}
}
\end{table*}

\section{Data analysis}\label{s:ana}

\subsection{Morphology}

In Figures~\ref{fig_data_i12112}, \ref{fig_data_i14348}, and \ref{fig_data_i22491}, we present the CO(2--1) and 248\,GHz continuum emission maps of the three ULIRGs. 

\subsubsection{Molecular gas}\label{ss:molgas}

\begin{figure*}
\centering
\includegraphics[width=\textwidth]{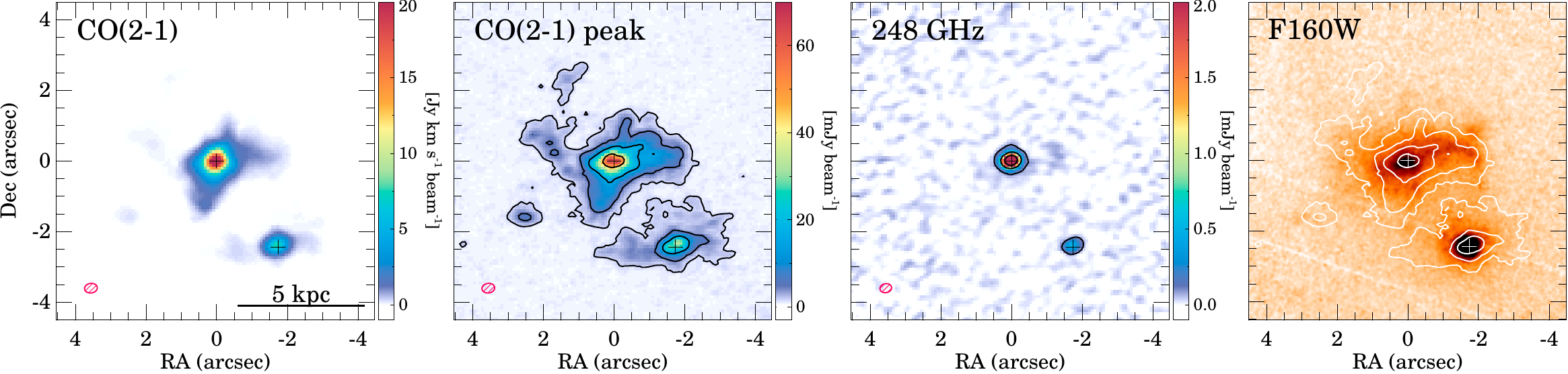}
\caption{ALMA and \textit{HST} maps for IRAS~12112+0305. The first and second panels are the CO(2--1) integrated flux and { peak intensity for $\sim$10\,km\,s$^{-1}$ channels}, respectively. The contour levels in the second panel correspond to (6, 18, 54, 162, 484)$\times\sigma$, where $\sigma$ is the line sensitivity (Table \ref{tbl_alma_log}). The third panel is the ALMA 248\,GHz continuum. The contours in this panel are (3, 27, 81)$\times\sigma$ where $\sigma$ is the continuum sensitivity (Table \ref{tbl_alma_log}). The fourth panel shows the near-IR \textit{HST}\slash NICMOS F160W map with the CO(2--1) peak contours. The position of the two nuclei is marked with a cross in all the panels. The red hatched ellipse represents the FWHM { and PA} of the ALMA beam (Table \ref{tbl_alma_log}).
\label{fig_data_i12112}}
\end{figure*}

\begin{figure*}
\centering
\includegraphics[width=\textwidth]{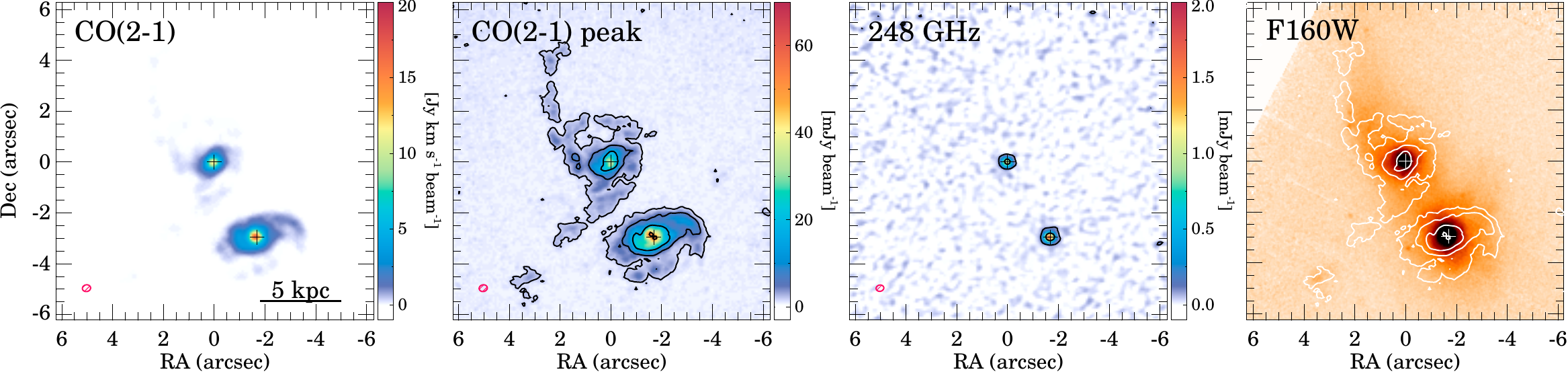}
\caption{Same as Figure~\ref{fig_data_i12112} but for IRAS~14348$-$1447.\label{fig_data_i14348}}
\end{figure*}

\begin{figure*}
\centering
\includegraphics[width=\textwidth]{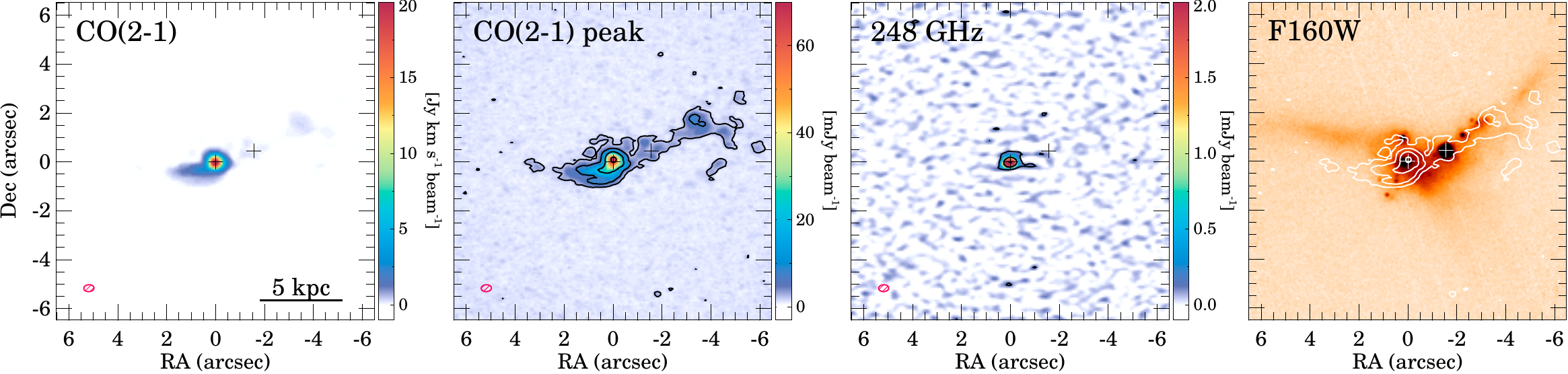}
\caption{Same as Figure~\ref{fig_data_i12112} but for IRAS~22491$-$1808.\label{fig_data_i22491}}
\end{figure*}

The molecular gas traced by the CO(2--1) transition, which is dominated by the emission from the central $\sim1-2$\,kpc, has an irregular morphology with multiple large scale tidal tails (up to $\sim$10\,kpc) and isolated clumps. These characteristics are very likely connected to the ongoing galaxy interactions taking place in these systems.
Similar tidal tails are observed in the stellar component in the near-IR {\it HST}\slash NICMOS NIC2 images (right hand panel of Figures~\ref{fig_data_i12112}--\ref{fig_data_i22491}). However, there are noticeable offsets, $\sim1-2$\,kpc, between the position of these stellar and molecular tidal tails. 

To measure the total CO(2--1) emission of each system, we first defined the extent of this emission by selecting all the contiguous pixels where the CO(2--1) line peak is above 6$\sigma$ (see second panel of Figures~\ref{fig_data_i12112}, \ref{fig_data_i14348}, and \ref{fig_data_i22491}). Then, we integrated the line flux in this area. The resulting flux densities are presented in Table \ref{tbl_integrated}.
The flux strongly peaks at the nuclei of these objects, so we calculate an effective radius based on the area, $A$, which encloses half of the total CO(2--1) emission as $R_{\rm eff} = \sqrt{A/\pi}$. This $R_{\rm eff}$ provides a better estimate of the actual size of the CO(2--1) emission. For these galaxies, the effective radius varies between { 400\,pc and 1\,kpc} (see Table \ref{tbl_integrated}).

Both in \IRAS~12112 and \IRAS~22491, the CO(2--1) emission is completely dominated by one of the galaxies which produces 80\%, and 90\%, respectively, of the total flux of the merging system. In \IRAS~14348, the CO(2--1) emission is also dominated by one of the nuclei (SW), but this one is only two times brighter than the NE nucleus.

\begin{table}[t]
\caption{Integrated CO(2--1) emission}
\label{tbl_integrated}
\centering
\begin{small}
\begin{tabular}{lcccccccccccc}
\hline \hline
\\
Object & $S_{\rm CO}$\tablefootmark{a} & Total size\tablefootmark{b} & \multicolumn{2}{c}{$R_{\rm eff}$\tablefootmark{c}} \\
& (Jy\,km\,s$^{-1}$) & (kpc$^2$) & (arcsec) & (pc) \\
\hline\\[-2ex]

I12112 SW & 24.5 & 9.3 $\pm$ 0.3 & 0.31 & 430 $\pm$ 30 \\
I12112 NE & 117.2 & 16.5 $\pm$ 0.4 & 0.41 & 570 $\pm$ 30 \\
I14348 SW & 105.9 & 18.9 $\pm$ 0.5 & 0.50 & 780 $\pm$ 40 \\
I14348 NE & 53.3 & 19.6 $\pm$ 0.5 & 0.38 & 590 $\pm$ 40 \\
I22491 E & 59.4 & 8.3 $\pm$ 0.3 & 0.30 & 450 $\pm$ 30 \\
I22491 W & 4.1 & 7.0 $\pm$ 0.3 & 0.81 & 1100 $\pm$ 40 \\
\hline
\end{tabular}
\end{small}
\tablefoot{
\tablefoottext{a}{Total CO(2--1) flux. { The absolute flux uncertainty is $\sim$10\%}.}
\tablefoottext{b}{Size of the area where the CO(2--1) emission is detected at $>6\sigma$. { This is calculated as the number of pixels with emission $>6\sigma$ multiplied by the projected pixel area  on the sky. The uncertainties are the square root of the number of pixels times the projected pixel area.}}
\tablefoottext{c}{Effective radius of the region which encloses 50\%\ of the total CO(2--1) emission defined as $R_{\rm eff} = \sqrt{A/\pi}$ where $A$ is the area of this region.}
}
\end{table}

\subsubsection{248\,GHz continuum}\label{ss:continuum}

Except for the western nucleus of \IRAS~22491, which is not seen at 248\,GHz, the remaining nuclei are clearly detected in both the CO(2--1) and continuum images. In all the cases, the 248\,GHz continuum emission is produced by a relatively compact source. To accurately measure the continuum properties, we used the {\sc uvmultifit} library \citep{MartiVidal2014} within CASA. This library can simultaneously fit various models to the visibility data. First, we tried a 2D circular Gaussian model which provided a good fit for all the sources except two (I12112 NE and I22491 E). For these two sources, we added a delta function with the same center as a second component to account for the unresolved continuum emission. This second unresolved component represents $70-80$\% of the total continuum emission in these objects. The results of the fits are presented in Table~\ref{tbl_continuum} and the center coordinates listed in Table~\ref{tbl_sample} (see also Appendix~\ref{apx_uv_fit}). 

The resolved continuum emission (2D circular Gaussian component of the model) has a FWHM between 260 and 1000\,pc, which is more compact than the CO(2--1) emission. For comparison, the CO(2--1) effective radius is 3 to 6 times larger than the FWHM\slash 2 of this 2D Gaussian component. Only in I22491 E, both have similar sizes, although, in this galaxy, the continuum emission is dominated by the unresolved component. Therefore, in all these ULIRGs, the 248\,GHz continuum emission is considerably more compact than the molecular CO(2--1) emission. { This is similar to what is observed in other local LIRGs and ULIRGs (e.g., \citealt{Wilson2008, Sakamoto2014, Saito2017}).}

In \IRAS~12112 and \IRAS~22491, the continuum emission is dominated by the same nucleus that dominates the CO(2--1) emission (see Section \ref{ss:molgas}). The fraction of the 248\,GHz continuum produced by these dominant nuclei are 90\%\ and $>$95\%, respectively, which are slightly higher than their contributions to the total CO(2--1) luminosity of their systems (80\%\ and 90\%, respectively). In \IRAS~14348, the SW nucleus produces 60\%\ of the continuum emission and the NW the remaining 40\%. These fractions are similar to those of the CO(2--1) produced in each nucleus (65\%\ and 35\%, respectively).

The 248\,GHz continuum emission is possibly produced by a combination of thermal dust continuum, free-free radio continuum, and synchrotron emission. The latter can be dominant at this frequency in the case of AGN, and, as discussed in Section~\ref{s:sample}, AGN emission is detected in the three ULIRGs.
To determine the non-thermal contribution to the measured 248\,GHz fluxes, we use the available interferometric radio (1.49 and 8.44\,GHz) observations for these systems \citep{Condon1990, Condon1991}.
The position of the 248\,GHz continuum sources is compatible with the location of the 1.49 and 8.44\,GHz radio continuum emission within 0\farcs15 ($\sim$half of the beam FWHM). Therefore, we assume that the radio and the 248\,GHz emissions are produced in the same regions. We also note that the western nucleus of \IRAS~22491 is undetected as well at radio wavelengths.
For the rest of the sources, we fit a power-law to the 1.49 and 8.44\,GHz fluxes and obtain spectral indexes between 0.42 and 0.72. Then, we use these spectral indexes to extrapolate the non-thermal emission at 248\,GHz. On average, this represents 20\%\ of the 248\,GHz emission for these ULIRGs (see Table~\ref{tbl_continuum}), with a minimum (maximum) contribution of 14\% (43\%). Therefore, most of the 248\,GHz emission is likely due to thermal dust emission { and free-free radio continuum} produced in the compact nuclear region.

\begin{table*}[ht]
\caption{ALMA continuum models}
\label{tbl_continuum}
\centering
\begin{small}
\begin{tabular}{lcccccccccccc}
\hline \hline
\\
Object & Obs freq.\tablefootmark{a} & Rest freq.\tablefootmark{a} & Total flux\tablefootmark{b} & Delta\tablefootmark{c} & Gaussian\tablefootmark{c} & \multicolumn{2}{c}{FWHM\tablefootmark{d}} & Non-thermal\\
& (GHz) & (GHz) &  (mJy) & (mJy) & (mJy) & (arcsec) & (pc) & fraction\tablefootmark{e} \\
\hline\\[-2ex]
I12112 SW & 231.12 & 247.99 & 0.69 $\pm$ 0.05 & \nodata & 0.69 $\pm$ 0.05 & 0.19 $\pm$ 0.02 & 260 & 0.43 \\
I12112 NE &   &   & 6.81 $\pm$ 0.14 & 4.60 $\pm$ 0.10 & 2.21 $\pm$ 0.10 & 0.35 $\pm$ 0.02 & 480 & 0.20 \\
I14348 SW & 229.08 & 247.98 & 2.42 $\pm$ 0.05 & \nodata & 2.42 $\pm$ 0.05 & 0.17 $\pm$ 0.01 & 260 & 0.21 \\
I14348 NE &   &   & 1.63 $\pm$ 0.05 & \nodata & 1.63 $\pm$ 0.05 & 0.17 $\pm$ 0.01 & 270 & 0.21 \\
I22491 E & 229.64 & 247.50 & 5.16 $\pm$ 0.11 & 4.09 $\pm$ 0.06 & 1.07 $\pm$ 0.10 & 0.68 $\pm$ 0.08 & 1000 & 0.14 \\
I22491 W & & & $<$0.14$^{f}$ & \nodata  & \nodata & \nodata & \nodata & \nodata \\
\hline
\end{tabular}
\end{small}
\tablefoot{The flux uncertainties are statistical uncertainties from the fit. The absolute flux calibration uncertainty is about 10\%.
\tablefoottext{a}{Observed and rest frame continuum frequencies.}
\tablefoottext{b}{Total flux of the continuum model.}
\tablefoottext{c}{Flux of the delta (unresolved) and Gaussian components of the models.}
\tablefoottext{d}{Deconvolved FWHM of the Gaussian component.}
\tablefoottext{e}{Non-thermal emission fraction at 248\,GHz estimated from the radio 1.49 and 8.44\,GHz fluxes (\citealt{Condon1990, Condon1991}; see Section \ref{ss:continuum}).}
\tablefoottext{f}{3$\sigma$ flux upper limit for an unresolved source.}
}
\end{table*}

\subsection{Molecular gas kinematics}

\begin{figure*}
\centering
\includegraphics[width=0.48\textwidth]{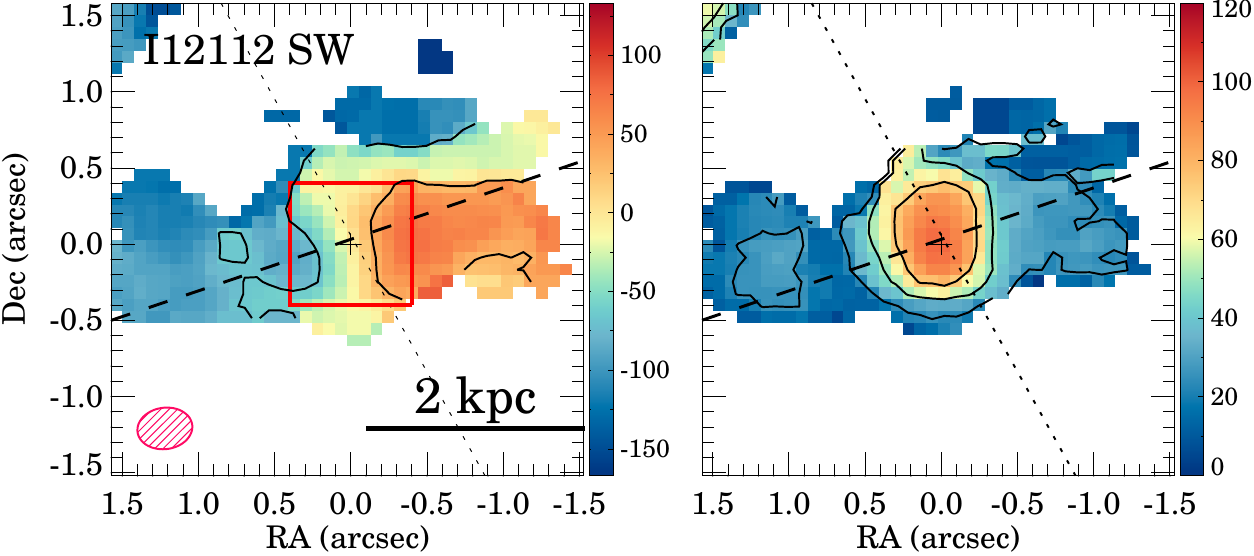}
\includegraphics[width=0.48\textwidth]{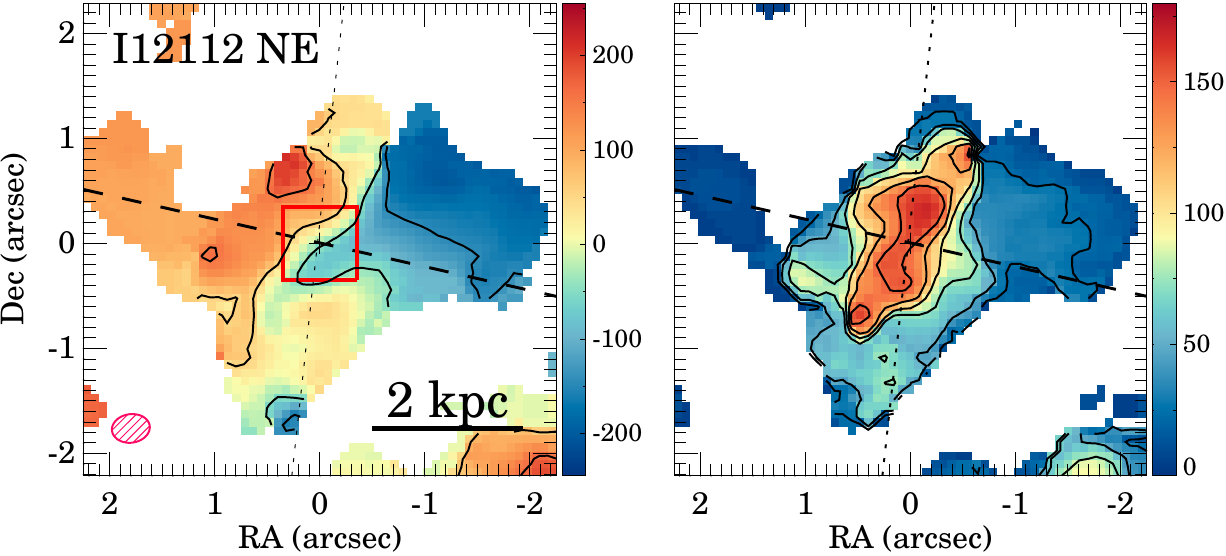}
\includegraphics[width=0.48\textwidth]{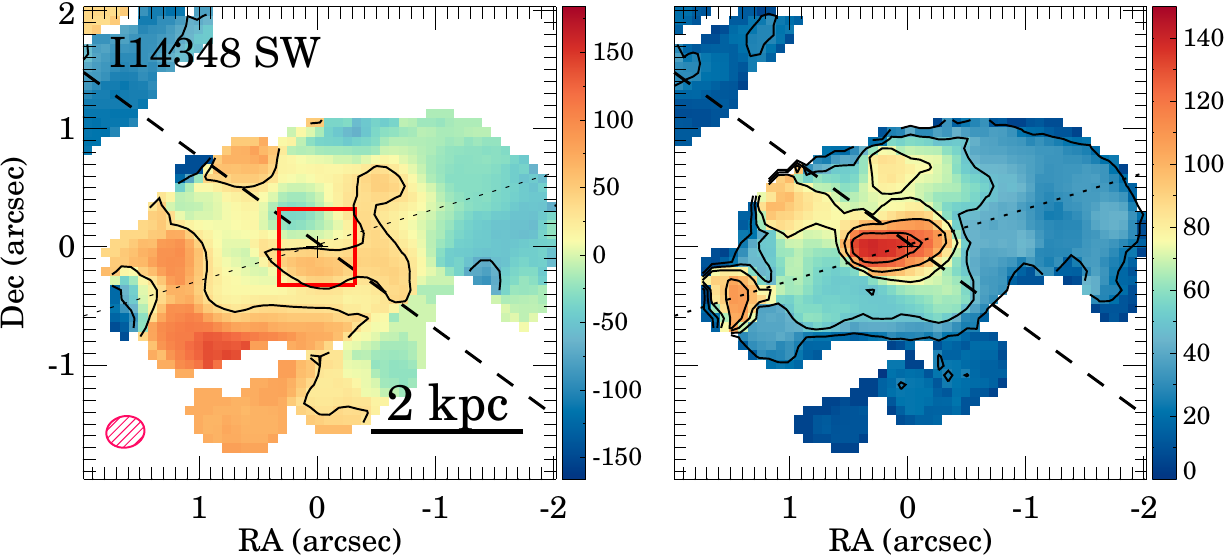}
\includegraphics[width=0.48\textwidth]{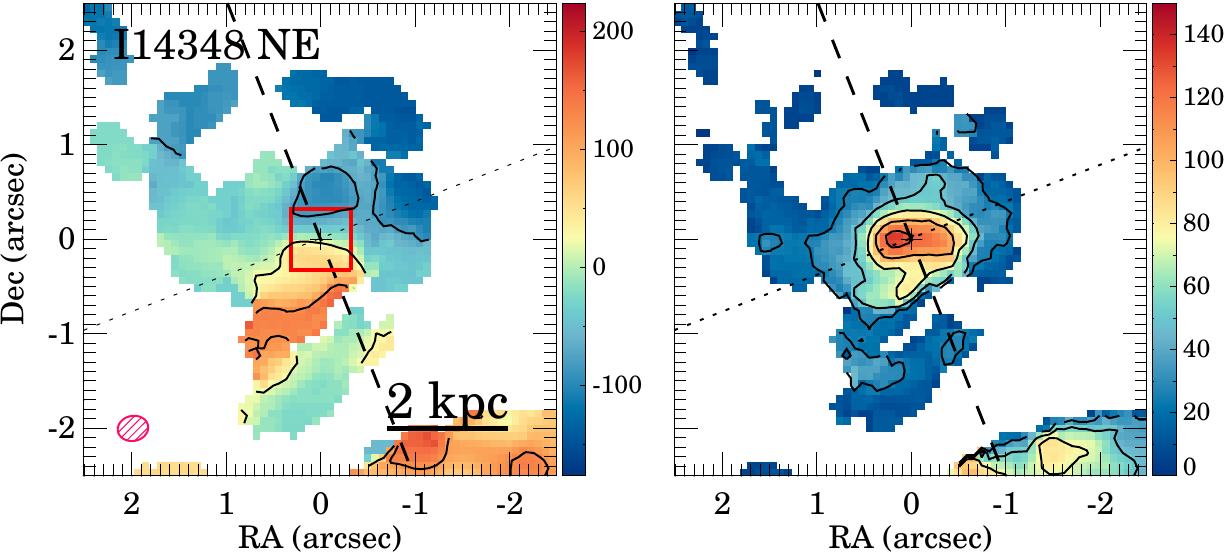}
\includegraphics[width=0.48\textwidth]{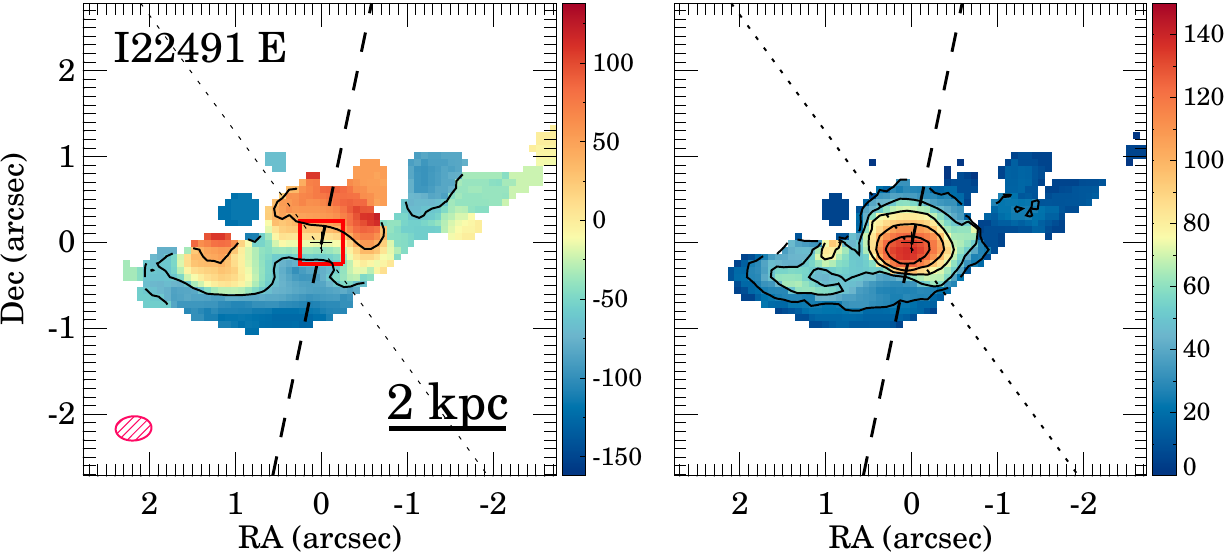}
\hspace{0.48\textwidth}
\caption{First (left panel) and second (right panel) moments of the CO(2--1) emission for each component of the ULIRGs. The spacings between the contour levels in the first and second moment maps are 100\,km\,s$^{-1}$ and 25\,km\,s$^{-1}$, respectively. For the first moment maps, the velocities are relative to the systemic velocity (see Table~\ref{tbl_sample}). The red box in this panel indicates the field of view presented in Figure~\ref{fig_alma_posdiagram} for each object. The dashed and dotted lines mark the kinematic major axis and the outflow axis, respectively, defined in Section \ref{ss:nuclear_kin} (see also Figure~\ref{fig_alma_posdiagram}). 
The black cross marks the position of the 248\,GHz continuum peak. The red hatched ellipse shows the beam FWHM { and PA}.
\label{fig_alma_vel}}
\end{figure*}

In Figure~\ref{fig_alma_vel}, we show the first and second moments of the CO(2--1) emission for each galaxy of the three ULIRG systems and indicate the outflow axis (dotted line) and the kinematic major axis of the nuclear disk (dashed line) { defined in  Section~\ref{ss:nuclear_kin} (see Figure~\ref{fig_alma_posdiagram})}.
The first moment maps indicate a complex velocity field, although a rotating disk pattern is present in all the systems. The second moment maps show that the velocity dispersion maximum ($120-170$\,km\,s$^{-1}$) is almost coincident with the location of the nucleus and that it is enhanced more or less along the molecular outflow axis (dotted line). The latter is expected since the high-velocity outflowing gas produces broad wings in the CO(2--1) line profile which enhance the observed second moment.

\subsubsection{Nuclear disks and molecular outflows}\label{ss:nuclear_kin}

\begin{table*}[ht]
\caption{Nuclear molecular { emission}}
\label{tbl_gaskin}
\centering
\begin{small}
\begin{tabular}{lcccccccccccc}
\hline \hline
\\
Object & PA\tablefootmark{a} & PA$_{\rm out}$\tablefootmark{b} & PA - PA$_{\rm out}$\tablefootmark{c} & v\tablefootmark{~d} & $\sigma$\tablefootmark{~e}  & v\slash $\sigma$\tablefootmark{~f}  & $i$\tablefootmark{~g} \\[0.1ex]
& (deg) & (deg) &  (deg) & (km\,s$^{-1}$) & (km\,s$^{-1}$) & & (deg) \\
\hline\\[-2ex]
I12112 SW & 289 $\pm$ 2 & 213 $\pm$ 10 & 75 $\pm$ 11 & 81 $\pm$ 6 & 130 $\pm$ 9 & 0.62 $\pm$ 0.09 & 25 $\pm$ 14 \\
I12112 NE & 80 $\pm$ 2 & 353 $\pm$ 5 & 87 $\pm$ 6 & 120 $\pm$ 8 & 168 $\pm$ 3 & 0.71 $\pm$ 0.06 & 28 $\pm$ 15 \\
I14348 SW & 232 $\pm$ 4 & 107 $\pm$ 8 & 126 $\pm$ 9 & 60 $\pm$ 10 & 148 $\pm$ 4 & 0.40 $\pm$ 0.09 & 15 $\pm$ 10 \\
I14348 NE & 202 $\pm$ 5 & 112 $\pm$ 7 & 90 $\pm$ 9 & 120 $\pm$ 40 & 138 $\pm$ 6 & 0.89 $\pm$ 0.30 & 36 $\pm$ 31 \\
I22491 E & 348 $\pm$ 2 & 36 $\pm$ 20 & 133 $\pm$ 20 & 110 $\pm$ 10 & 122 $\pm$ 5 & 0.88 $\pm$ 0.12 & 36 $\pm$ 23 \\
\hline
\end{tabular}
\end{small}
\tablefoot{
\tablefoottext{a, b}{Position angle (East of North) of the receding part of the kinematic major axis and the high-velocity outflowing gas, respectively (see Section \ref{ss:nuclear_kin} and Figure~\ref{fig_alma_posdiagram}).}
\tablefoottext{c}{Difference between the position angles of the outflow and the kinematic major axis.}
\tablefoottext{d}{Semi-amplitude of the observed CO(2--1) rotation curve.}
\tablefoottext{e}{Second moment of the nuclear CO(2--1) emission profile (Figures~\ref{fig_outflow_i12112_SW}--\ref{fig_outflow_i22491_E}).}
\tablefoottext{f}{Observed dynamical ratio.}
\tablefoottext{g}{{ Disk} inclination assuming an intrinsic dynamical ratio for local ULIRGs of 1.5 $\pm$ 0.6 (See Section \ref{ss:nuclear_kin}; \citealt{GarciaMarin06, Bellocchi2013}).}
}
\end{table*}

\begin{figure*}
\centering
\includegraphics[width=0.32\textwidth]{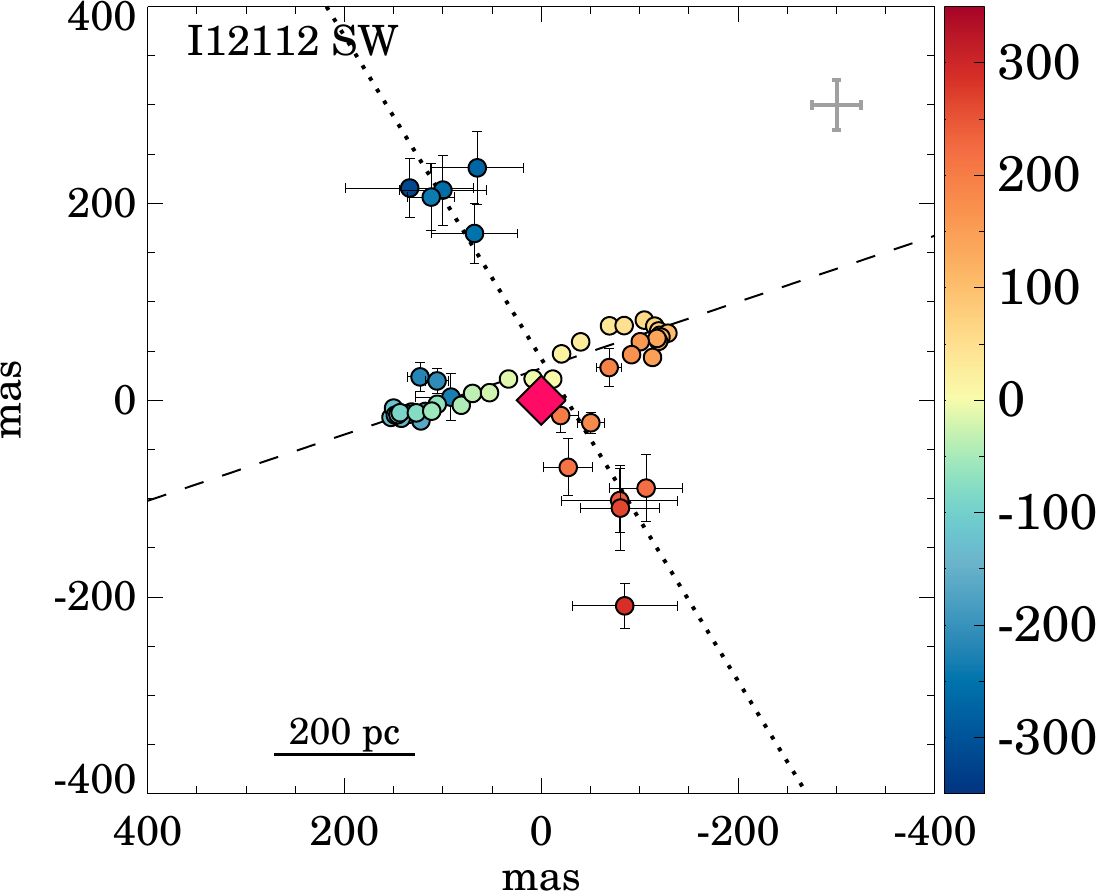}
\includegraphics[width=0.32\textwidth]{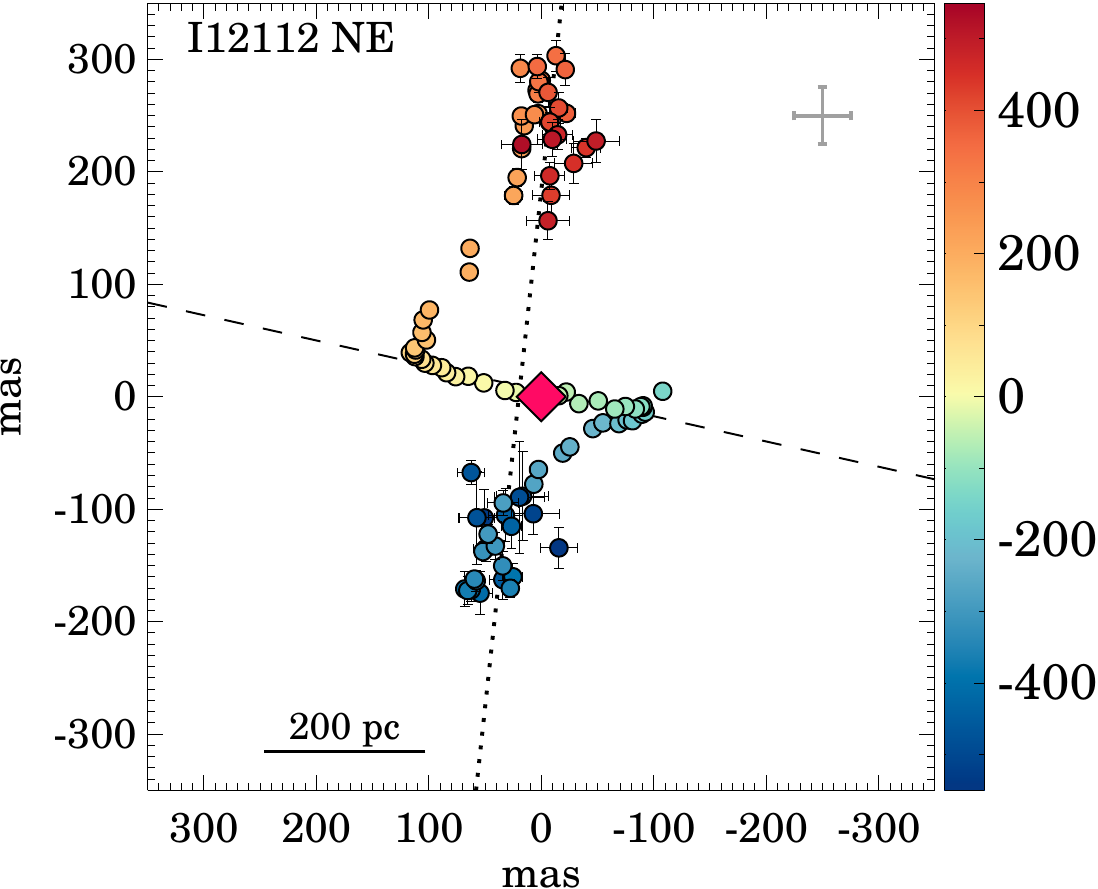}
\includegraphics[width=0.32\textwidth]{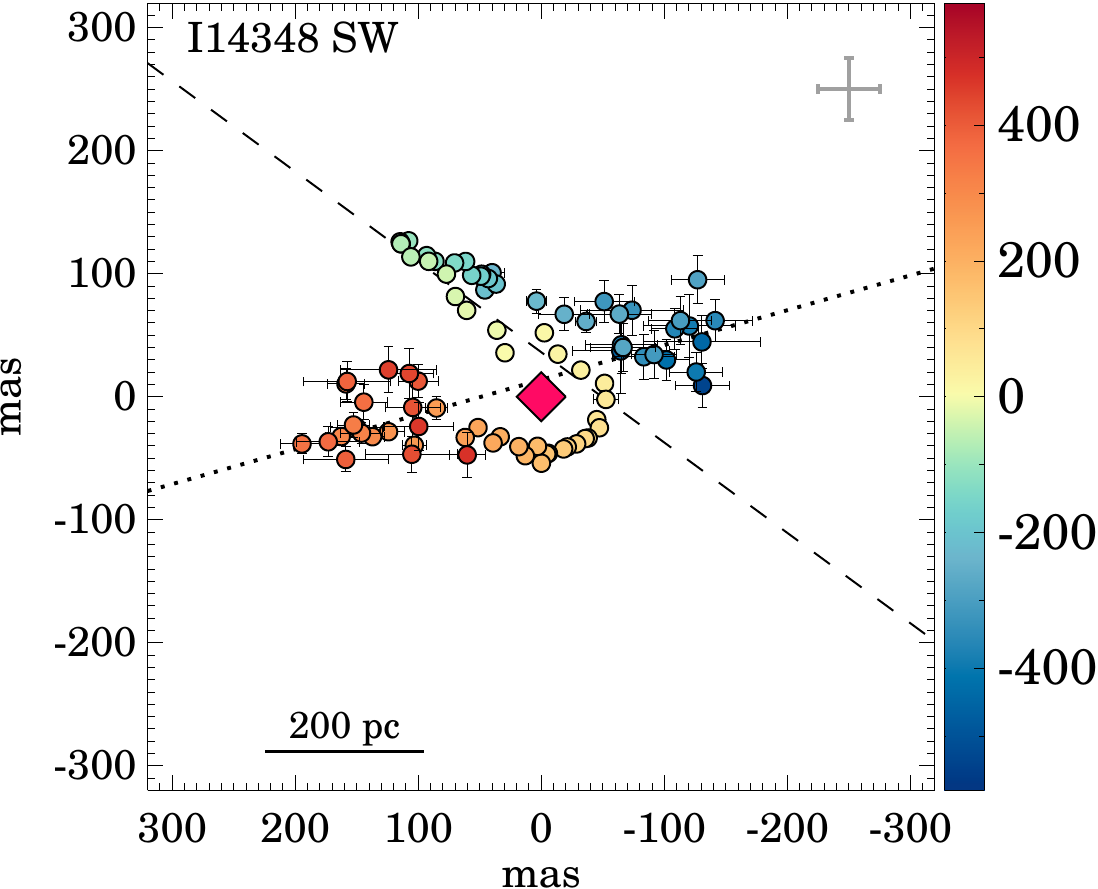}
\includegraphics[width=0.32\textwidth]{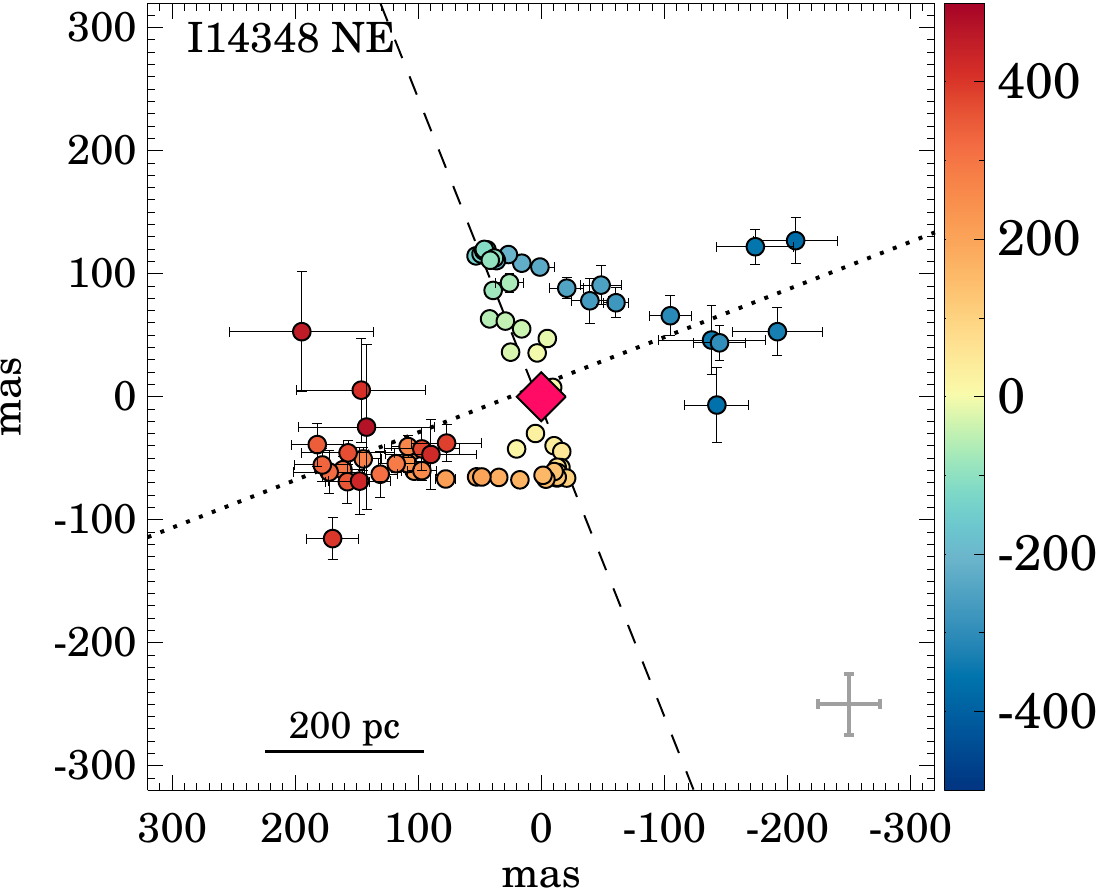}
\includegraphics[width=0.32\textwidth]{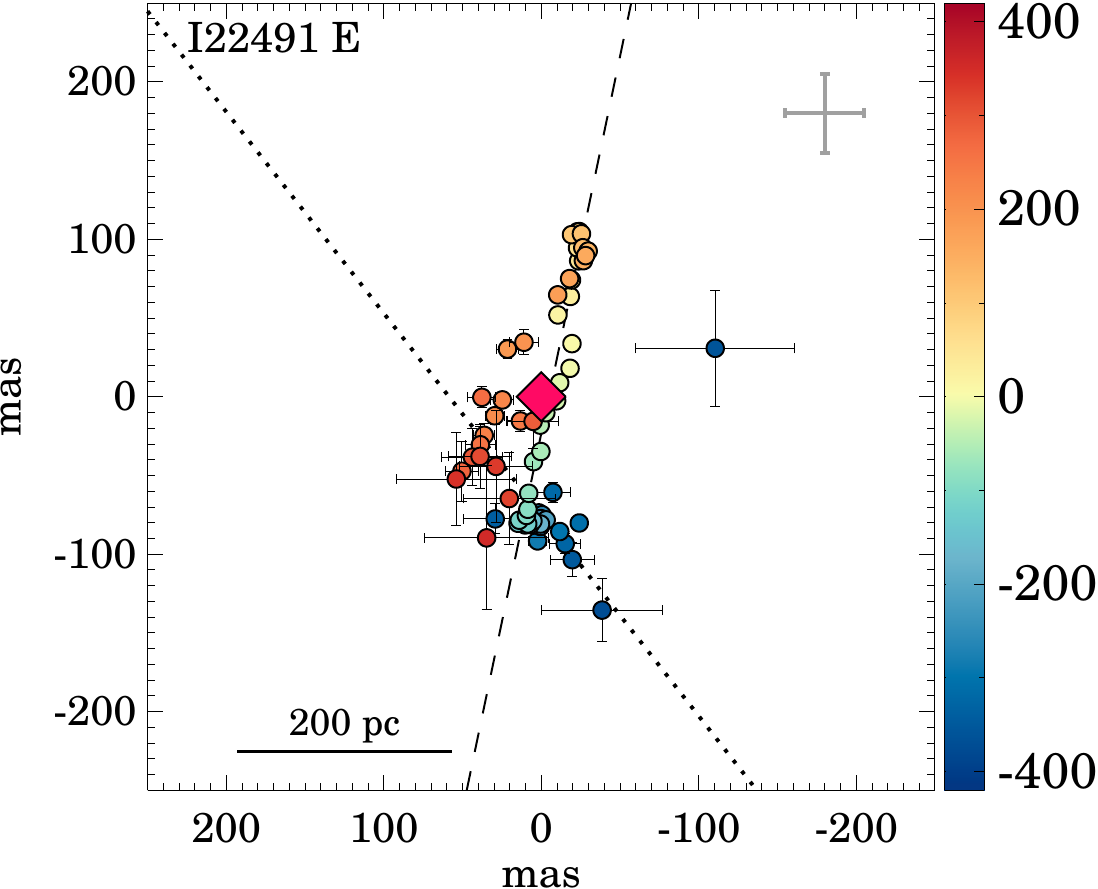}
\hspace{0.32\textwidth}
\caption{Centroid of the CO(2--1) emission measured in each { $\sim$10\,km\,s$^{-1}$} velocity channel. The color of the points indicates the CO(2--1) velocity with respect to the systemic velocity not corrected for inclination. The { rose} diamond marks the position of the 248\,GHz continuum peak. The dashed line is the linear fit to the low velocity gas and corresponds to the kinematic major axis of the rotating disk. The dotted line is the linear fit to the high-velocity red- and blue-shifted gas which traces the projection of the outflow axis in the sky. The error bars in each point indicate the statistical uncertainty in the centroid position. The gray error bars represent the astrometric accuracy of these observations for channels with SNR$>$10 (see Section~\ref{ss:nuclear_kin}).
\label{fig_alma_posdiagram}}
\end{figure*}

\begin{figure*}
\centering
\includegraphics[width=0.32\textwidth]{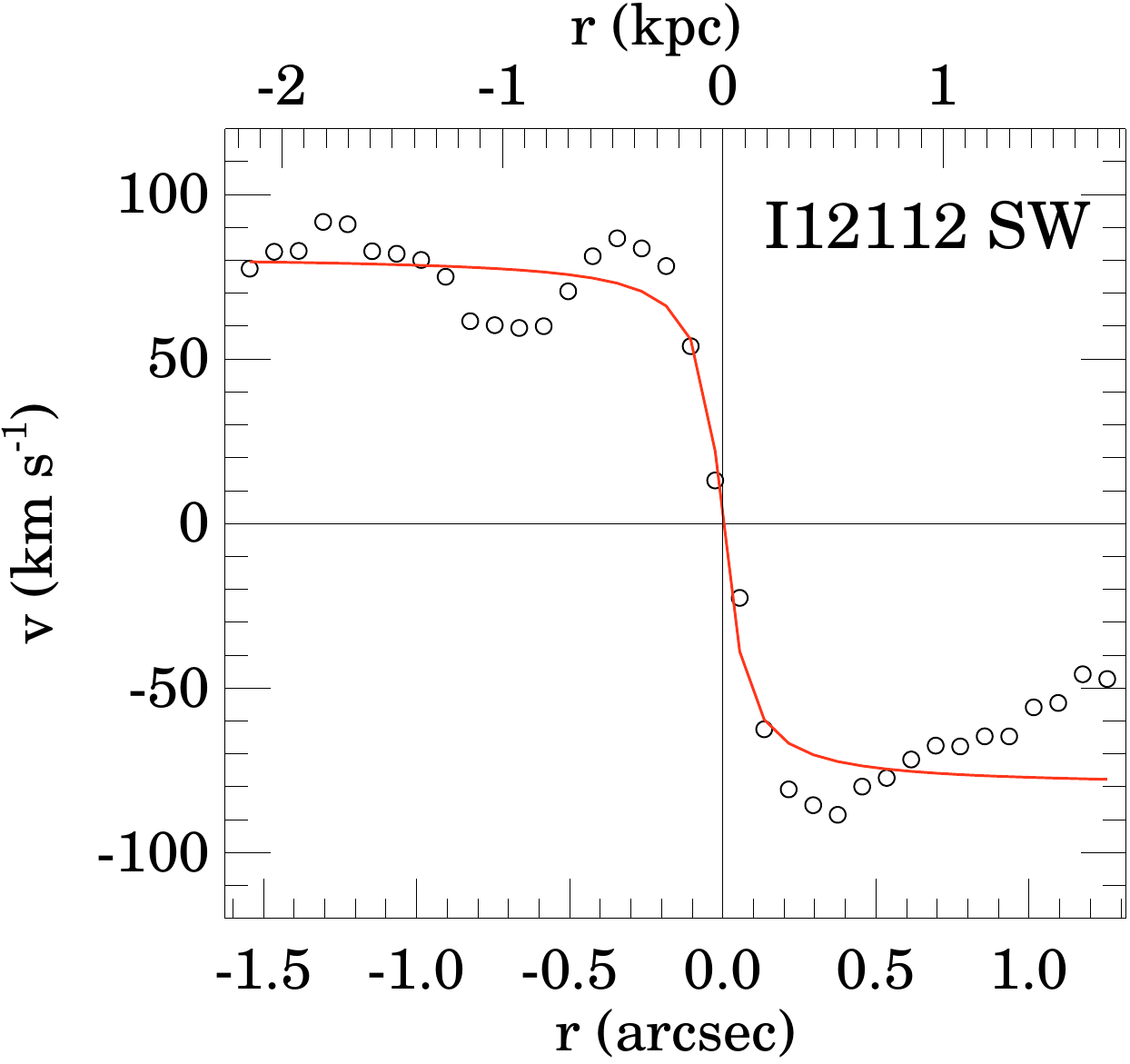}
\includegraphics[width=0.32\textwidth]{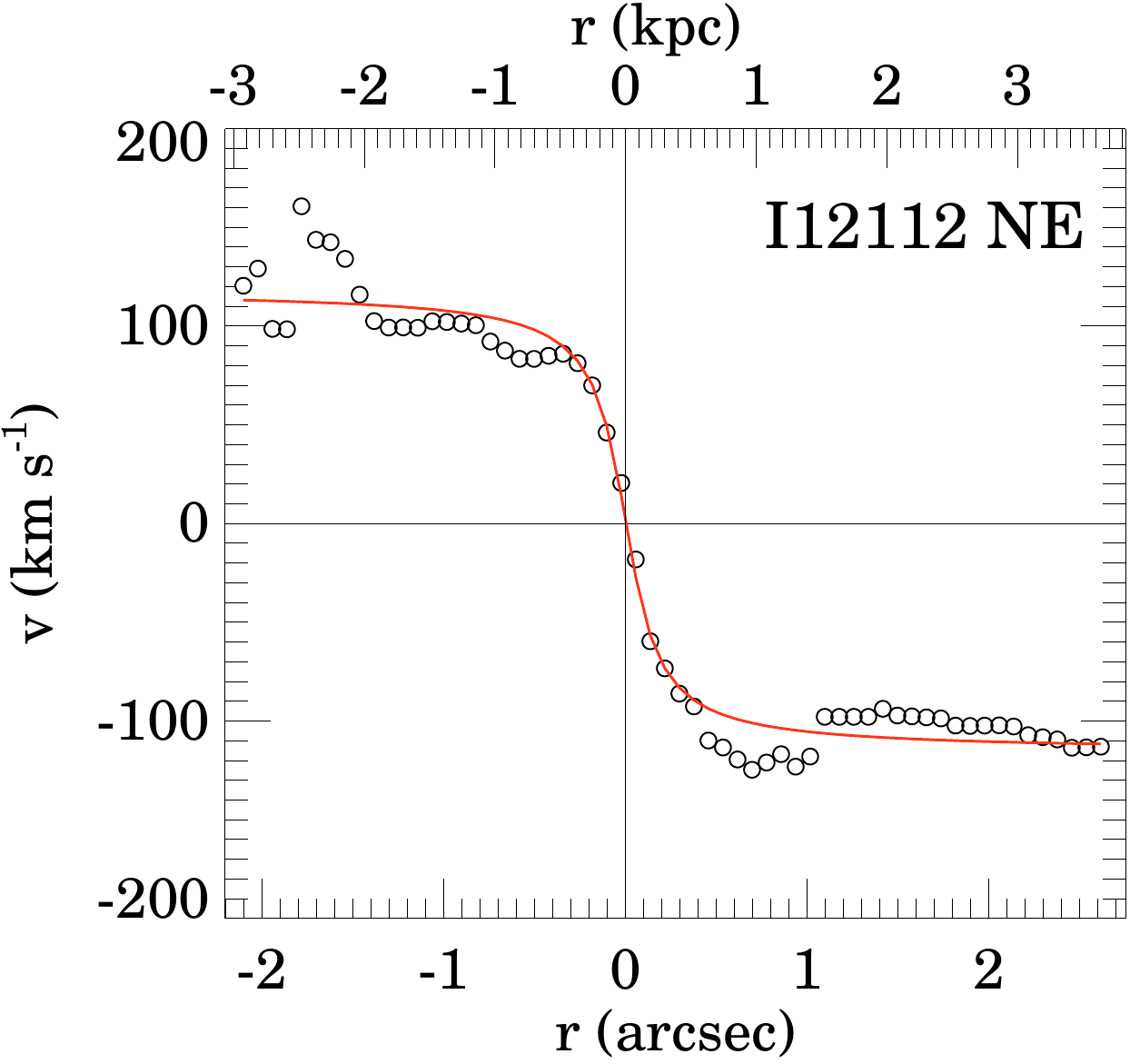}
\includegraphics[width=0.32\textwidth]{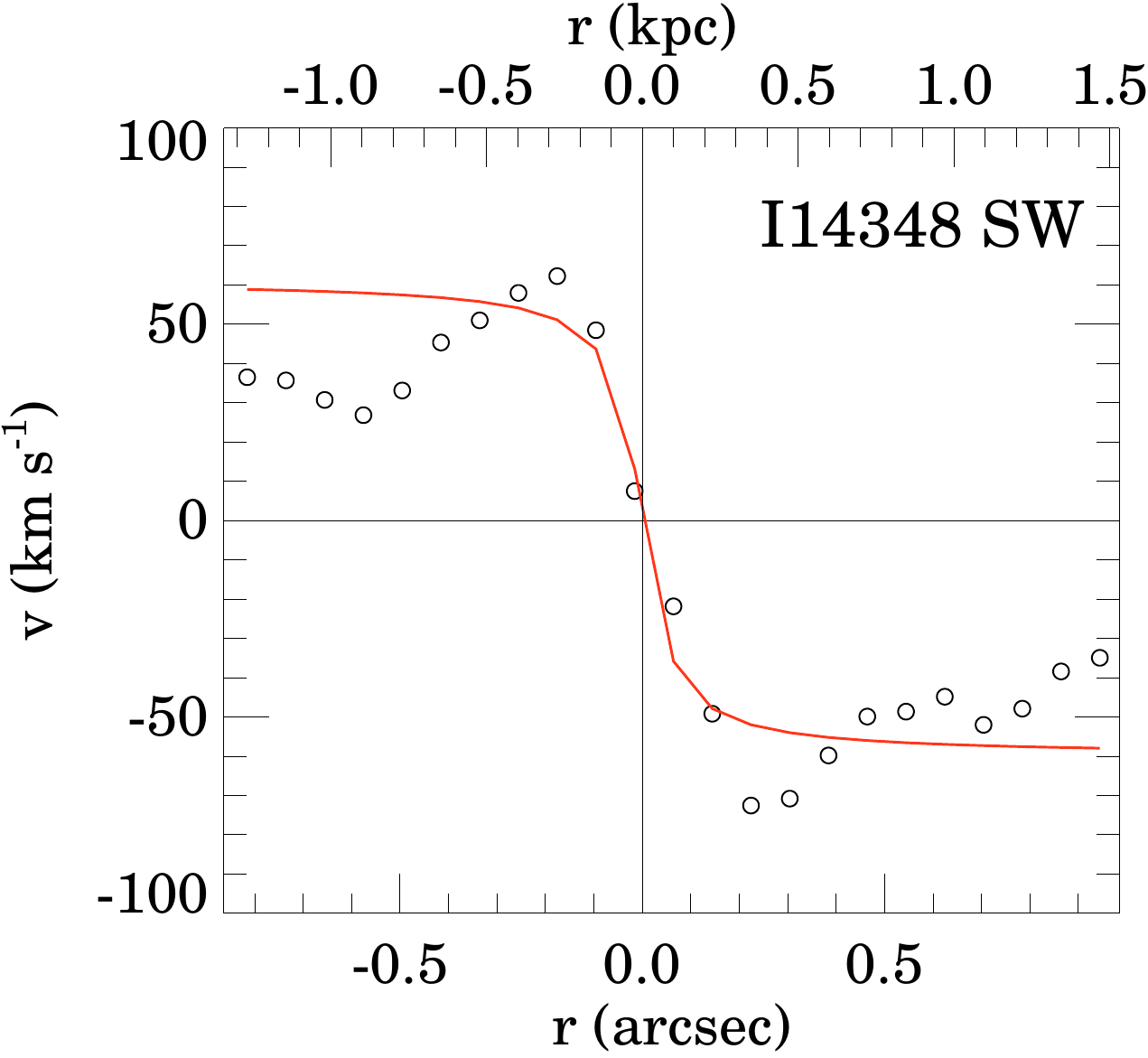}
\includegraphics[width=0.32\textwidth]{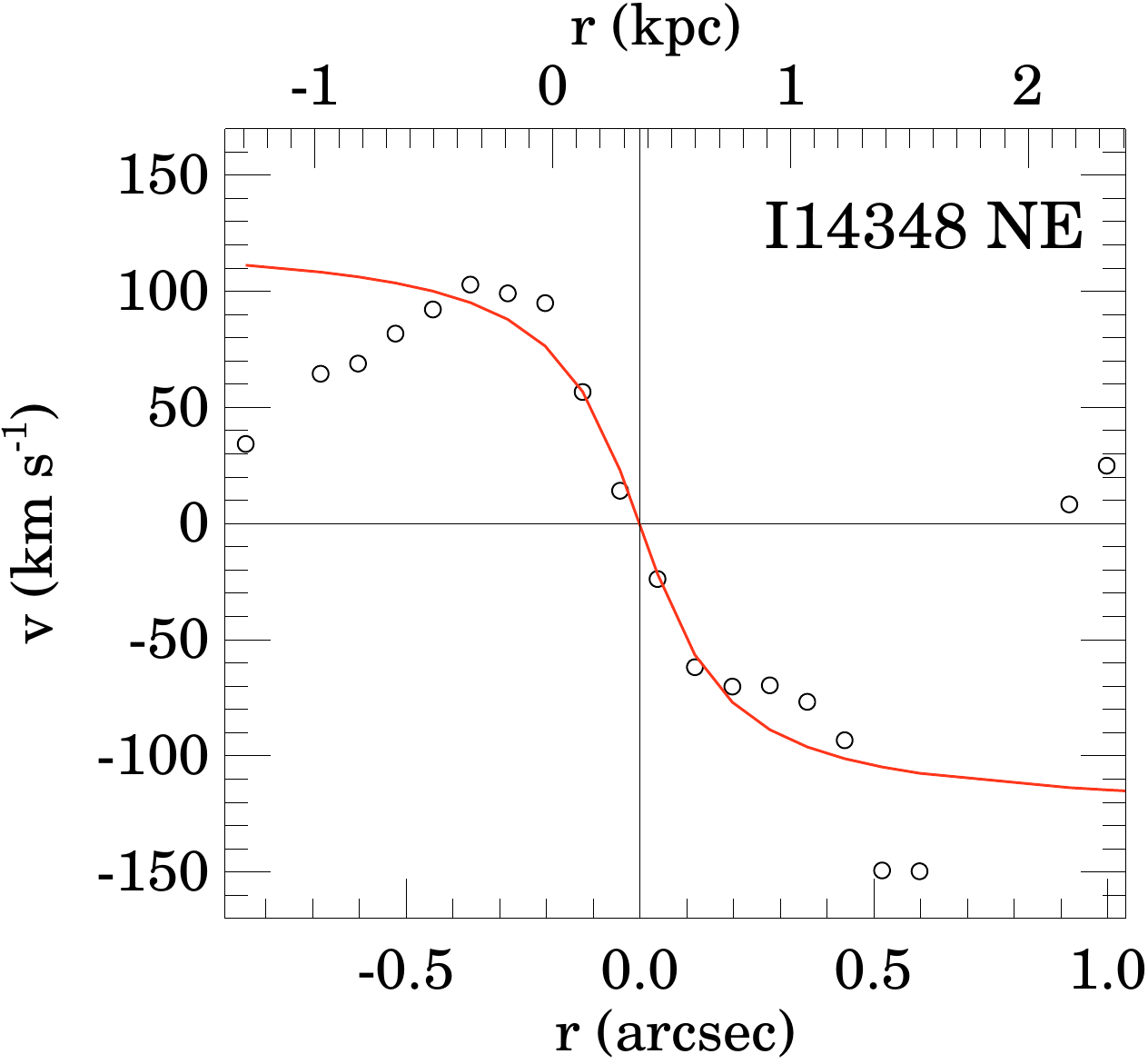}
\includegraphics[width=0.32\textwidth]{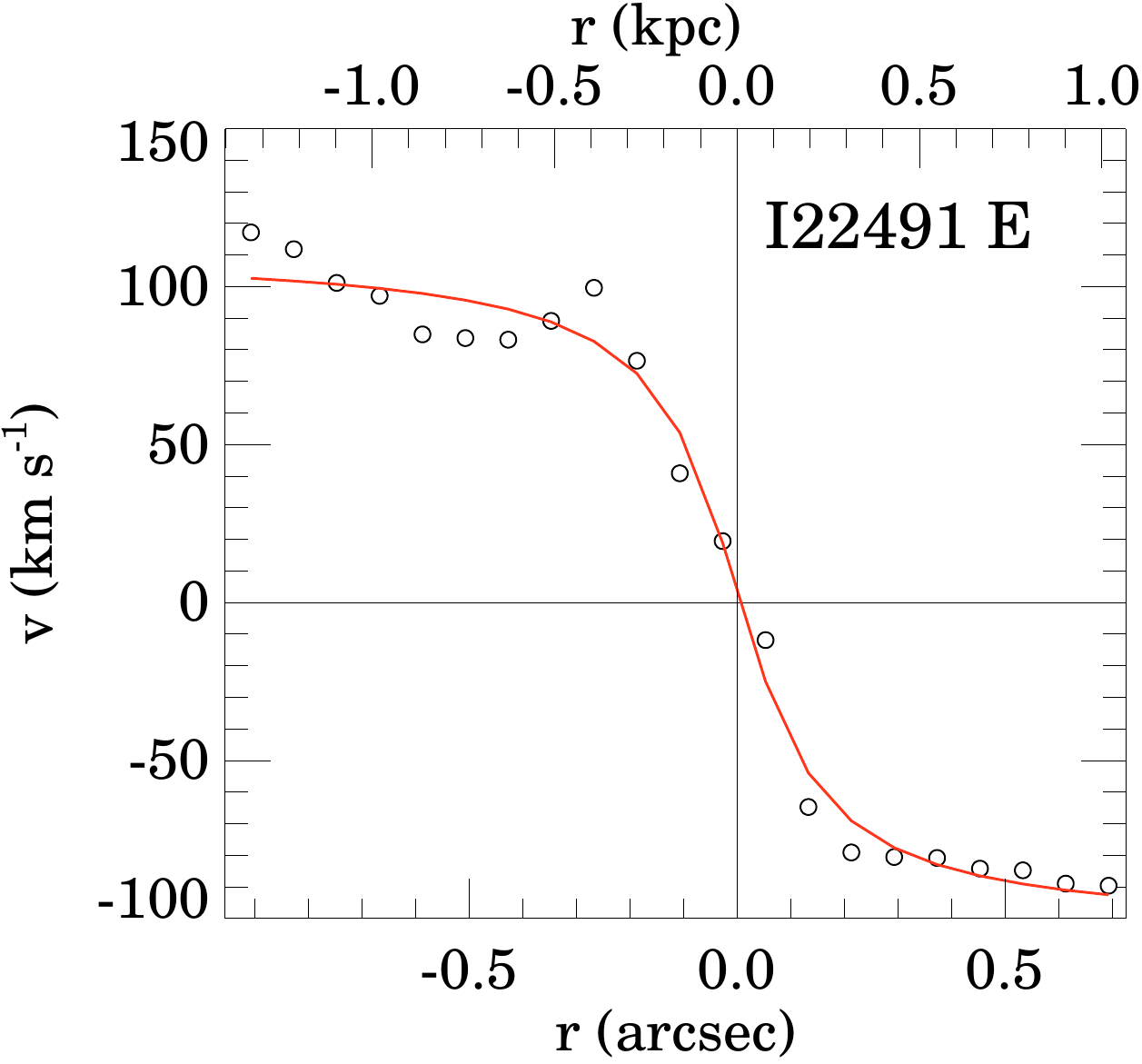}
\hspace{0.32\textwidth}
\caption{Rotation curves of the ULIRGs extracted along the kinematic major axis. The red line shows the best fit arctan model to the data. The fit results are presented in Table~\ref{tbl_gaskin}. \label{fig_rotcurve}}
\end{figure*}

Similar to \citet{GarciaBurillo2015}, we derive the centroid of the CO(2--1) emission in each velocity channel to study the nuclear gas kinematics and identify high-velocity gas decoupled from the rotating disks. The results are presented in Figure~\ref{fig_alma_posdiagram}. 
{ Thanks to the high SNR of the data, we are able to determine the centroid positions with statistical uncertainties $<$10\,mas for most of the channels. The astrometric accuracy for the frequency and array configuration of these observations is $\sim$25\,mas for channels with a SNR higher than 10. For channels with a SNR of $\sim$3, this accuracy is reduced to $\sim$80\,mas\footnote{ see Section 10.5.2 of the ALMA Cycle 6 Technical Handbook.}. Therefore, the shifts of the centroid positions shown in Figure~\ref{fig_alma_posdiagram} are expected to be real.}

For all the objects, the low-velocity emission centroids follow a straight line. This is consistent with the emission from a rotating disk which is not completely resolved. The direction of this line traces the position angle (PA) of the kinematic major axis of the rotating disk. Therefore, we did a linear fit to these points and derived the disk PA (Table~\ref{tbl_gaskin}). Also, using these fits, we determined the systemic velocity as the velocity of the point along the major axis closest to the continuum peak (Table~\ref{tbl_sample}).

By contrast, the centroids of the high-velocity gas do not lie on the kinematic major axis and they cluster at two positions, one for the red-shifted and the other for the blue-shifted gas. These two positions are approximately symmetric with respect to the nucleus. 
This is strong evidence of the decoupling of the high-velocity gas from the global disk rotation and, as we discuss in Section \ref{ss:mol_outflow}, this is compatible with the expected gas distribution of a massive molecular outflow originating at the nucleus. Alternatively, if we assume a coplanar geometry for the high-velocity gas, these PA twists could be explained by a strong nuclear bar-like structure. However, because of the extremely high radial velocities implied by this geometry, up to 400\,km\,s$^{-1}$, in the following, we only discuss the out-of plane outflow possibility.

The inclination of the disks is an important parameter to derive accurate outflow properties. It is commonly derived from the ratio between the photometric major and minor axes assuming that the galaxy is circular. However, in these merger systems, it is not obvious how to define these axes and also the circular morphology assumption might be incorrect.
Therefore, we use an alternative approach based on the kinematic properties of the nuclear disks to estimate the inclination. First, we extract the rotation curve along the kinematic major axis and fit an arctan model (e.g., \citealt{Courteau1997}) to determine the curve semi-amplitude, v (Figure~\ref{fig_rotcurve} and Table~\ref{tbl_gaskin}). Then, we measure the velocity dispersion, $\sigma$, of the nuclear region (1--2\,kpc) and calculate the observed dynamical ratio v$\slash\sigma$.
\citet{GarciaMarin06} and \citet{Bellocchi2013} measured the v$\slash\sigma$ ratios in a sample of 25 ULIRGs (34 individual galaxies) with H$\alpha$ integral field spectroscopy. Assuming a mean inclination of 57\degree ($\sin i = 0.79$; see \citealt{Law2009}), we can correct their v values for inclination and determine an intrinsic v$\slash\sigma$ ratio of 1.5$\pm$0.6 for ULIRGs.
Then, we compare the observed dynamical ratios in each galaxy with this intrinsic ratio to estimate their inclinations ($15-40$\degree).

\subsection{Properties of the molecular outflows}\label{ss:mol_outflow}

\subsubsection{Observed properties: PA, size, flux, and velocity}\label{ss:observed_properties}

\begin{table*}[ht]
\caption{Nuclear CO(2--1) emission and observed outflow properties}
\label{tbl_outflow_obs}
\centering
\begin{small}
\begin{tabular}{lcccccccccccc}
\hline \hline
\\
Object & $r$\tablefootmark{~a} & v range\tablefootmark{~b} & \multicolumn{3}{c}{$S_{\rm CO}$} & $|$v$_{\rm high}|$\tablefootmark{e} & v$_{\rm max}$\tablefootmark{f} & $R_{\rm c}$\tablefootmark{~g} & $R_{\rm max}$\tablefootmark{~h} \\[0.1ex]
\cline{4-6} \\[-1.9ex]
 & & & Total\tablefootmark{c} & Blue-shifted\tablefootmark{~d} & Red-shifted\tablefootmark{~d}  & & & &  \\
& (arcsec [kpc]) &(km\,s$^{-1}$) &  (Jy\,km\,s$^{-1}$) & (Jy\,km\,s$^{-1}$) & (Jy\,km\,s$^{-1}$) & (km\,s$^{-1}$) & (km\,s$^{-1}$) & (arcsec) & (arcsec) \\
\hline\\[-2ex]
I12112 SW& 0.55\,[0.8] & [230, 550]& 19.5 $\pm$ 0.1 & 0.25 $\pm$ 0.04 & 0.32 $\pm$ 0.04 & 324 $\pm$ 12 & 470 $\pm$ 20 & 0.24 $\pm$ 0.03 & 0.55 $\pm$ 0.05 \\
I12112 NE& 1.3\,[1.8] & [220, 900]& 115.9 $\pm$ 0.2 & { 5.4} $\pm$ 0.1 & 5.3 $\pm$ 0.1 & 465 $\pm$ 30 & 800 $\pm$ 90 & 0.24 $\pm$ 0.03 & 1.15 $\pm$ 0.30 \\
I14348 SW& 1.1\,[1.7] & [230, 800]& 97.7 $\pm$ 0.2 & 3.5 $\pm$ 0.1 & 4.1 $\pm$ 0.1 & 419 $\pm$ 38 & 740 $\pm$ 30 & 0.18 $\pm$ 0.03 & 1.05 $\pm$ 0.05 \\
I14348 NE& 0.7\,[1.1] & [240, 560]& 42.1 $\pm$ 0.1 & 0.48 $\pm$ 0.05 & 1.0 $\pm$ 0.1 & 373 $\pm$ 5 & 520 $\pm$ 110 & 0.22 $\pm$ 0.03 & 0.60 $\pm$ 0.05 \\
I22491 E& 1.0\,[1.5] & [210, 600]& 57.4 $\pm$ 0.1 & 1.1 $\pm$ 0.1 & 0.9 $\pm$ 0.1 & 325 $\pm$ 33 & 400 $\pm$ 110 & 0.10 $\pm$ 0.01 & 0.35 $\pm$ 0.05 \\
\hline
\end{tabular}
\end{small}
\tablefoot{
\tablefoottext{a}{Radius of the aperture used to measure the CO(2--1) emission in arcsec and kpc.}
\tablefoottext{b}{Absolute value of the velocity range considered to measure the blue- and red-shifted wings of the CO(2--1) profile with respect to the systemic velocity.}
\tablefoottext{c}{Total CO(2--1) flux measured within an aperture of radius $r$ centered on the object nucleus.}
\tablefoottext{d}{Blue- and red-shifted high-velocity CO(2--1) emission measured in the indicated velocity range { after subtracting the best-fit 2-Gaussian model (see Section~\ref{ss:observed_properties} and Figures~\ref{fig_outflow_i12112_SW}$-$\ref{fig_outflow_i22491_E})}.}
\tablefoottext{e}{Absolute value of the intensity weighted average velocity of the high-velocity gas with respect to the systemic velocity.}
\tablefoottext{f}{Maximum velocity at which CO(2--1) emission is detected at more than 3$\sigma$.}
\tablefoottext{g}{Half of the maximum separation between the centroids of the blue- and red-shifted high-velocity emission with the same $|$v$-$v$_{\rm sys}|$ (see Figure~\ref{fig_alma_posdiagram}).}
\tablefoottext{h}{Largest radius at which high-velocity CO(2--1) emission is detected.}
}
\end{table*}

\begin{figure*}
\centering
\hfill
\includegraphics[width=0.46\textwidth]{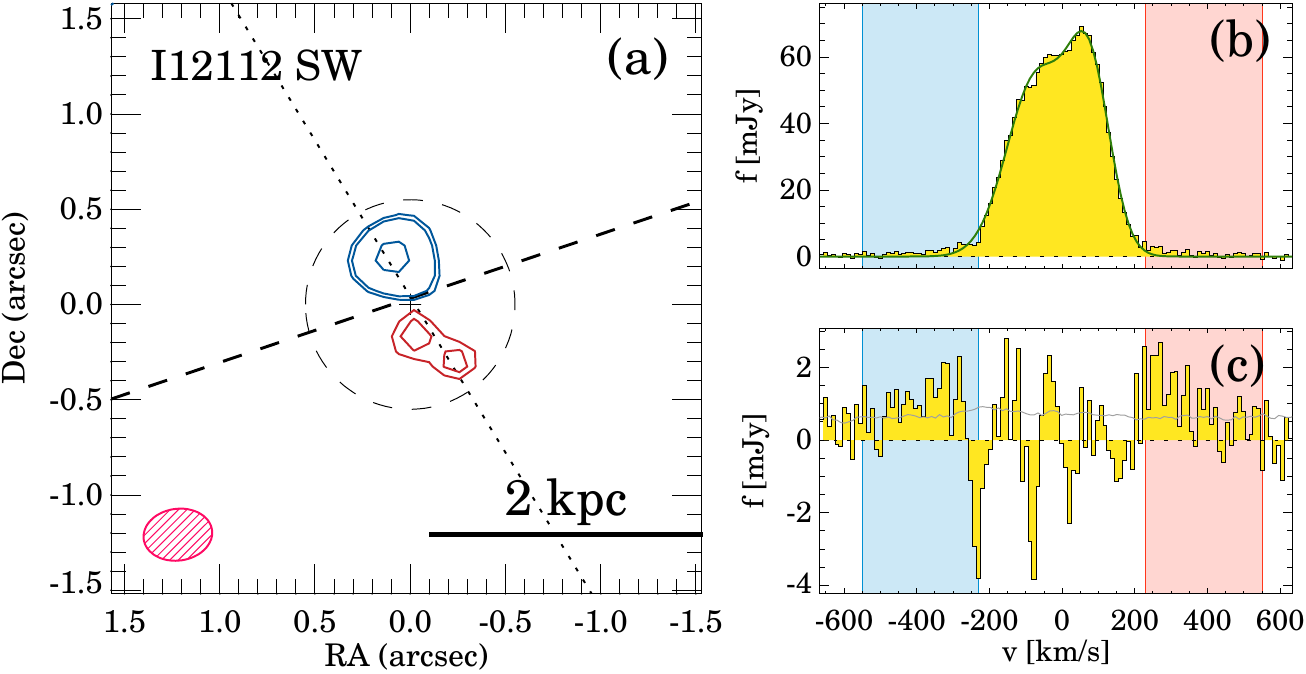}
\hfill
\includegraphics[width=0.48\textwidth]{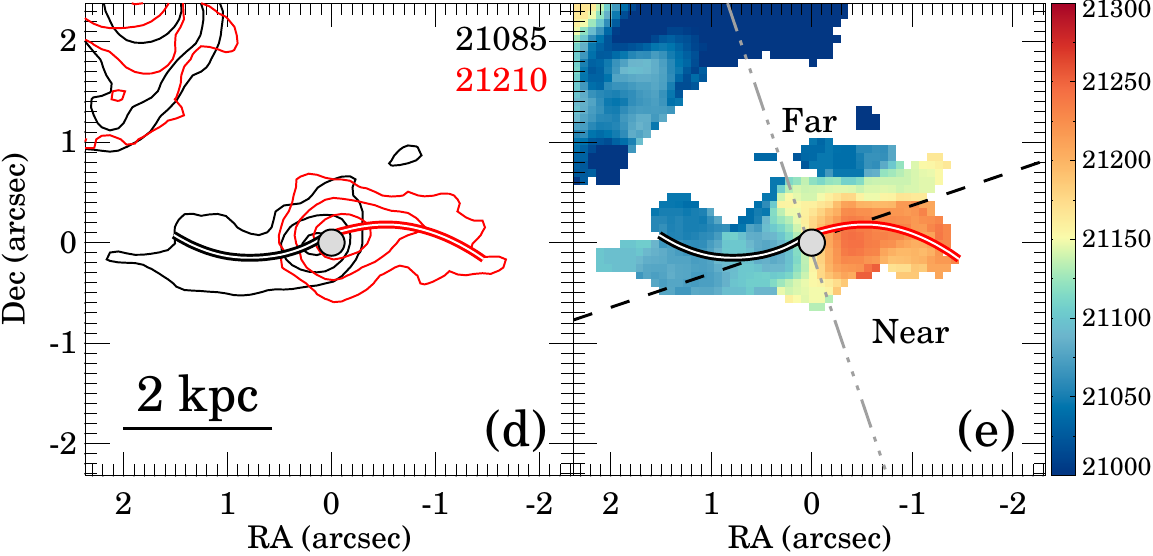}
\hfill
\caption{The blue and red contours in panel \textit{a} represent the integrated blue- and red-shifted high-velocity CO(2--1) emission, respectively. The specific velocity range is listed in Table~\ref{tbl_outflow_obs}. { The lowest contour corresponds to the 3$\sigma$ level. The next contour levels are (0.5, 0.9)$\times$ the peak of the high-velocity emission when these are above the 3$\sigma$ level. For I12112~SW, $\sigma=$\,30\,mJy\,km\,s$^{-1}$\,beam$^{-1}$ and the red and blue peaks are at 110 and 240\,mJy\,km\,s$^{-1}$\,beam$^{-1}$, respectively.} The dotted and the dashed lines are the outflow axis and the kinematic major axis, respectively (see Table~\ref{tbl_gaskin}). The red hatched ellipse represents the beam FWHM { and PA}. The dashed circle indicates the region from which the nuclear spectrum was extracted. 
Panel \textit{b} shows the nuclear spectrum in yellow and the { best-fit} model in gray. 
Panel \textit{c} shows the difference between the observed spectrum and the { best-fit} model. The shaded blue and red velocity ranges in these panels correspond to the velocity ranges used for the contours of panel \textit{a}. The gray line in panel \textit{c} marks the 3$\sigma$ noise level per channel.
Panel \textit{d} shows the CO(2--1) emission at the velocities indicated by the numbers at the top-right corner of the panel as black and red contours, respectively. The black and red double lines trace the morphological features (spiral arms, tidal tails) observed at those velocities. { Panel \textit{e} shows the CO(2--1) mean velocity field (same as in Figure~\ref{fig_alma_vel})}. The black dashed line is the kinematic major axis and the gray dot-dashed line the kinematic minor axis. The far and near sides of the rotating disk are indicated, assuming that the observed morphological features are trailing.\label{fig_outflow_i12112_SW}}
\end{figure*}

\begin{figure*}
\centering
\hfill
\includegraphics[width=0.46\textwidth]{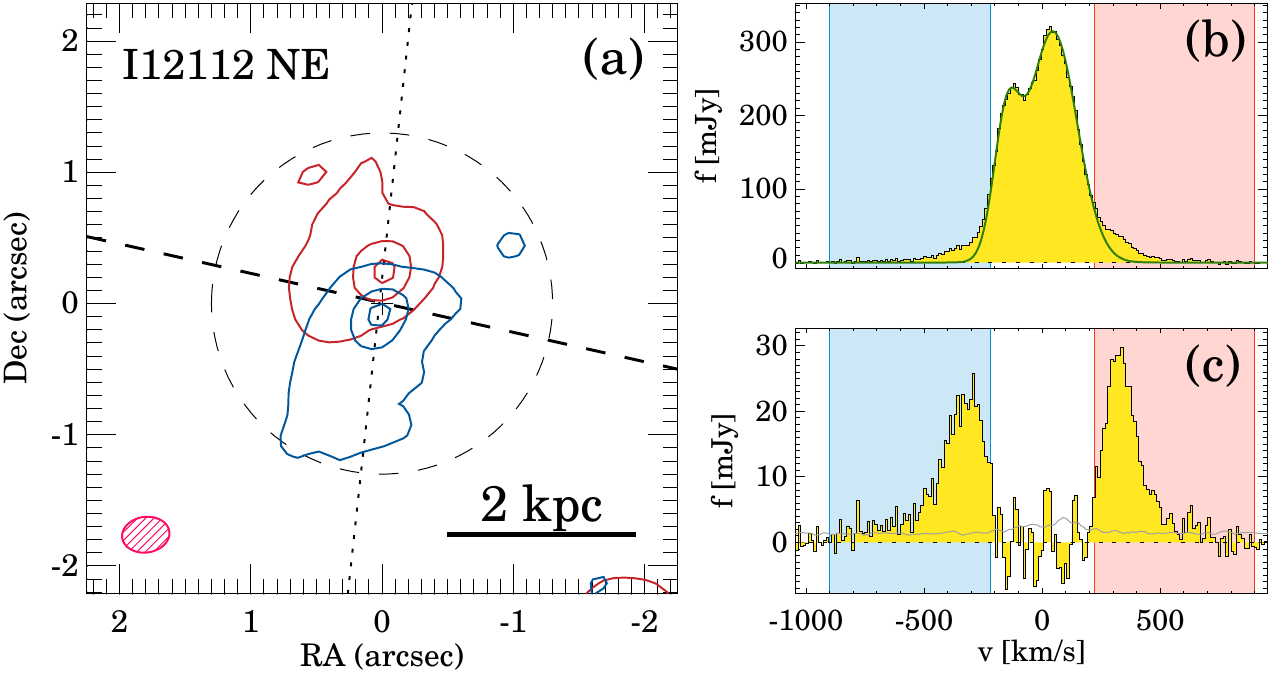}
\hfill
\includegraphics[width=0.48\textwidth]{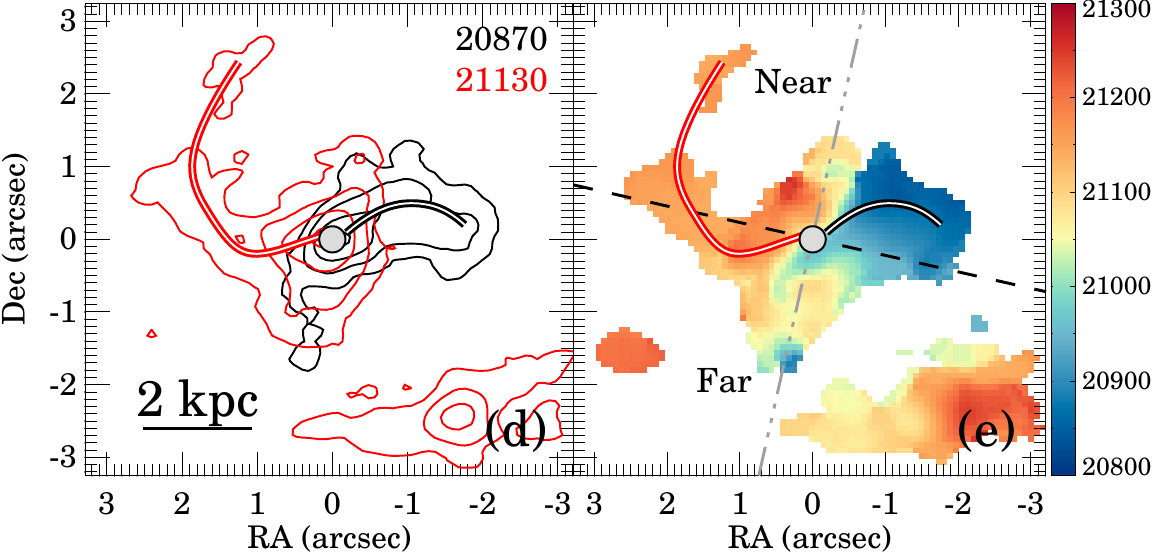}
\hfill
\caption{Same as Figure~\ref{fig_outflow_i12112_SW} but for I12112 NE. 
{ In panel \textit{a}, $\sigma=$\,45\,mJy\,km\,s$^{-1}$\,beam$^{-1}$ and the red and blue peaks are at 2.4 and 2.9\,Jy\,km\,s$^{-1}$\,beam$^{-1}$, respectively.}
\label{fig_outflow_i12112_NE}}
\end{figure*}

\begin{figure*}
\centering
\hfill
\includegraphics[width=0.46\textwidth]{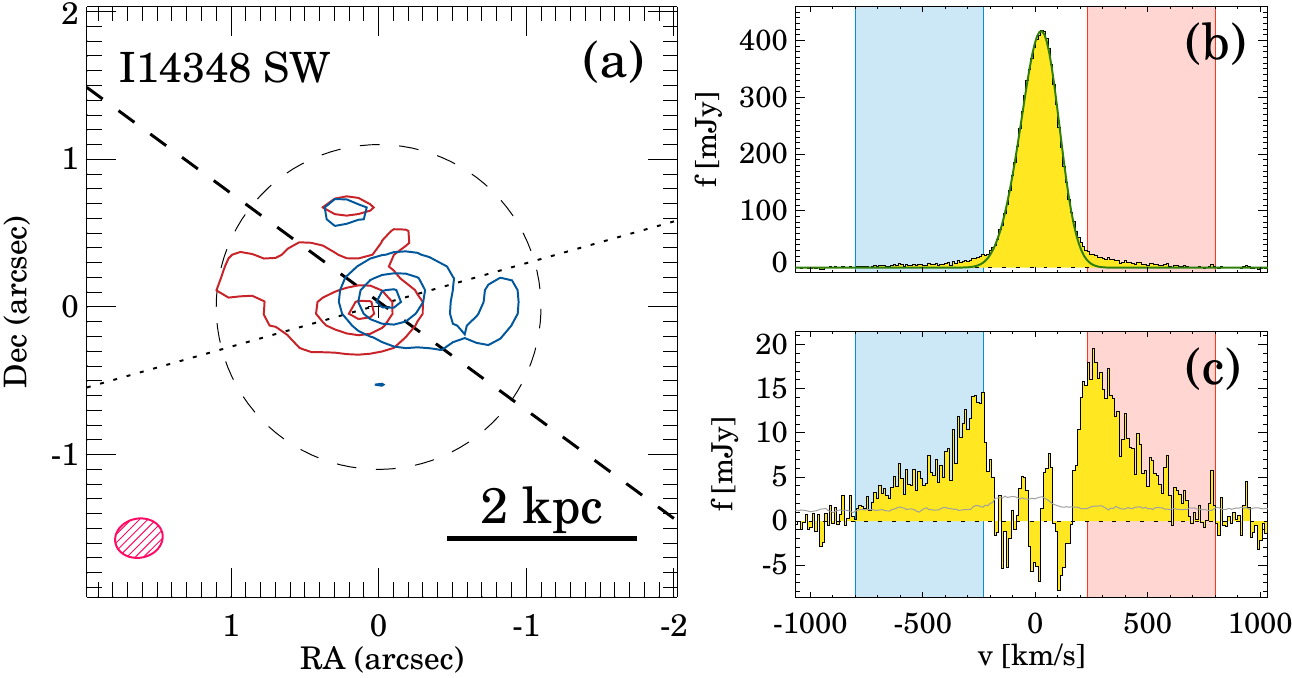}
\hfill
\includegraphics[width=0.48\textwidth]{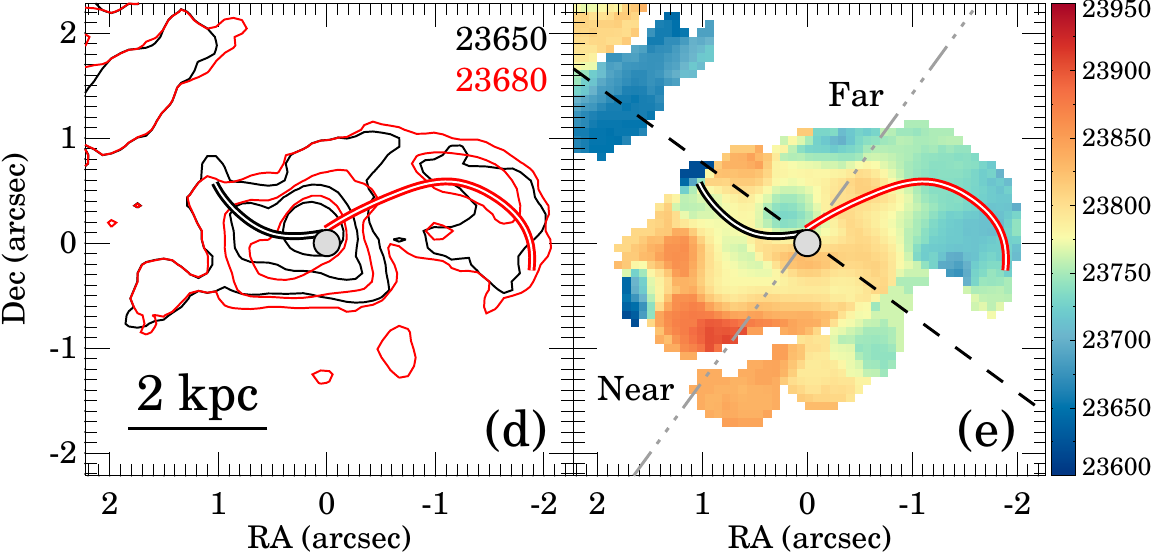}
\hfill
\caption{Same as Figure~\ref{fig_outflow_i12112_SW} but for I14348 SW. 
{ In panel \textit{a}, $\sigma=$\,46\,mJy\,km\,s$^{-1}$\,beam$^{-1}$ and the red and blue peaks are at 1.4 and 1.0\,Jy\,km\,s$^{-1}$\,beam$^{-1}$, respectively.}
\label{fig_outflow_i14348_SW}}
\end{figure*}

\begin{figure*}
\centering
\hfill
\includegraphics[width=0.46\textwidth]{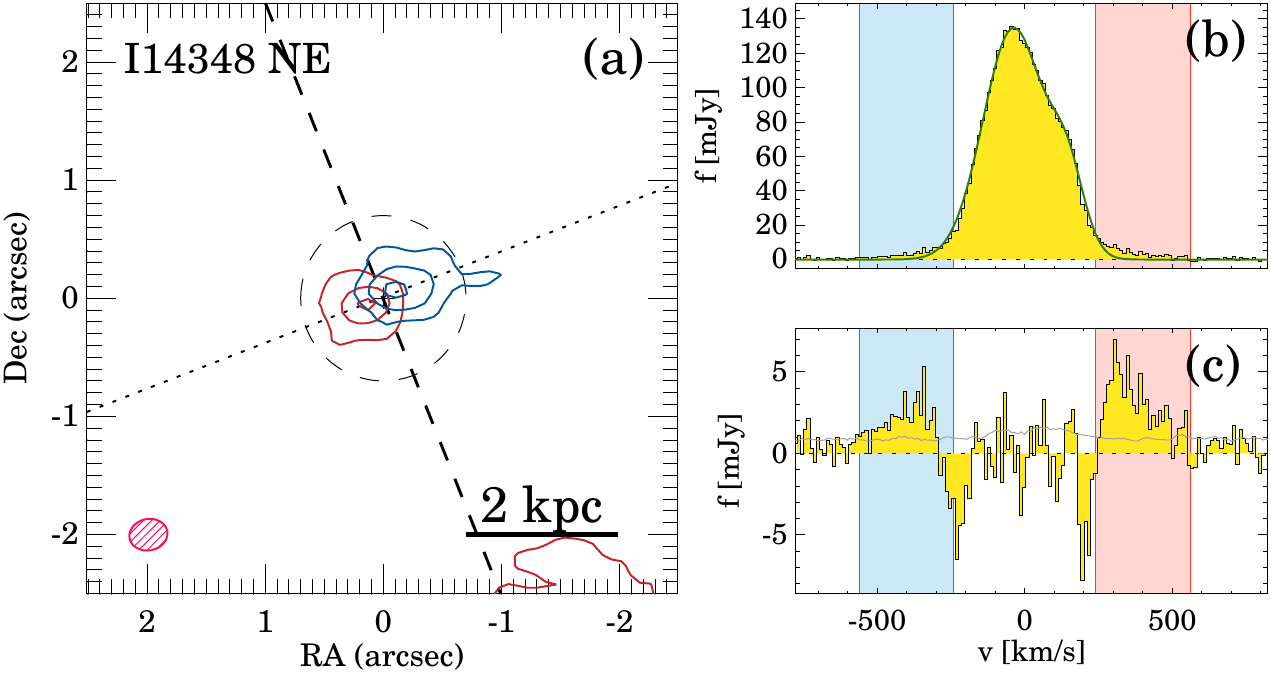}
\hfill
\includegraphics[width=0.48\textwidth]{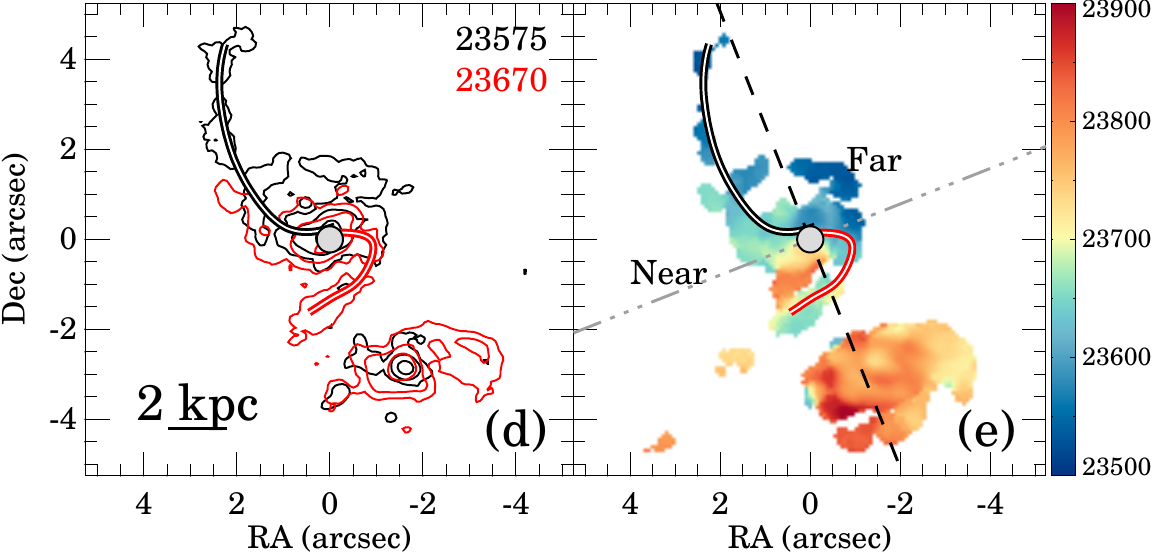}
\hfill
\caption{Same as Figure~\ref{fig_outflow_i12112_SW} but for I14348 NE. 
{ In panel \textit{a}, $\sigma=$\,32\,mJy\,km\,s$^{-1}$\,beam$^{-1}$ and the red and blue peaks are at 0.74 and 0.56\,Jy\,km\,s$^{-1}$\,beam$^{-1}$, respectively.}
\label{fig_outflow_i14348_NE}}
\end{figure*}

\begin{figure*}
\centering
\hfill
\includegraphics[width=0.46\textwidth]{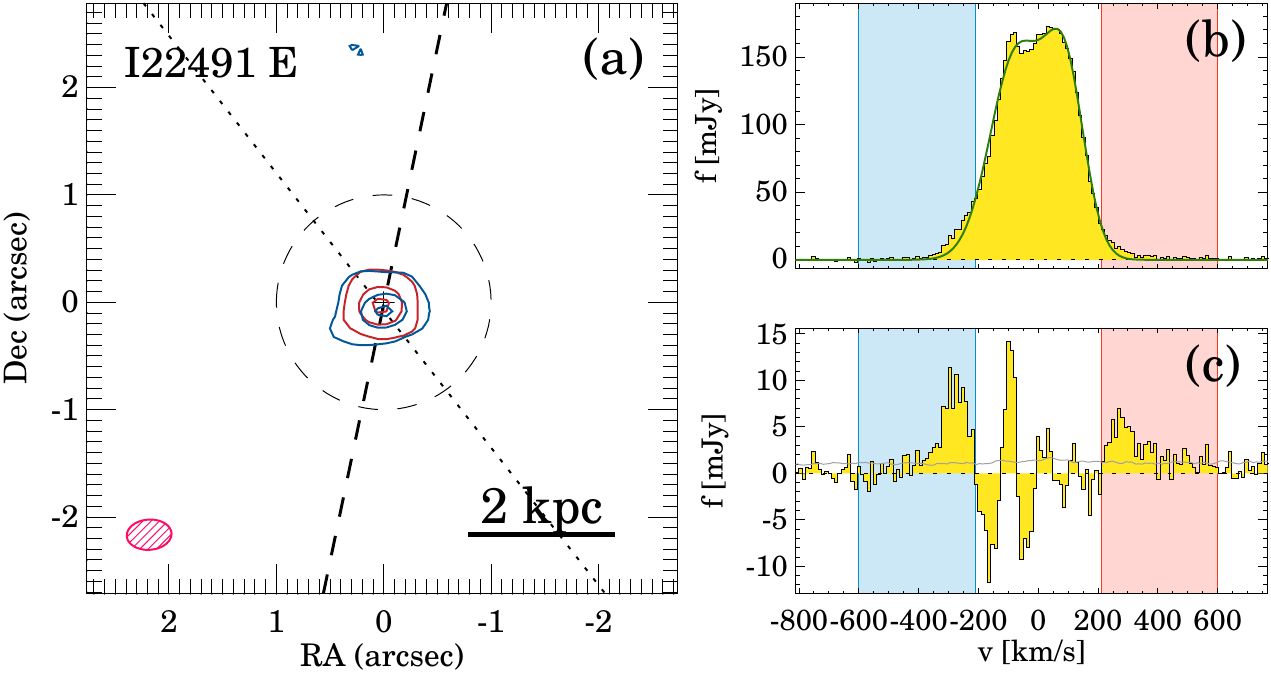}
\hfill
\includegraphics[width=0.48\textwidth]{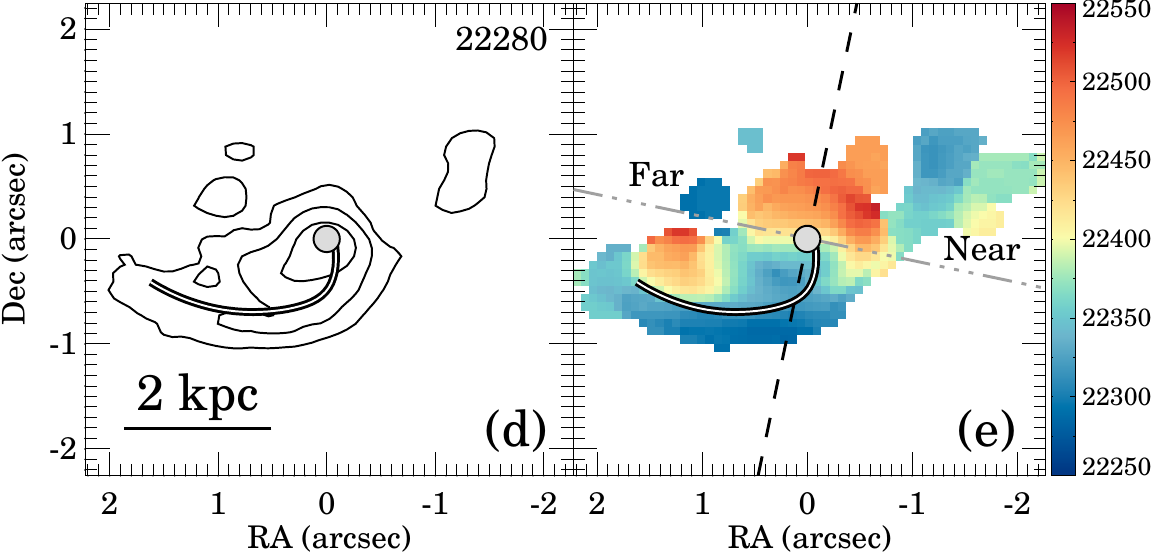}
\hfill
\caption{Same as Figure~\ref{fig_outflow_i12112_SW} but for I22491 E. 
{ In panel \textit{a}, $\sigma=$\,43\,mJy\,km\,s$^{-1}$\,beam$^{-1}$ and the red and blue peaks are at 1.1 and 2.2\,Jy\,km\,s$^{-1}$\,beam$^{-1}$, respectively.}
\label{fig_outflow_i22491_E}}
\end{figure*}

In the previous section, we presented the detection of high-velocity gas in 5 out of 6 ULIRG nuclei which is compatible with the presence of massive molecular outflows. But depending on which side of the rotating disk is closest to us, this high-velocity gas can be interpreted as an inflow or as an outflow. To investigate this, we plot the morphological features (spiral arms or tidal tails) we identified in the CO(2--1) channel maps (panels $d$ of Figures~\ref{fig_outflow_i12112_SW}--\ref{fig_outflow_i22491_E} { and Appendix~\ref{apx_channels}}). Then, in panels $e$ of these figures, we plot the identified morphological features over the velocity fields and, assuming that these features are trailing, we can determine the near- and far-side of the rotating disk. For all the galaxies where the high-velocity emission is spatially resolved, the blue-shifted high-velocity emission appears on the far side of the disk and the red-shifted emission on the near side. This is a clear signature of outflowing gas.

In addition, we can measure the PA of these outflows by fitting the position of the red and blue centroid clusters with a straight line (Figure~\ref{fig_alma_posdiagram} and Table~\ref{tbl_gaskin}). 
We calculated the difference between the PA of the high-velocity gas and that of the kinematic major axis of the disk  (Table~\ref{tbl_gaskin}) and found values around 90\degree\ for 3 cases. This PA difference is the expected value for an outflow perpendicular to the rotating disk. For \IRAS~14348 SW, the PA difference is $\sim$126$\pm$8\degree\ which suggests a different outflow orientation. Finally, the outflow PA of \IRAS~22491 E seems to deviate from a perpendicular orientation although with less significance due to the large uncertainty ($\sim$120$\pm$20\degree).

In panel {\it a} of Figures~\ref{fig_outflow_i12112_SW}--\ref{fig_outflow_i22491_E}, we show the spatial distribution of the high-velocity gas emission. This emission is spatially resolved, except for \IRAS~22491 E, and reaches projected distances, $R_{\rm max}$, up to 0\farcs4--1\farcs2 ($0.5-1.8$\,kpc; see Table~\ref{tbl_outflow_obs}) from the nucleus.  
We note that these sizes are a factor of $3-5$ larger than the sizes derived from the separation between the blue- and red-shifted emission centroids ($R_{\rm c}$). In the following, we use the $R_{\rm c}$ as the outflow size because, as a flux-weighted estimate of the outflow size, it is a better estimate of the extent of the region where most of the outflowing molecular gas is located. On the other hand, $R_{\rm max}$ is dominated by the faint CO(2--1) emission at larger radii and it is also likely dependent on the sensitivity of the observations.

The outflows are clearly spatially resolved ($2\times R_{\rm max}> $ FWHM of the beam). However, the angular resolution is not high enough to allow us to measure the outflow properties as function of radius. For this reason, we only consider the integrated outflow emission and measure a total outflow flux. To do so, we extracted the spectrum of the regions where high-velocity gas is detected (panels $b$ of Figures~\ref{fig_outflow_i12112_SW}--\ref{fig_outflow_i22491_E}). We fitted a two Gaussian model to the CO(2--1) line profile. { This model reproduces well the core of the observed line profile}. The residual blue and red wings (i.e., the outflow emission) are shown in panels $c$ of Figures~\ref{fig_outflow_i12112_SW}--\ref{fig_outflow_i22491_E} for each galaxy. From these spectra, we also estimate the flux-weighted average velocity of the outflowing gas ($320-460$\,km\,s$^{-1}$) and the maximum velocity at which we detect CO(2--1) emission ($400-800$\,km\,s$^{-1}$). The total CO(2--1) flux, as well as the flux in the high-velocity wings, are presented in Table~\ref{tbl_outflow_obs}.

\subsubsection{Derived properties}\label{ss:derived_properties}

\begin{table*}[ht]
\caption{Derived molecular outflow properties}
\label{tbl_outflow_derived}
\centering
\begin{small}
\begin{tabular}{lcccccccccccc}
\hline \hline
\\
Object & $M_{\rm out}$\tablefootmark{a} & $M_{\rm tot}$\tablefootmark{b} & v$_{\rm out}$\tablefootmark{c} & v$_{\rm max}$\tablefootmark{d} & $R_{\rm out}$\tablefootmark{e} & $R_{\rm max}$\tablefootmark{f} & $\log t_{\rm dyn}$\tablefootmark{g} & $\log  \dot{M}_{\rm out}$\tablefootmark{h} & $\log  t_{\rm dep}$\tablefootmark{i} & $\log  \dot{P}_{\rm out}$\tablefootmark{j} & $\log  L_{\rm out}$\tablefootmark{k}  \\
& (10$^8$\,\Msun) & (10$^9$\,\Msun) & (km\,s$^{-1}$) & (km\,s$^{-1}$) & (pc) & (kpc) & (yr) & (\Msun\,yr$^{-1}$) & (yr) & (g\,cm\,s$^{-2}$) & (erg\,cm$^{-2}$\,s) \\
\hline\\[-2ex]
I12112 SW & 0.31 & 1.0& 360 & 520 & 810 & 1.8 & 6.3 & 1.1 & 7.9 & 34.49 & 41.74 \\
I12112 NE & { 5.7} & 6.1 & 530 & 910 & 710  & 3.4 & 6.1 & 2.6 & 7.2 & 36.13 & 43.55 \\
I14348 SW & 5.2 & 6.7 & 430 & 770 & 1000 & 6.0 & 6.4 & 2.4 & 7.5 & 35.79 & 43.13 \\
I14348 NE & 1.0 & 2.9 & 460 & 650 & 590 &  1.5 & 6.1 & 1.9 & 7.5 & 35.38 & 42.75 \\
I22491 E & 1.2 & 3.5  & 400 & 500 & 250 &   0.9 & 5.8 & 2.3 & 7.2 & 35.72 & 43.02 \\
\hline
\end{tabular}
\end{small}
\tablefoot{
\tablefoottext{a,b}{Outflow and integrated molecular gas masses, respectively, assuming a ULIRG-like conversion factor $\alpha_{\rm CO}$ of 0.78 \Msun (K\,km\,s$^{-1}$\,pc$^{-2}$)$^{-1}$ and $r_{21}$ ratio of 0.91 \citep{Bolatto2013}.}
\tablefoottext{c}{Inclination corrected outflow velocity $|{\rm v}_{\rm high}|$\slash $\cos i$ (see Table~\ref{tbl_outflow_obs}).}
\tablefoottext{d}{{ Inclination corrected maximum outflow velocity $|{\rm v}_{\rm max}|$\slash $\cos i$ (see Table~\ref{tbl_outflow_obs}).}}
\tablefoottext{e}{Inclination corrected outflow radius range estimated from $R_{\rm c}$\slash $\sin i$ (see Table~\ref{tbl_outflow_obs}).}
\tablefoottext{f}{{ Inclination corrected outflow maximum radius derived using $R_{\rm max}$\slash $\sin i$ (see Table~\ref{tbl_outflow_obs}).}}
\tablefoottext{g}{Outflow dynamical time $t_{\rm dyn}=R_{\rm out}$\slash v$_{\rm out}$ (see Table~\ref{tbl_outflow_obs}).}
\tablefoottext{h}{Outflow rate $\dot{M}_{\rm out} = {\rm v}_{\rm out} \times M_{\rm out}$\slash $R_{\rm out}$. The uncertainty is { $\sim$0.4\,dex (see Section~\ref{ss:derived_properties}).}}
\tablefoottext{i}{Depletion time $t_{\rm dep}=M_{\rm tot}$\slash $\dot{M}_{\rm out}$.}
\tablefoottext{j}{Outflow momentum rate $\dot{P}_{\rm out} = \dot{M}_{\rm out}\times {\rm v}_{\rm out}$.}
\tablefoottext{k}{Outflow kinetic luminosity $L_{\rm out} = \hbox{1 \slash 2}\times \dot{M}_{\rm out}\times {\rm v}_{\rm out}^2$.}
}
\end{table*}

\begin{table*}[ht]
\caption{Escape Outflow}
\label{tbl_outflow_escape}
\centering
\begin{small}
\begin{tabular}{lcccccccccccc}
\hline \hline
\\
Object & v$_{\rm esc}$\tablefootmark{a} & v$_{\rm range}$\tablefootmark{b} & $S_{\rm CO}^{\rm esc}$\tablefootmark{c} & $S_{\rm CO}^{\rm esc}$\slash  $S_{\rm CO}^{\rm out}$\tablefootmark{d} & $\log \dot{M}_{\rm esc}$ \\
& (km\,s$^{-1}$) & (km\,s$^{-1}$) & (Jy\,km\,s$^{-1}$) & & (\Msun\,yr$^{-1}$)\\
\hline\\[-2ex]
I12112 SW & 465 & [425, 550] & 0.14 $\pm$ 0.03 & 0.24 & 0.49 \\
I12112 NE & 465 & [414, 900] & { 3.0} $\pm$ 0.1 & { 0.28} & 2.1 \\
I14348 SW & 590 & [570, 800] & 1.2 $\pm$ 0.1 & 0.16 & 1.6 \\
I14348 NE & 590 & [459, 560] & 0.28 $\pm$ 0.04 & 0.18 & 1.2 \\
I22491 E & 400 & [319, 600] & 0.7 $\pm$ 0.1 & 0.34 & 1.8 \\
\hline
\end{tabular}
\end{small}
\tablefoot{
\tablefoottext{a}{Escape velocity at 2\,kpc (see \citealt{Emonts2017}).}
\tablefoottext{b}{Observed velocity range used to measure the molecular gas with ${\rm v}>{\rm v_{\rm esc}}$ taking into account the inclination of the object.}
\tablefoottext{c}{CO(2--1) emission with ${\rm v}>{\rm v_{\rm esc}}$.}
\tablefoottext{d}{Ratio between the CO(2--1) emission from molecular gas with ${\rm v}>{\rm v_{\rm esc}}$ and the total outflowing gas from Table \ref{tbl_outflow_obs}.}
\tablefoottext{e}{Molecular gas escape rate.}
}
\end{table*}

In Table~\ref{tbl_outflow_derived}, we present the derived properties for these outflows based on the observations and assuming that they are perpendicular to the rotating disk. The latter is consistent with the $\sim$90\degree\ PA difference between the kinematic major axis and the outflow axis measured in 3 of the galaxies. For the other 2 cases (PA difference $\sim$120\degree), { this assumption might introduce a factor of $\sim$2 uncertainty in the derived outflow rates.}

To convert the CO(2--1) fluxes into molecular masses, we assume a ULIRG-like $\alpha_{\rm CO}$ conversion factor of 0.78 \Msun (K\,km\,s$^{-1}$\,pc$^{-2}$)$^{-1}$ and a ratio between the CO(2--1) and CO(1--0) transitions, $r_{21}$, of 0.91 \citep{Bolatto2013}.
The outflow velocity is corrected for the inclination by dividing by $\cos i$, where $i$ is the disk inclination (Table~\ref{tbl_gaskin}). Similarly, the outflow radius is corrected by dividing by $\sin i$. 
The average corrected outflow velocity is $\sim$440\,km\,s$^{-1}$ and the average deprojected radius $\sim$700\,pc.

Using these quantities, we calculate the dynamical time as $t_{\rm dyn} = R_{\rm out}\slash {\rm v}_{\rm out}$ (about $\sim$1\,Myr for these outflows). Then, we estimate the outflow rate using $\dot{M}_{\rm out}=M_{\rm out}\slash t_{\rm dyn}$. We find $\dot{M}_{\rm out}$ values between $\sim$12 and $\sim$400\,\Msun\,yr$^{-1}$. From these estimates, we can derive the depletion time ($t_{\rm dep}$), outflow momentum rate ($\dot{P}_{\rm out}$), and the outflow kinetic  luminosity ($L_{\rm out}$; see e.g., \citealt{GarciaBurillo2015}).

{ The uncertainties on the outflow rate, and the quantities derived from it, are dominated by the uncertainty in the value of the $\alpha_{\rm CO}$ conversion factor which is not well established for the outflowing gas (e.g., \citealt{Aalto2015}) and the outflow geometry (inclination). For the conversion factor, we assume a ULIRG-like value ($\alpha_{\rm CO}=0.78$\Msun (K\,km\,s$^{-1}$\,pc$^{-2}$)$^{-1}$), although depending on the gas conditions, this factor can vary within 0.3\,dex (e.g., \citealt{Papadopoulos2012b, Bolatto2013}). Therefore, from the uncertainty in the inclination and the conversion factor, we assume a 0.4\,dex uncertainty for these values.}

Finally, in Table~\ref{tbl_outflow_escape}, we estimate the fraction of the outflowing gas that would escape the gravitational potential of these galaxies. We use the escape velocities at 2\,kpc calculated by \citet{Emonts2017} for these systems which range from $\sim$400 to 600\,km\,s$^{-1}$. We integrate the CO(2--1) emission with velocities higher than these escape ones (taking into account the inclination of the outflows) and obtain that $15-30$\%\ of the high-velocity gas will escape to the intergalactic medium. The escape outflow rates are $3-120$\,\Msun\,yr$^{-1}$.
However, these escape rates can be lower if the velocity of the outflowing gas is decreased due to dynamical friction.

\subsection{Nuclear SFR}\label{ss:sfrrate}

\begin{table*}[ht]
\caption{Outflows and nuclear SFR}
\label{tbl_outflow_agn_sfr}
\centering
\begin{small}
\begin{tabular}{lcccccccccccc}
\hline \hline
\\
Object & $\log$\,SFR(IR)\tablefootmark{a} & $\log$\,SFR(1.4\,GHz)\tablefootmark{b} & $\log \eta$\,\tablefootmark{c}  & $\log \frac{\dot{P}_{\rm out}}{P_{\rm SNe}}$\tablefootmark{d}  & $\log \frac{L_{\rm out}}{L_{\rm SNe}}$\tablefootmark{e}\\
& (\Msun\,yr$^{-1}$) & (\Msun\,yr$^{-1}$) \\
\hline\\[-2ex]
I12112 SW & 1.12 & 1.57 &  --0.02 &  --0.65 &  --1.4  \\
I12112 NE & 2.26 & 2.21 & 0.34    &  --0.12 &  --0.67 \\
I14348 SW & 2.14 & 2.36 & 0.26    &  --0.28 &  --0.91 \\
I14348 NE & 1.97 & 2.12 &  --0.07 &  --0.58 &  --1.2  \\
I22491 E  & 2.04 & 1.76 & 0.26    &  --0.32 &  --0.99 \\

\hline
\end{tabular}
\end{small}
\tablefoot{
\tablefoottext{a}{SFR derived from the IR luminosity assigned to each nucleus (see Section~\ref{ss:ir_lum}) using the calibration of \citet{Kennicutt2012}. This is the adopted SFR for these nuclei.}
\tablefoottext{b}{SFR derived from the non-thermal radio continuum using the \citet{Murphy2011} SFR calibration (see Section~\ref{ss:radio_cont}).}
\tablefoottext{c}{Logarithm of the mass loading factor \hbox{$\eta = \dot{M}_{\rm out}\slash {\rm SFR}$(IR)}.}
\tablefoottext{d}{Ratio between the outflow momentum rate and the momentum  injected by supernova explosions. We assume that the $P_{\rm SN}$ per SN is {$1.3\times 10^{5}$\Msun\,km\,s$^{-1}\times(n_0/100$\,cm$^{-3})^{-0.17}$} \citep{Kim2015} using $n_0 = 100$\,cm$^{-3}$ and that the SN rate, $\nu_{\rm SN}$\,(yr$^{-1}$), is 0.012$\times$SFR(IR)(\Msun\,yr$^{-1}$) for the adopted IMF \citep{Leitherer1999}.}
\tablefoottext{e}{Ratio between the total kinetic luminosity of the molecular outflow and that injected by supernova explosions ($L_{\rm SNe}$(erg\,s$^{-1}$)=$9\times 10^{41} {\rm SFR}$(IR)(\Msun\,yr$^{-1}$); \citealt{Leitherer1999} adapted for a \citealt{Kroupa2001} IMF).}
}
\end{table*}

\begin{figure*}[t]
\centering
\includegraphics[height=0.25\textwidth]{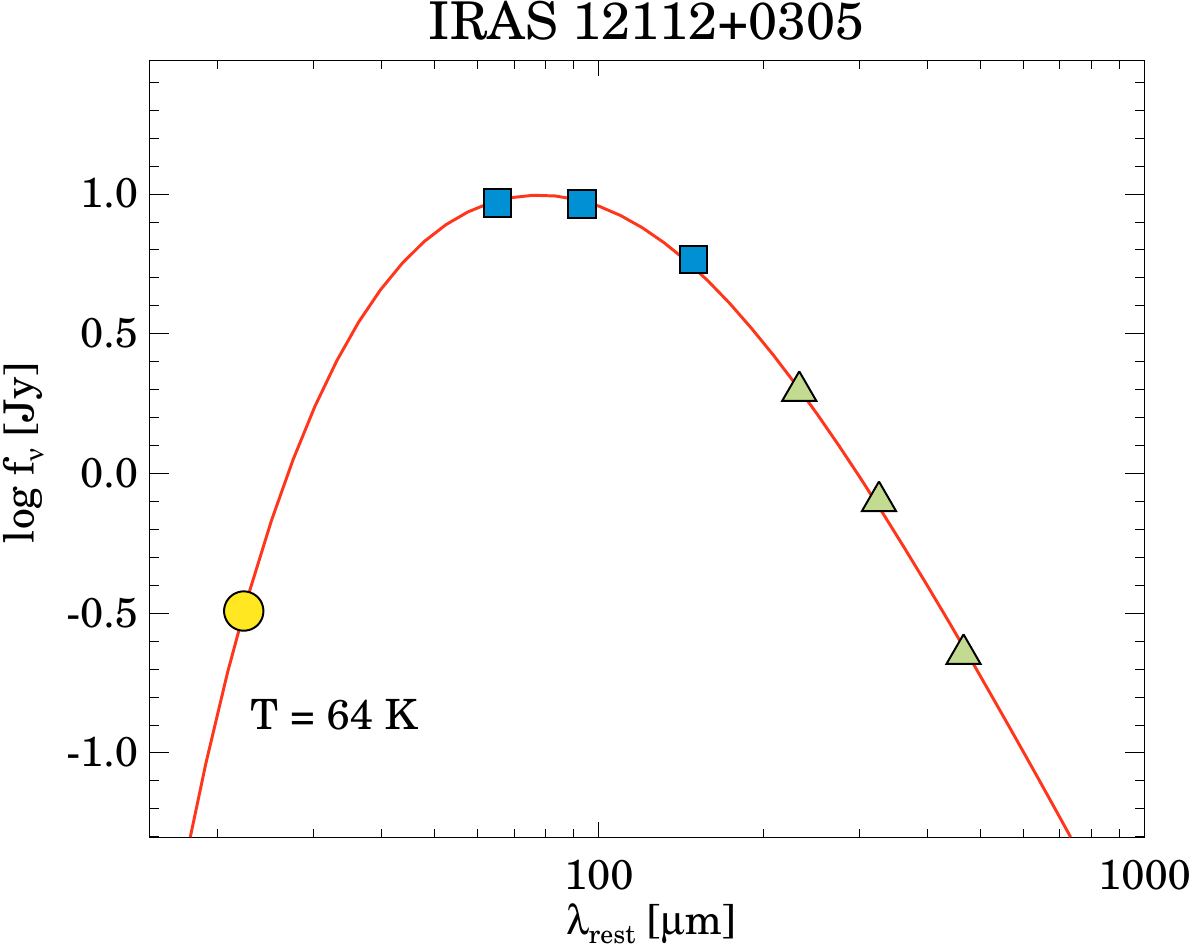}
\includegraphics[height=0.25\textwidth]{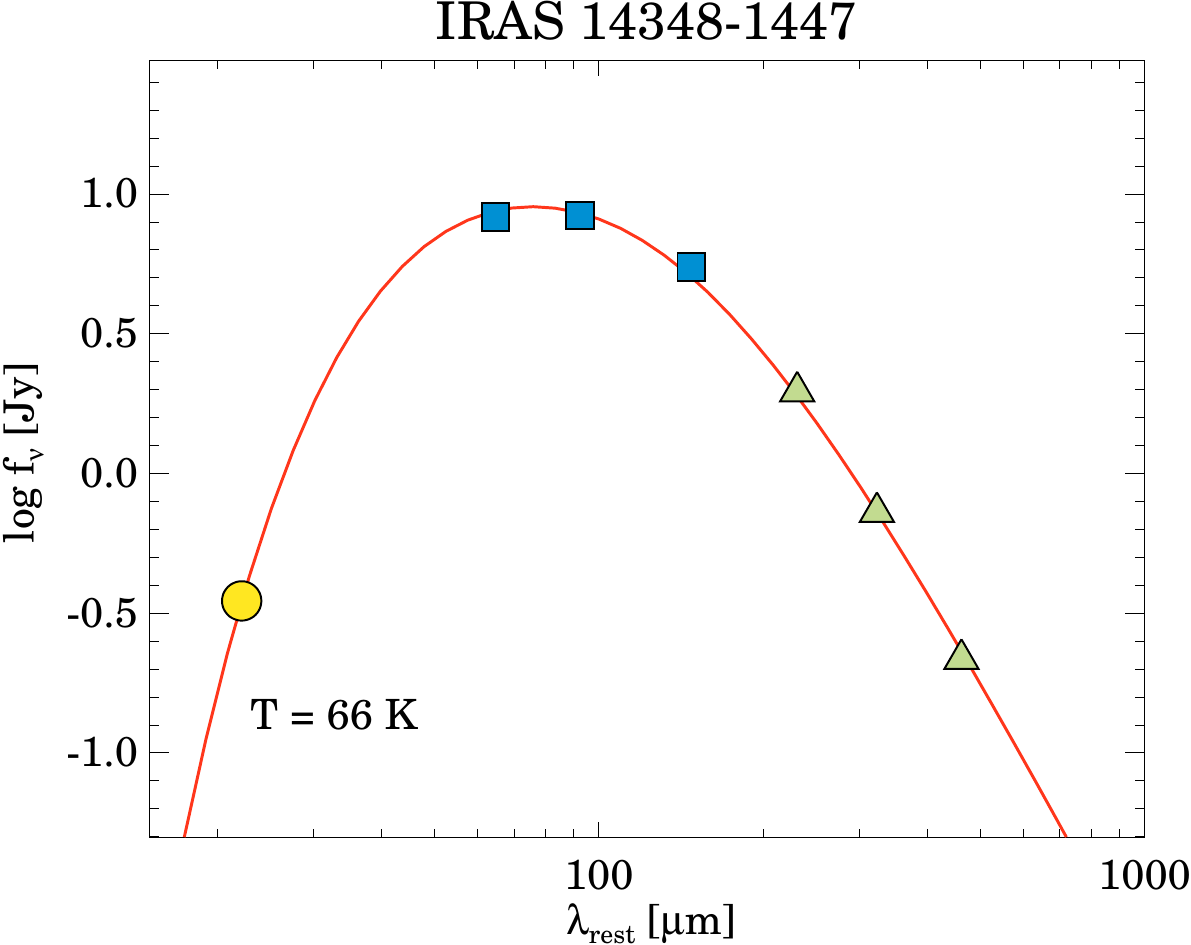}
\includegraphics[height=0.25\textwidth]{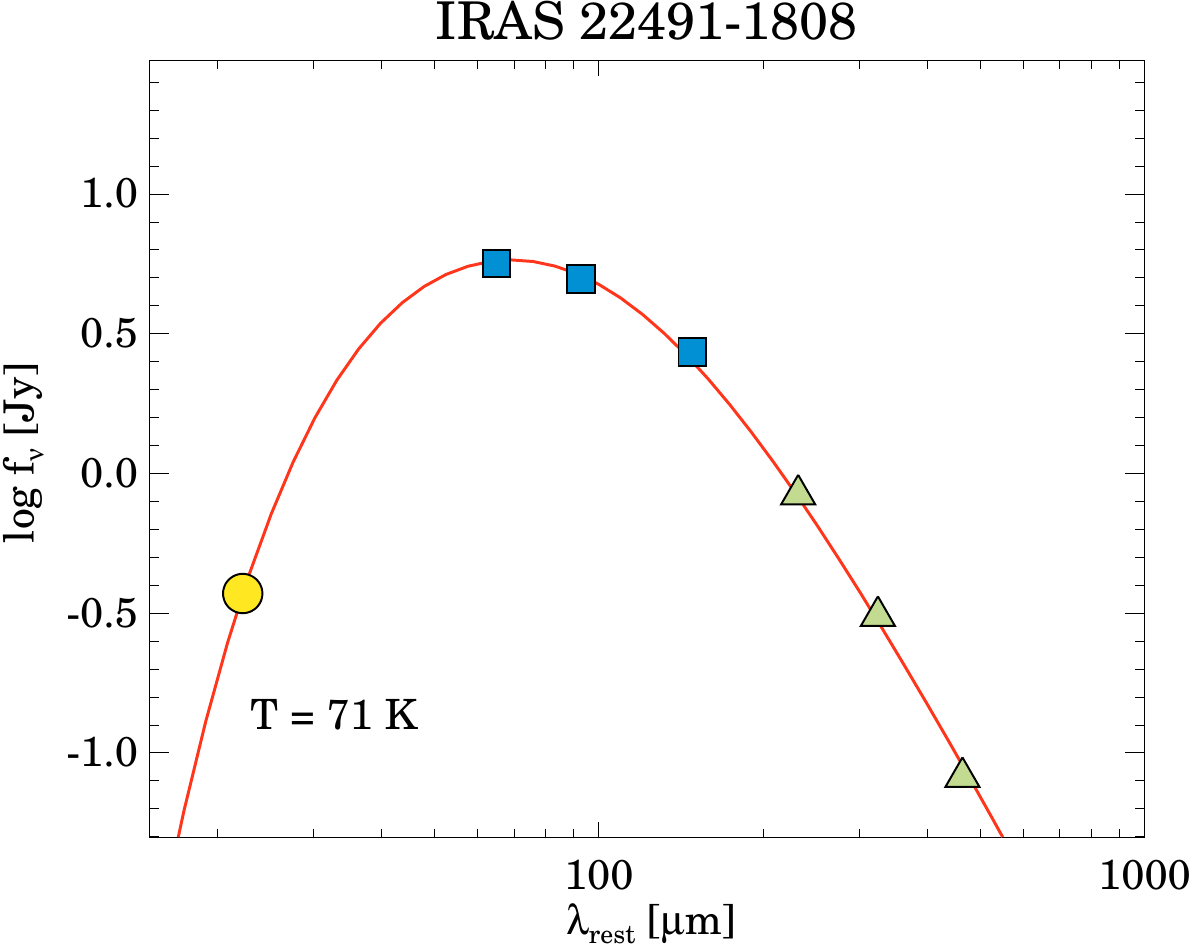}
\caption{Mid- and far-IR spectral energy distribution of the three ULIRGs. The yellow circle corresponds to the \Spitzer\slash MIPS 24\micron\ flux, the blue squares to the \Herschel\slash PACS 60, 100, and 160\micron\ fluxes, and the green triangles to the \Herschel\slash SPIRE 250, 350, and 500\micron\ fluxes. The solid red line is the best fit to the data using a single temperature gray body. \label{fig_sed_fits}}
\end{figure*}

Measuring the SFR in these ULIRGs is important to evaluate the impact of the molecular outflows. Most of the outflowing molecular gas is concentrated in the central $1-2$\,kpc, so to determine the local effect of the outflows, we must compare them with the nuclear SFR. However, the nuclei of local ULIRGs are extremely obscured regions (e.g., \citealt{GarciaMarin2009, Piqueras2013}) and estimating their SFR is not straightforward. In this section, we use two approaches to measure the nuclear SFR using the IR luminosity and the 248\,GHz continuum and the radio continuum which should not be heavily affected by extinction.

\subsubsection{IR luminosity}\label{ss:ir_lum}

The total IR luminosity ($L_{\rm IR}$) is a good tracer of the SF in dusty environments such as the nuclei of ULIRGs (e.g., \citealt{Kennicutt2012}). However, there are no far-IR observations with the two nuclei of the systems spatially resolved. For this reason, we first derived the integrated $L_{\rm IR}$ of each system. We fit a single temperature gray body model to the 24\micron\ to 500\micron\ fluxes from \Spitzer\ and \Herschel\ \citep{Piqueras2016, Chu2017} following \citet{Pereira2016b}. The resulting $L_{\rm IR}$ are $\sim$0.2\,dex lower than those derived using the {IRAS} fluxes, but we consider these new $L_{\rm IR}$ more accurate since we are using more data points which cover a wider wavelength range to fit the IR emission (7 points between $24-500$\micron\ vs. 4 points between $12-100$\micron) and also we avoid flux contamination from unrelated sources thanks to the higher angular resolution of the new data ($6\arcsec-35\arcsec$ vs. $0\farcm5-2\arcmin$). The \LIR\ are listed in Table~\ref{tbl_sample} and the best-fits shown in Figure~\ref{fig_sed_fits}.

Then, we assign a fraction of \LIR\ due to the star-formation (i.e., after subtracting the AGN luminosity from Table~\ref{tbl_sample}) to each nucleus which is proportional to their contribution to the total thermal continuum at 248\,GHz (dust emission plus free-free radio continuum; see Table~\ref{tbl_continuum}) of the system. By doing this, we assume that all the \LIR\ is produced in the central $300-1000$\,pc, which is consistent with the compact distribution of the molecular gas around the nuclei (Section~\ref{ss:molgas}), and that the 248\,GHz continuum scales with the \LIR. The latter is true for the free-free radio continuum contribution at this frequency which is proportional to the ionizing flux and, therefore, to the SFR and \LIR. The dust emission at 248\,GHz depends on the average dust temperature of each nucleus ($f_\nu\slash L\propto T^{-3}$ in the Rayleigh-Jeans tails of the black body). Although, given the similar temperatures we obtained for the integrated emission ($T\sim65-70$\,K; see Figure~\ref{fig_sed_fits}), our assumption seems to be reasonable.

Finally, we converted these nuclear IR luminosities into SFR using the \citet{Kennicutt2012} calibration (see Table~\ref{tbl_outflow_agn_sfr}). The SFRs range from 13 to 180\,\Msun\,yr$^{-1}$.

\subsubsection{Radio continuum}\label{ss:radio_cont}

We also estimated the SFR from the non-thermal radio continuum observations of these galaxies (see Section~\ref{ss:continuum}). Using the observed 1.49 and 8.44\,GHz fluxes and the derived spectral indexes, we estimated the rest-frame 1.4\,GHz continuum and applied the \citet{Murphy2011} SFR calibration. Here, we ignore any contribution from an AGN to the radio emission. These objects seem to be dominated by SF and just have small AGN contribution, but the radio emission of these AGN is uncertain.

The obtained radio SFR are listed in Table~\ref{tbl_outflow_agn_sfr}. These values are comparable to those obtained from the IR luminosity. The average difference between the two estimates is 0.2\,dex with a maximum of 0.4\,dex. Therefore, the two methods provide compatible SFR values and, in the following, we adopt the SFR(IR) with a 0.2\,dex uncertainty.

\section{Discussion}\label{s:discu}

\subsection{Outflow energy source}\label{ss:energy}

\begin{figure*}
\centering
\includegraphics[height=0.3\textwidth]{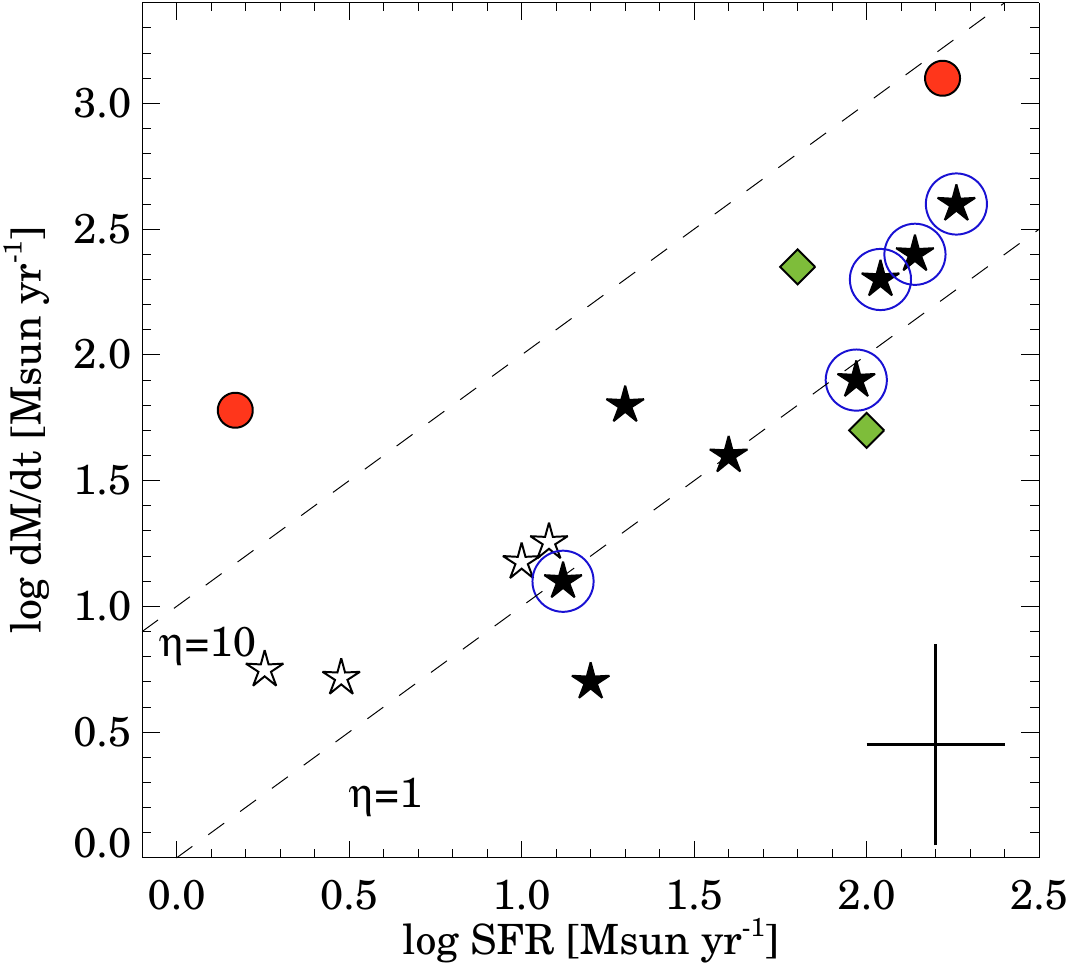}
\hfill
\includegraphics[height=0.3\textwidth]{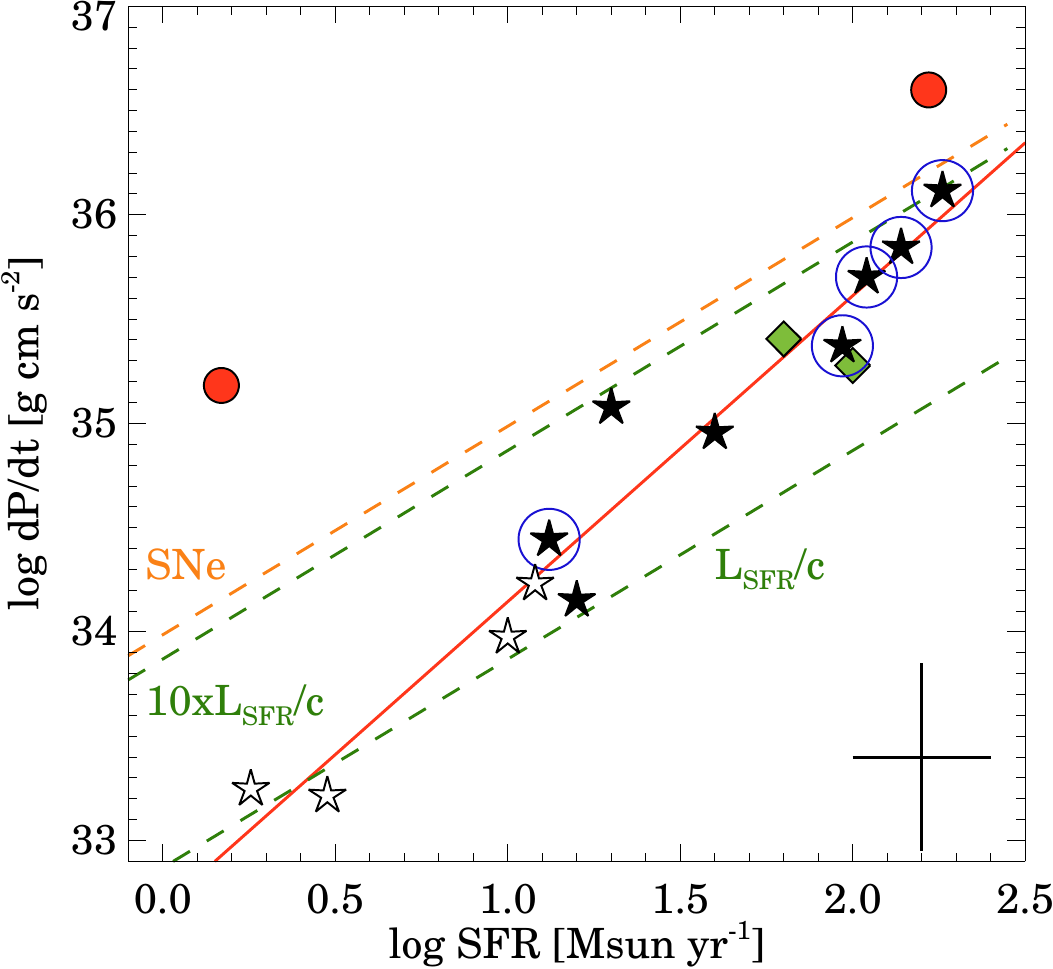}
\hfill
\includegraphics[height=0.3\textwidth]{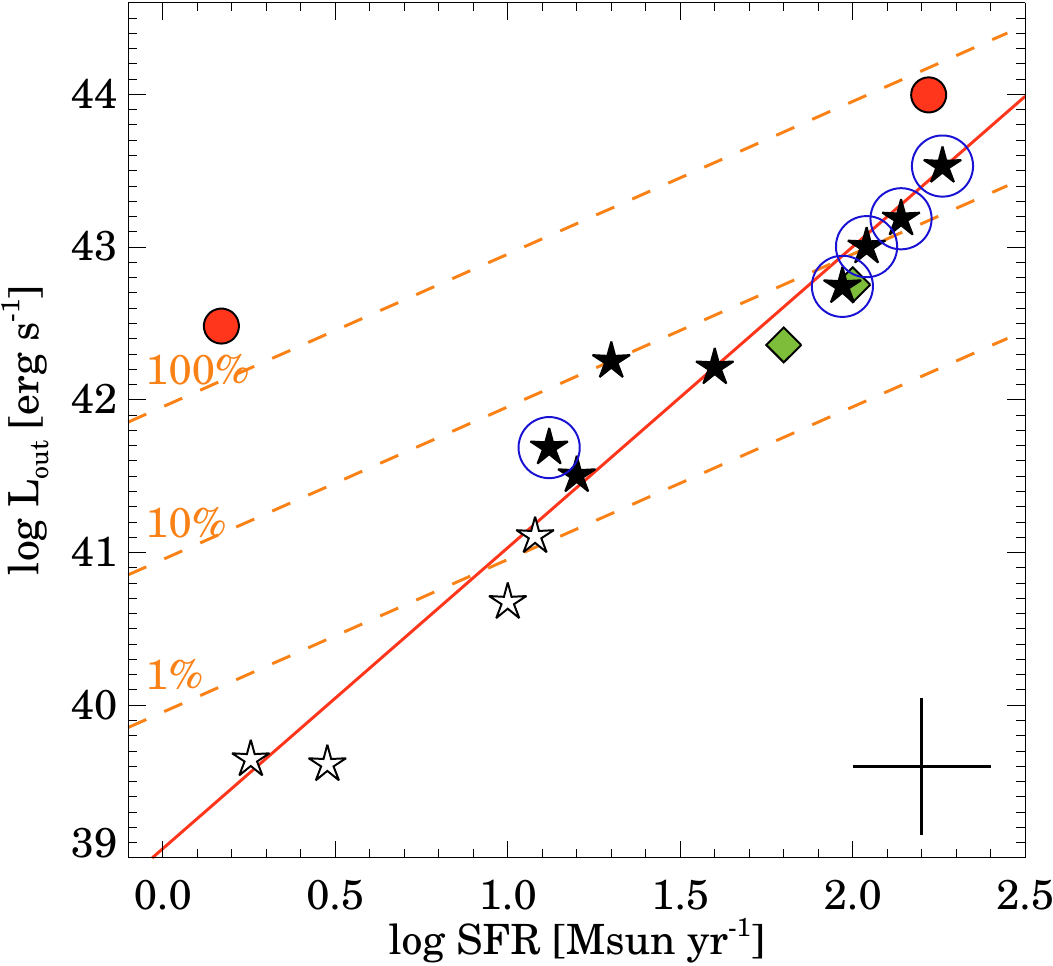}
\caption{{ Mass outflow rate} vs. nuclear SFR ({\it left}), outflow momentum rate vs. nuclear SFR ({\it middle}), and outflow kinetic luminosity vs. nuclear SFR ({\it right}). 
{ Red circles indicate nuclei with outflows launched by an AGN, green diamonds are objects hosting an AGN with molecular outflows of uncertain SF\slash AGN origin, and stars represent star-formation dominated nuclei}. The blue circles mark the ULIRGs presented in this work. The white stars are the lower luminosity starburts compiled by \citet{Cicone2014}. The remaining points correspond to local U\slash LIRGs from the literature: NGC~1614 and \IRAS~17208-0014 \citep{GarciaBurillo2015, Pereira2015_n1614, Piqueras2016}; NGC~3256 N and S \citep{Sakamoto2014, Emonts2014, Ohyama2015, Pereira2011, Lira2002}; ESO~320-G030 \citep{Pereira2016b}; { Arp~220 W \citep{BarcosMunoz2018}; and NGC~6240 \citep{Saito2018}.}
The crosses at the lower right corners represent the typical error bars of the points. 
The  black lines in the {\it left} panel correspond to mass loading factors, $\eta=\dot{M}$\slash SFR, of 1 and 10. The dashed orange line in the {\it middle panel} marks the total momentum injected by SNe as function of the SFR. The dashed green lines indicate the L(SFR)\slash c ratio and 10 times this value.
The dashed orange lines in the {\it right} panel indicate the $L_{\rm out}$ = $a\times L_{\rm SNe}$ with $a=$1, 0.1, and 0.01 as function of the SFR.
The solid red lines in the middle and right panels are the best linear fits to the star-forming objects.
\label{fig_sfr_mout}}
\end{figure*}

In the left panel of Figure~\ref{fig_sfr_mout}, we show the relation between the outflow rate and the nuclear SFR (i.e., the mass loading factor $\eta$). In this figure, we include local U\slash LIRGs with spatially resolved observations (filled symbols) as well as the lower luminosity starbursts compiled by \citet{Cicone2014}\footnote{For NGC~3256, we use the newer observations presented by \citet{Sakamoto2014} which distinguish between the Northern and Southern nuclei instead of the \citet{Sakamoto2006} data used by \citet{Cicone2014}.}. In total, we include observations for { 7 ULIRG nuclei, 5 LIRG nuclei, and 4 starbursts.}
{ For the 2 nuclei classified as AGN, we derive the nuclear SFR from their IR luminosity after subtracting the AGN contribution (see \citealt{GarciaBurillo2015, Ohyama2015}).}

{ The 5 new ULIRG nuclei (encircled stars in this figure) have mass loading factors}, $\eta$, $\sim0.8-2$ (see Table~\ref{tbl_outflow_agn_sfr}). These are similar to those observed in local starburst galaxies which are typically lower than $\sim2-3$ (e.g., \citealt{Bolatto2013, Cicone2014, Salak2016}). This suggests that the outflows in these ULIRGs are also powered by SF.

To further investigate the energy source, in Table~\ref{tbl_outflow_agn_sfr}, we list the ratios between the kinetic luminosity and momentum rates of the outflows and the total energy and momentum injected by supernovae (SNe), respectively. 
We assume that the SNe total energy and momentum are upper limits on the energy and momentum that the starburst can inject into the outflow (independent of the launching mechanism; see Section~\ref{ss:launch}).
For all the galaxies, both the energy and momentum in the molecular outflowing gas are lower than those produced by SNe. Although this does not imply an SF origin, we cannot rule out the SF origin based on the energy or momentum in of these outflows.

Molecular outflows from AGN usually have maximum velocities up to $\sim$1000\,km\,s$^{-1}$ { (e.g., \citealt{Cicone2014, Veilleux2017})} which are higher than those due to SF (few hundreds of km\,s$^{-1}$). We found the maximum outflow velocities in \IRAS~12112 NE and \IRAS~14348 SW ($\sim700-800$\,km\,s$^{-1}$). These are not as high as those observed in other AGN, but are $1.5-2$ times higher than in the rest of our sample and might indicate an AGN powered outflow in these objects. However, there are molecular outflows detected in more nearby starbursts which also reach these high velocities (e.g., \citealt{Sakamoto2014}). Therefore, the velocities of the outflows in these ULIRGs are not high enough to claim an AGN origin.

Similarly, the orientation of the outflow gives information on its origin. Outflows produced by starbursts tend to be perpendicular to the disk of the galaxy where it is easier for the gas to escape. On the contrary, the angle of AGN outflows is, in principle, independent of the disk orientation (e.g., \citealt{Pjanka2017}). We found that the PA of these outflows are compatible with being perpendicular to the disk (i.e., possible SF origin) except for \IRAS~14348 SW (i.e., possible AGN origin; see Table~\ref{tbl_gaskin}).

In summary, the mass, energy, momentum, velocity, and geometry of these outflows seem to be compatible with those expected for a SF powered outflow. The only exception could be the outflow of \IRAS~14348 SW. This outflow has a relatively high velocity compared to the others and also a different geometry, so it might be powered by an AGN. X-ray observations also suggest the presence of a Compton-thick AGN, although the bolometric luminosity of this AGN seems to be $<$10\%\ of the total IR luminosity \citep{Iwasawa2011} and would not be able to produce the observed outflow.
Therefore, since there is no clear evidence for an AGN origin, we assumed a SF origin in this case too.

\subsection{Outflow effects on the star-formation}

The nuclear outflow depletion times are $15-80$\,Myr which are comparable to those found in other ULIRGs \citep{Cicone2014, GarciaBurillo2015, GonzalezAlfonso2017}. These times do not include the possible inflow of molecular gas into the nuclear region. However, between 70\% and 90\% of the molecular gas is already in these central regions (see Tables~\ref{tbl_integrated} and \ref{tbl_outflow_obs}), so we do not expect significant molecular inflows to occur. Inflows of atomic gas might be present too, but there are no spatially resolved \ion{H}{i} observations available for these objects to infer the atomic gas distribution.
In addition, we have to take into account that most of the outflowing gas ($\sim60-80$\%; Table~\ref{tbl_outflow_escape}) will not escape the gravitational potential of these systems and will become available to form new stars in the future. 
We can estimate how long it will take for the outflowing gas to rain back into the system from the average outflow velocity, the outflow radius, and the escape velocity (Tables~\ref{tbl_outflow_derived} and \ref{tbl_outflow_escape}). From the escape velocity we obtain the gravitational parameter, $\mu=GM$, using the following relation:
\begin{equation}
 \mu = \frac{1}{2}\times r_{\rm esc} \times {\rm v_{esc}}^2
\end{equation}
where v$_{\rm esc}$ and r$_{\rm esc}$ are the escape velocity and the radius at which it is calculated, respectively. Then, assuming that the outflowing gas moves radially and that it is not affected by any dynamical friction, the equations of motion are:
\begin{equation}
\begin{aligned}
\frac{dr}{dt} &= {\rm v}
\\
\frac{d{\rm v}}{dt} &= -\frac{\mu}{r^2}
\end{aligned}
\end{equation}

with the initial conditions $t_0 = t_{\rm dyn}$, $r_0=R_{\rm out}$, and ${\rm v}_0={\rm v}_{\rm out}$. Integrating these equations numerically, we can determine when $r$ becomes 0, and obtain an estimate of the outflow cycle duration. By doing this, we find cycle durations of $5-10$\,Myr (these can be shorter if the dynamical friction is important). 
Therefore, even if the outflow depletion times are slightly shorter than the SF depletion times ({ $M_{\rm tot}\slash $SFR$\sim30-80$\,Myr)}, the outflowing gas will return to the starburst region after few Myr where it will be available again to form stars. In consequence, the main effects of these outflows are to delay the formation of stars in the nuclear starbursts and to expel a fraction of the total molecular gas ($\sim15-30$\%) into the intergalactic medium. However they will not completely quench the nuclear star-formation.

{ \citet{Walter2017} suggested that the molecular outflow detected in the low-luminosity starburst galaxy NGC~253 is accelerating at a rate of 1 km\,s$^{-1}$\,pc$^{-1}$ when observed at 30\,pc resolution. For these ULIRGs, we find that the higher velocity outflowing molecular gas is not located farther from the nucleus than the lower velocity gas (see Figure~\ref{fig_alma_posdiagram} and Appendix~\ref{apx_channels}). Therefore, outflow acceleration does not seem to be important for these outflows at $\sim$500\,pc scale and will likely not affect the cycle duration and outflow effects discussed above.}

\subsection{Outflow launching mechanism in starbursts}\label{ss:launch}

There are two main mechanisms that can launch outflows in starbursts. Radiation pressure from young stars can deposit momentum into dust grains. Dust and gas are assumed to be dynamically coupled and, therefore, this process can increase the gas outward velocity and produce an outflow. This class of outflows is known as momentum-driven (e.g., \citealt{Murray2005, Thompson2015}). The second mechanism is related to the energy injection into the interstellar medium (ISM) by SNe. If the gas does not cool efficiently, this energy increase translates into an adiabatic expansion of the gas which drives the outflow. These outflows are known as energy-driven (e.g., \citealt{Chevalier1985, Costa2014}).

\begin{figure}
\centering
\includegraphics[height=0.31\textwidth]{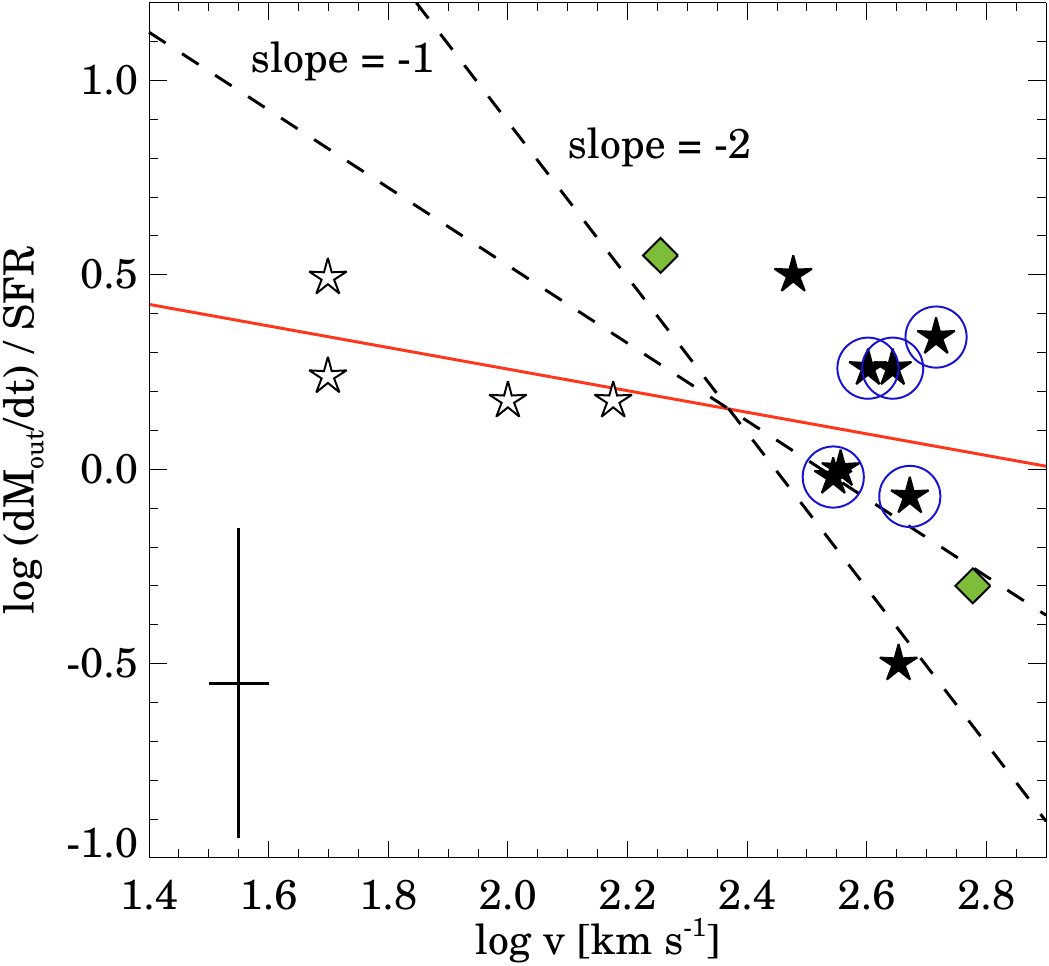}
\caption{Mass loading factor vs. outflow velocity. Only outflows powered by SF are plotted in this figure. Galaxy symbols are as in Figure~\ref{fig_sfr_mout}. The dashed black lines are the best fits with fixed slopes of --1 and --2 (see Section~\ref{ss:launch}). The red line is the best linear fit.
\label{fig_eta_vout}}
\end{figure}

The scaling relation between the mass loading factor and the outflow velocity is different for energy- and momentum-driven outflows ($\eta\sim {\rm v^{-2}}$ for energy-driven and $\eta\sim {\rm v^{-1}}$ for momentum-driven; e.g., \citealt{Murray2005}). \citet{Cicone2014} found a slope of --1 for this relation and suggested that the molecular phase of outflows are possibly momentum-driven. However, Figure~\ref{fig_eta_vout} shows that, after adding the new data points, the slope of the $\eta$ vs. v relation is shallower than --1. The best linear fit is:

\begin{equation}
\log\,\eta = (0.8 \pm 0.4) - (0.3 \pm 0.2) \log {\rm v_{out}} ({\rm km\,s^{-1}})
\end{equation}

This does not necessarily imply that these outflows are not momentum-driven. Actually, the --1 slope for momentum-driven outflows implicitly ignores the dependency of the outflow velocity on the optical depth, $\tau_{\rm FIR}$, of the launching region. When the FIR optical depth increases, the momentum transfer from the radiation to the dust\slash gas can be considerably more efficient \citep{Thompson2015, Hopkins2013}. For instance, if $\tau_{\rm FIR}>1$, the momentum boost factor, $\dot{P}_{\rm out}\slash (L\slash c)$, can significantly exceed $\sim$2 \citep{Thompson2015}. 

To test this, in the middle panel of Figure~\ref{fig_sfr_mout}, we plot the outflow momentum rate as a function of the SFR. The best linear fit to the starbursts data is:
\begin{equation}
\log \dot{P}_{\rm out} ({\rm g\,cm\,s^{-2}}) = (32.7 \pm 0.3) +  (1.5 \pm 0.2) \log {\rm SFR} (M_\odot {\rm yr^{-1}})
\end{equation}
which has a slope $>$1. That is, for those starbursts with the lower SFR, the momentum boost factor is $\sim$1 (see also \citealt{Cicone2014}). But this factor increases for objects with higher SFR up to $\sim$8. For one of these starbursts, ESO~320-G030, we measured a very high optical depth in the region launching the outflow $\gtrsim$8 at 100$\micron$ \citep{Pereira2017Water}. Therefore, higher dust opacities in the more vigorous starbursts could explain these momentum boost factors $>2$.

We also explore the possible role of SNe in the launching of these outflows. We plot the momentum injected by SNe in the middle panel of Figure~\ref{fig_sfr_mout}, which is more than a factor of 10 higher than the radiation pressure. For all the starbursts the momentum rate of their outflows is lower than the momentum due to SN explosions. Therefore, these outflows could be launched by SNe. If this is the case, the momentum coupling between the SNe and the ISM seems to be more efficient at higher SFR. While for the low SFR objects the outflows carry less than 10\% of the SNe momentum, the outflows in higher SFR objects carry up to 75\% of the momentum injected by SNe.

Similarly, in the right panel of Figure~\ref{fig_sfr_mout}, we compare the kinetic luminosity of the outflows with the energy produced by SNe. The outflow kinetic luminosity represents $4-20$\% of the energy produced by SNe for the U\slash LIRGs, whereas for the lower luminosity starbursts, this fraction is $\lesssim$1\%. Therefore, if these outflows are driven by SNe, this suggests that the coupling efficiency between the SNe and the ISM increases with increasing SFR. The best linear fit is:
\begin{equation}
\log L_{\rm out} ({\rm erg\,s^{-1}}) = (39.0 \pm 0.3) +  (2.0 \pm 0.2) \log {\rm SFR} (M_\odot {\rm yr^{-1}})
\end{equation}

We also note that, for the { AGN U\slash LIRGs}, the observed kinetic luminosities of the outflows are $1-5$\% of the AGN luminosity \citep{Cicone2014, GarciaBurillo2015}. Thus, if SNe are the main drivers of outflows in starbursts, the coupling between the SN explosions and the ISM must be more efficient than for AGN, at least, when the SFR is sufficiently high.

\subsection{Multi-phase outflows}\label{ss:multiphase}

\begin{table}[t]
\caption{Hot-molecular outflow phase}
\label{tbl_multiphase}
\centering
\begin{small}
\begin{tabular}{lcccccccccccc}
\hline \hline
\\
Object & v$_{\rm cold\,H_2}$\tablefootmark{a} & v$_{\rm hot\,H_2}$\tablefootmark{b} & $M_{\rm hot\,H_2}$\tablefootmark{c} & $M_{\rm hot\,H_2}\slash M_{\rm cold\,H_2}$\tablefootmark{d}\\
& (km\,s$^{-1}$) & (km\,s$^{-1}$) &  (10$^3$\Msun) & (10$^{-5}$) \\
\hline\\[-2ex]
I12112 NE & 465 & 430 & 6.8$\pm$3.7 & 1.3$\pm$0.7 \\
I14348 SW & 419 & 520 & 8.4$\pm$2.2 & 1.6$\pm$0.5 \\ 
I22491 E  & 325 & 320 & 5.9$\pm$1.9 & 4.9$\pm$1.6 \\
\hline
\end{tabular}
\end{small}
\tablefoot{
\tablefoottext{a}{Cold molecular outflow velocity (see Table~\ref{tbl_outflow_obs}).}
\tablefoottext{b,c}{Velocity and mass of the hot-molecular outflowing gas \citep{Emonts2017}.}
\tablefoottext{d}{Hot-to-cold molecular gas ratio in the outflows.}
}
\end{table}

We measure similar outflow dynamical times, around 1\,Myr, in all the galaxies. These are much shorter than the age of the star-formation burst expected in ULIRGs ($\sim$60-100\,Myr; \citealt{RodriguezZaurin2010}) and also much shorter than the outflow depletion times ($\sim$15-80\,Myr; see Section~\ref{ss:sfrrate}).

This might be connected to the evolution of the gas within the outflow. For instance, if the molecular gas is swept from the nuclear ISM, it might be able to survive only $\sim$1\,Myr in the hot gas outflow environment before the molecular gas dissociates and becomes neutral atomic gas (e.g., \citealt{Decataldo2017}). These dynamical times are also consistent with those measured in the molecular outflow of a local starburst observed at much higher spatial resolution \citep{Pereira2016b, Aalto2016}. Alternatively, if the outflow has a bi-conical geometry, its projected area increases proportionally to $r^2$ as it expands. Therefore, even if the molecular gas is not dissociated, its column density  rapidly decreases with increasing $r$ and, eventually, the CO emission will be below the detection limit of the observations. The present data do not allow us to distinguish between these possibilities because the outflow structure is just barely spatially resolved and, therefore, it is not possible to the accurately measure the radial dependency of the outflow properties. It has been suggested that molecular gas forms in the outflow (e.g., \citealt{Ferrara2016, Richings2018}). If so, these observations indicate that molecular gas does not efficiently form in outflows, at least, beyond 1\,kpc or after 1\,Myr.

\subsubsection{Hot and cold molecular phase}

There are observations of the ionized and hot molecular phases of the outflows in I12112, I14348, and I22491 that demonstrate their multi-phase structure and suggest that transitions between the different phases are possible (\citealt{Arribas2014, Emonts2017}). { For these galaxies, a direct comparison of the CO(2--1) data with the observations of the ionized phase (H$\alpha$) are not possible due to the relatively low angular resolution of the H$\alpha$ data ($\gtrsim$1\arcsec). However, the detection of a broad H$\alpha$ component indicates the presence of ionized gas in the outflow. The comparison between the cold molecular and the ionized phases of the outflow in NGC~6240 is presented by \citet{Saito2018}. They show that the outflow mass is dominated by the cold molecular phase in that object.}

For the hot molecular phase, we have maps at higher angular resolution \citep{Emonts2017}. This hot phase is traced by the near-IR ro-vibrational H$_2$ transitions and is detected in three cases (I12112 NE, I14348 SW, and I22491 E). The two cases where no outflow was detected in the hot phase, I12112 SW and I14348 NE, contain the least massive of the CO outflows in our ALMA sample, and may therefore have been below the detection limit of the near-IR data. In general, there is a good agreement between the outflow velocity structures (see figures 2, 3, and 4 of \citealt{Emonts2017}). Also, there is a good agreement between the average outflow velocities (see Table~\ref{tbl_multiphase}). Interestingly, for the hot molecular H$_{2}$ gas, only the blueshifted part of the outflows was unambiguously detected. The redshifted part of the outflows, as seen in CO, may have suffered from very high obscuration in the near-IR H$_{2}$ lines, although the poorer spectral resolution and lower sensitivity of the near-IR data compared to the ALMA data makes this difficult to verify. 

The average hot-to-cold molecular gas mass ratio is (2.6$\pm$1.0)$\times$10$^{-5}$. If we only consider the blueshifted part of the outflows, this ratio would be higher by up to a factor of about two. These estimates are slightly lower but comparable to the ratio of 6$-$7$\times$10$^{-5}$ observed in the outflows of local LIRGs  \citep{Emonts2014, Pereira2016b}, and well within the $10^{-7}-10^{-5}$ range found for molecular gas in starburst galaxies and AGN \citep{Dale2005}. This ratio provides information on the temperature distribution of molecular gas (e.g., \citealt{Pereira2014}) and the excitation of the outflowing gas (e.g., \citealt{Emonts2014, Dasyra2014}). 

The hot-to-cold molecular gas mass ratio can also be used to obtain a rough estimate of the total outflowing mass of molecular gas when only near-IR H$_2$ data are available. This method was used in \citet{Emonts2017} to extrapolate total molecular mass outflow rates in I12112 NE, I14348 SW, and I22491 E, as based on the near-IR H$_{2}$ data alone. This resulted in mass outflow estimates that were significantly lower than those found in CO and OH surveys of starburst galaxies and ULIRGs
\citep{Sturm2011, Spoon2013, Veilleux2013, Cicone2014, GonzalezAlfonso2017}. However, our new ALMA results reveal higher molecular mass outflow rates, bringing them back in agreement with these earlier surveys. This shows the importance of directly observing of the cold component of molecular outflows, and it highlights the synergy between ALMA and the James Webb Space Telescope for studying the role of molecular outflows in the evolution of galaxies.

\section{Conclusions}\label{s:conclusions}

We have analyzed new ALMA CO(2--1) observations of 3 \hbox{low-$z$} ULIRG systems ($d\sim 350$\,Mpc). Thanks to the high SNR and spatial resolution of these data, we have been able to study the physical properties and kinematics of the molecular gas around 5 out of 6 nuclei of these 3 ULIRGs. Then, we have used data from the literature to investigate the properties of these outflows and their impact on the evolution of the ULIRG systems. The main results of this paper are the following:
\begin{enumerate}

\item We have detected fast { (deprojected v$_{\rm out}\sim 350-550$\,km\,s$^{-1}$; v$_{\rm max}\sim 500-900$\,km\,s$^{-1}$)} massive molecular outflows ($M_{\rm out}\sim(0.3-5)\times$10$^{8}$\,\Msun) in the 5 well detected nuclei of these 3 low-$z$ ULIRGs. The outflow emission is spatially resolved and we measure { deprojected outflow effective radii} between 250\,pc and 1\,kpc.  The PA of the outflow emission is compatible with an outflow perpendicular to the rotating molecular disk in 3 cases. { Only in one case, the outflow PA is clearly not along the kinematic minor axis and suggests a different outflow orientation.}

\item The outflow dynamical times are between 0.5 and 3\,Myr and the outflow rates between 12 and 400\,\Msun\,yr$^{-1}$. Taking into account the nuclear SFR, the mass loading factors are 0.8 to $\sim$2. { These values are similar to those found in other local ULIRGs}.
The total molecular gas mass in the regions where the outflows originate is $(1-7)\times$10$^9$\,\Msun. Therefore, the outflow depletion times are $15-80$\,Myr. { We also estimate that only $15-30$\% of the outflowing gas has ${\rm v}>{\rm v_{\rm esc}}$ and will escape the gravitational potential of the nucleus.}

\item We use multiple indicators to determine the power source of these molecular outflows (e.g, mass loading factor, outflow { energy and momentum vs. those injected by SNe}, maximum outflow velocity, geometry, etc.). For all the nuclei, { the observed molecular outflows} are compatible with being powered by the strong nuclear starburst.

\item The outflow depletion times are slightly shorter than the SF depletion times { ($30-80$\,Myr)}. However, we find that most of the outflowing molecular does not have enough velocity to escape the gravitational potential of the nucleus. Assuming that the outflowing gas is not affected by any dynamical friction, we estimate that most of this outflowing material will return to the molecular disk after $5-10$\,Myr and become available to form new stars. Therefore, the main effects of these outflows are to expel part of the total molecular gas ($\sim15-30\%$) into the intergalactic medium and delay the formation of stars { but, possibly, they are not completely quenching the nuclear star-formation.}

\item \citet{Cicone2014} suggested that outflows in starbursts are driven by the radiation pressure due to young stars (i.e., momentum-driven) based on the --1 slope of the mass loading factor { vs.} outflow velocity relation. After adding more points to this relation, we find a shallower slope --(0.3$\pm$0.2). For momentum-driven outflows, this shallower slope can be explained if the dust optical depth increases for higher luminosity starbursts enhancing the momentum boost factor. One of the nuclear starbursts in our sample has an optical depth $\gtrsim$8 at 100$\micron$ and might support this scenario. Alternatively, these outflows might be launched by SNe. If so, the coupling efficiency between the ISM and SNe increases with increasing SFR. For the stronger starbursts, { these molecular} outflows carry up to 75\% and 20\% of the momentum and energy injected by SNe, respectively.

\item We explore the possible evolution of the cold molecular gas in the outflow. The relatively small sizes ($<$1\,kpc) and short dynamical times ($<$3\,Myr) of the outflows suggest that molecular gas cannot survive longer in the outflow environment or that it cannot form efficiently beyond these distances or times. The detection of other outflow phases, hot molecular and ionized, for these galaxies suggests that transformation between the different outflow gas phases might exist. Alternatively, in a uniform bi-conical outflow geometry, the CO column density will eventually be below the detection limit { and explain the non-detection of the outflowing molecular gas beyond $\sim$1\,kpc. New high-spatial resolution observations of similar outflows will help to distinguish between these possibilities. }

\end{enumerate}

\begin{acknowledgements}
{ We thank the anonymous referee for useful comments and suggestions.}
MPS acknowledges support from STFC through grant ST/N000919/1.
LC, SGB, and AL acknowledge financial support by the Spanish MEC under grants ESP2015-68964 and AYA2016-76682-C3-2-P.
This paper makes use of the following ALMA data: ADS/JAO.ALMA\#2015.1.00263.S, ADS/JAO.ALMA\#2016.1.00170.S. ALMA is a partnership of ESO (representing its member states), NSF (USA) and NINS (Japan), together with NRC (Canada) and NSC and ASIAA (Taiwan) and KASI (Republic of Korea), in cooperation with the Republic of Chile. The Joint ALMA Observatory is operated by ESO, AUI/NRAO and NAOJ.
The National Radio Astronomy Observatory is a facility of the National Science Foundation operated under cooperative agreement by Associated Universities, Inc.
\end{acknowledgements}

\clearpage

\appendix
\onecolumn
\section{Continuum visibility fits}\label{apx_uv_fit}

\begin{figure}[h]
\centering
\includegraphics[height=0.25\textwidth]{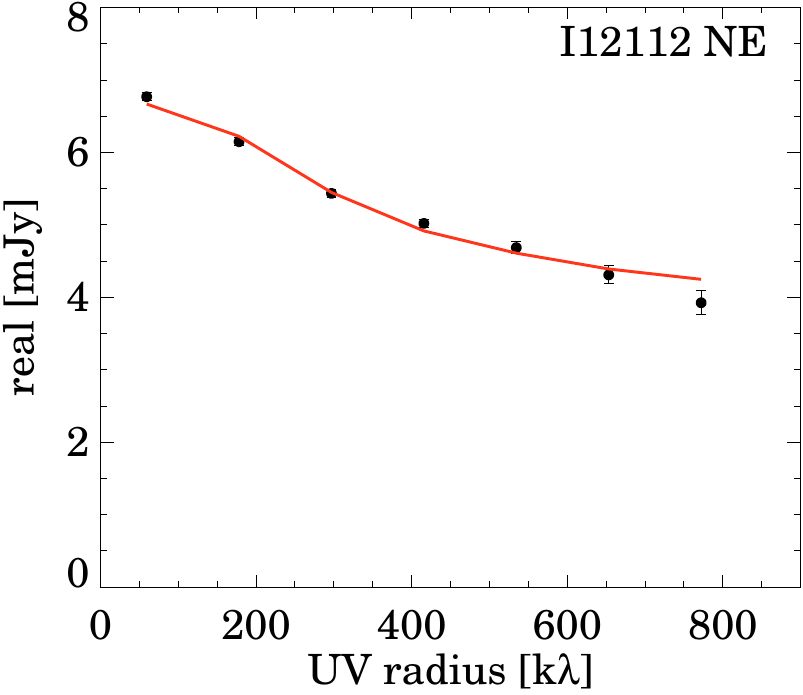}
\includegraphics[height=0.25\textwidth]{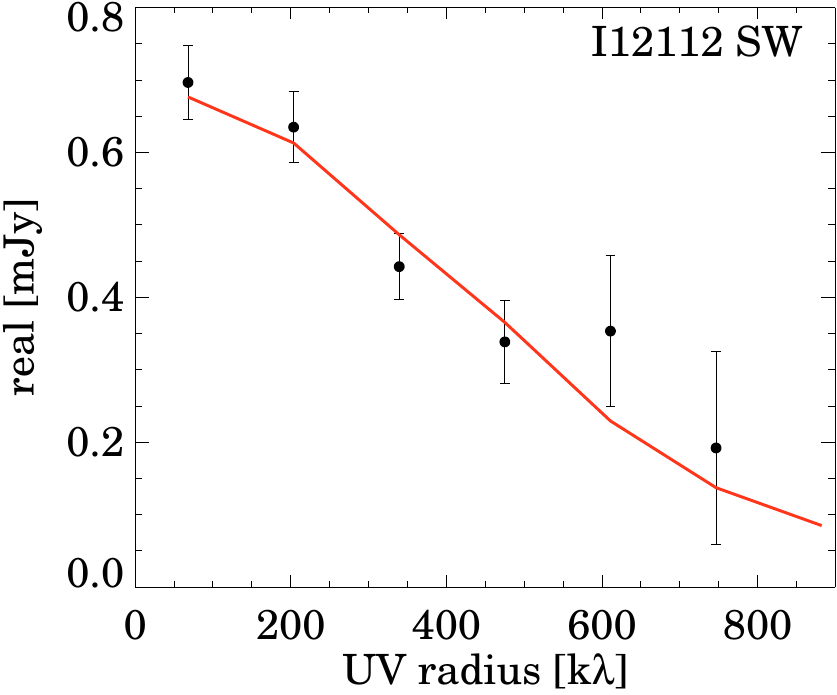}
\includegraphics[height=0.25\textwidth]{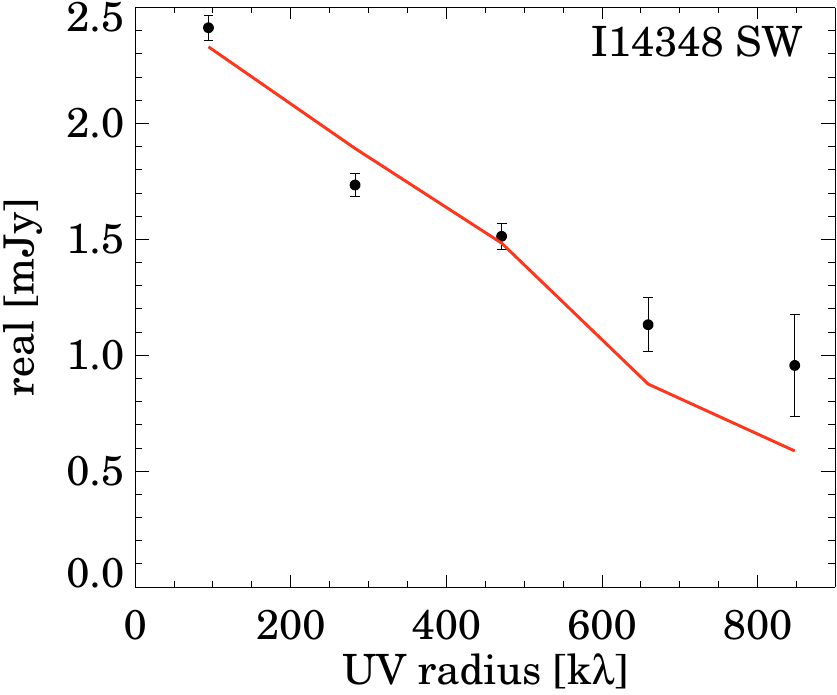}
\includegraphics[height=0.25\textwidth]{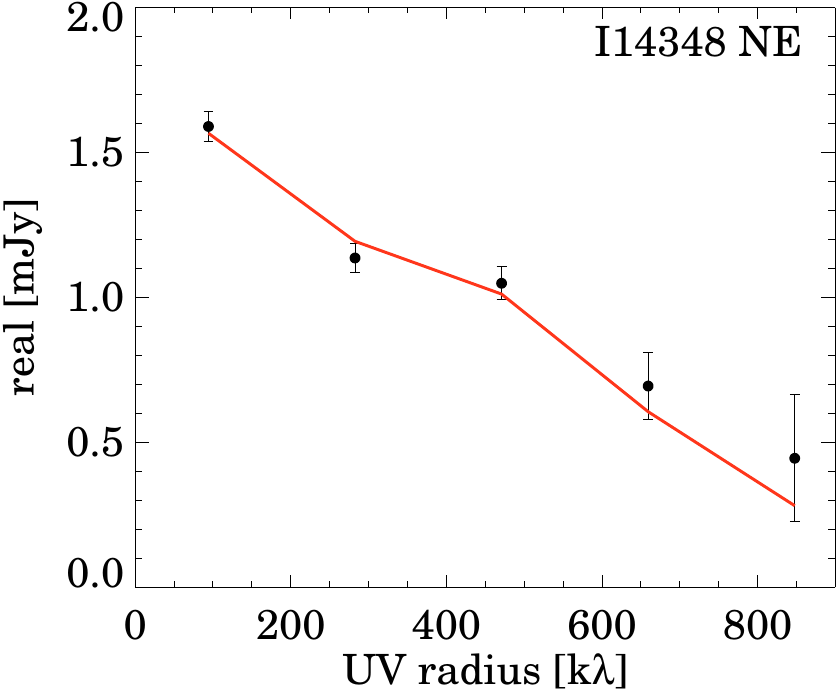}
\includegraphics[height=0.25\textwidth]{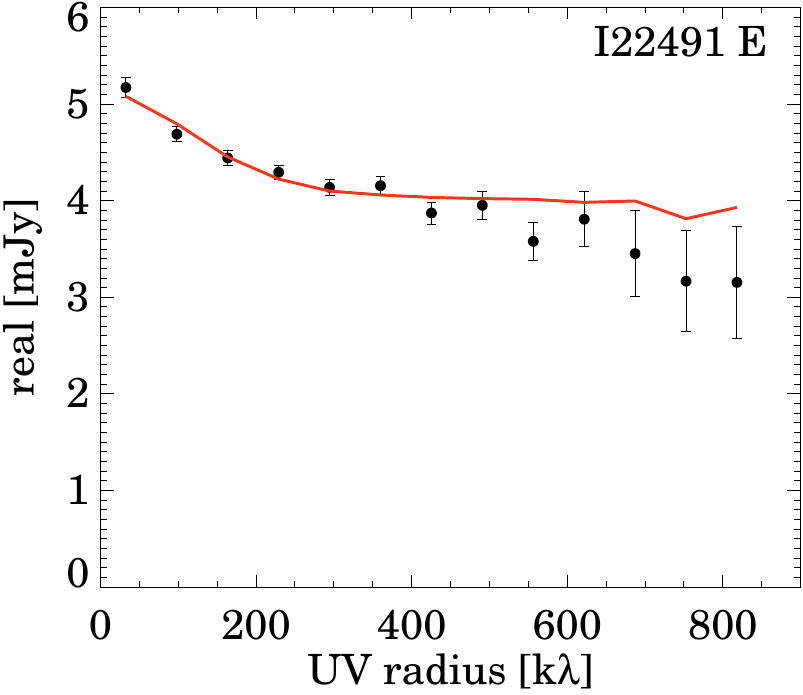}
\caption{Real part of the 248\,GHz continuum visibilities as function of the $uv$ radius. The red line is the best-fit model discussed in Section \ref{ss:continuum} (see also Table \ref{tbl_continuum}). \label{fig_uv_fits}}
\end{figure}

Figure~\ref{fig_uv_fits} compares the real part of the continuum visibilities for each source with the best { fit} model discussed in Section~\ref{ss:continuum}. To obtain these visibilities, we shifted the phase center to the coordinates obtained by {\sc uvmultifit}. For the objects with two continuum sources in the field of view, we subtracted the model of the source that is not plotted in that panel.

\clearpage
\section{CO(2--1) channel maps}\label{apx_channels}

\begin{figure}[h]
\centering
\includegraphics[width=0.96\textwidth]{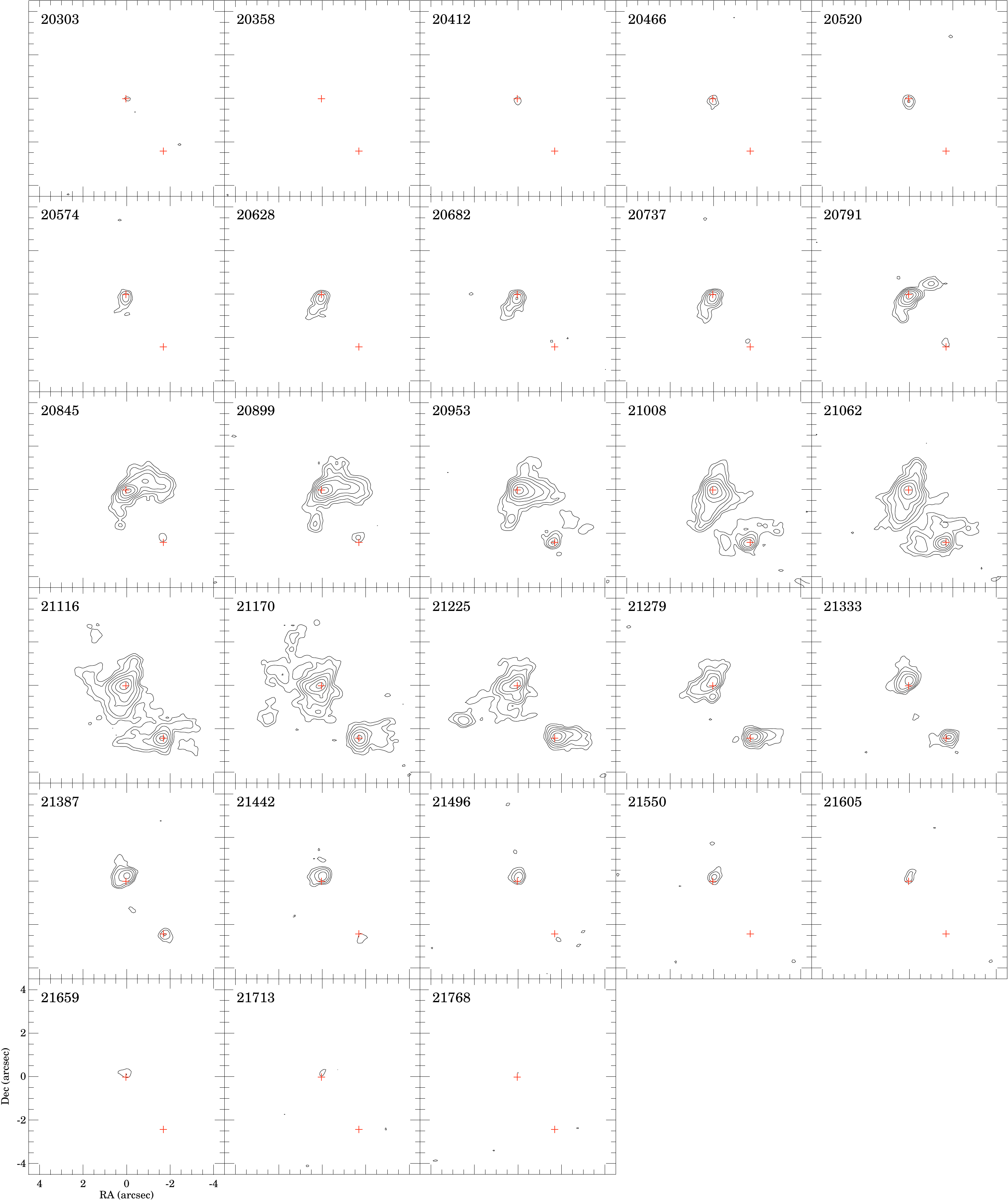}
\caption{Channel maps showing the CO(2--1) 230.5\,GHz emission in {IRAS}~12112+0305. Each panel shows the emission averaged over 39\,MHz ($\sim$50\,km\,s$^{-1}$) channels. The contours correspond to (3, 6, 12, 24, 48, 96, 192)$\times\sigma$ and $\sigma$ is the rms measured in each channel ($180-260$\,$\mu$Jy\,beam$^{-1}$) for this system. The relativistic LSRK velocity is indicated in each panel. The red crosses mark the location of the nuclei listed in Table~\ref{tbl_sample}.\label{fig_channels_i12112}
}
\end{figure}

\begin{figure}[h]
\centering
\includegraphics[width=0.96\textwidth]{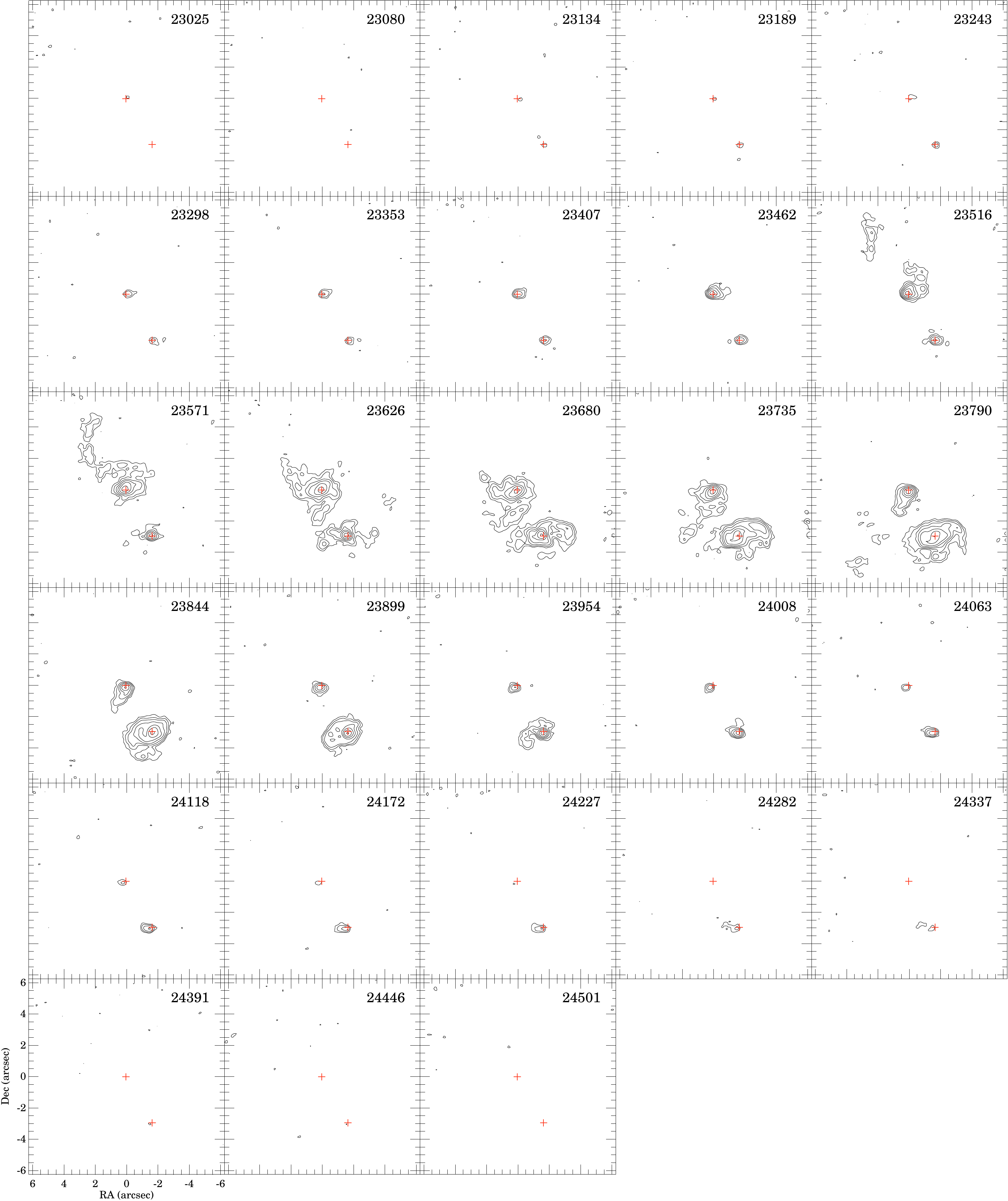}
\caption{Same as Figure~\ref{fig_channels_i12112} but for {IRAS}~14348$-$1447. For this system $\sigma=200-300$\,$\mu$Jy\,beam$^{-1}$ depending on the channel.}
\end{figure}

\begin{figure}[h]
\centering
\includegraphics[width=0.96\textwidth]{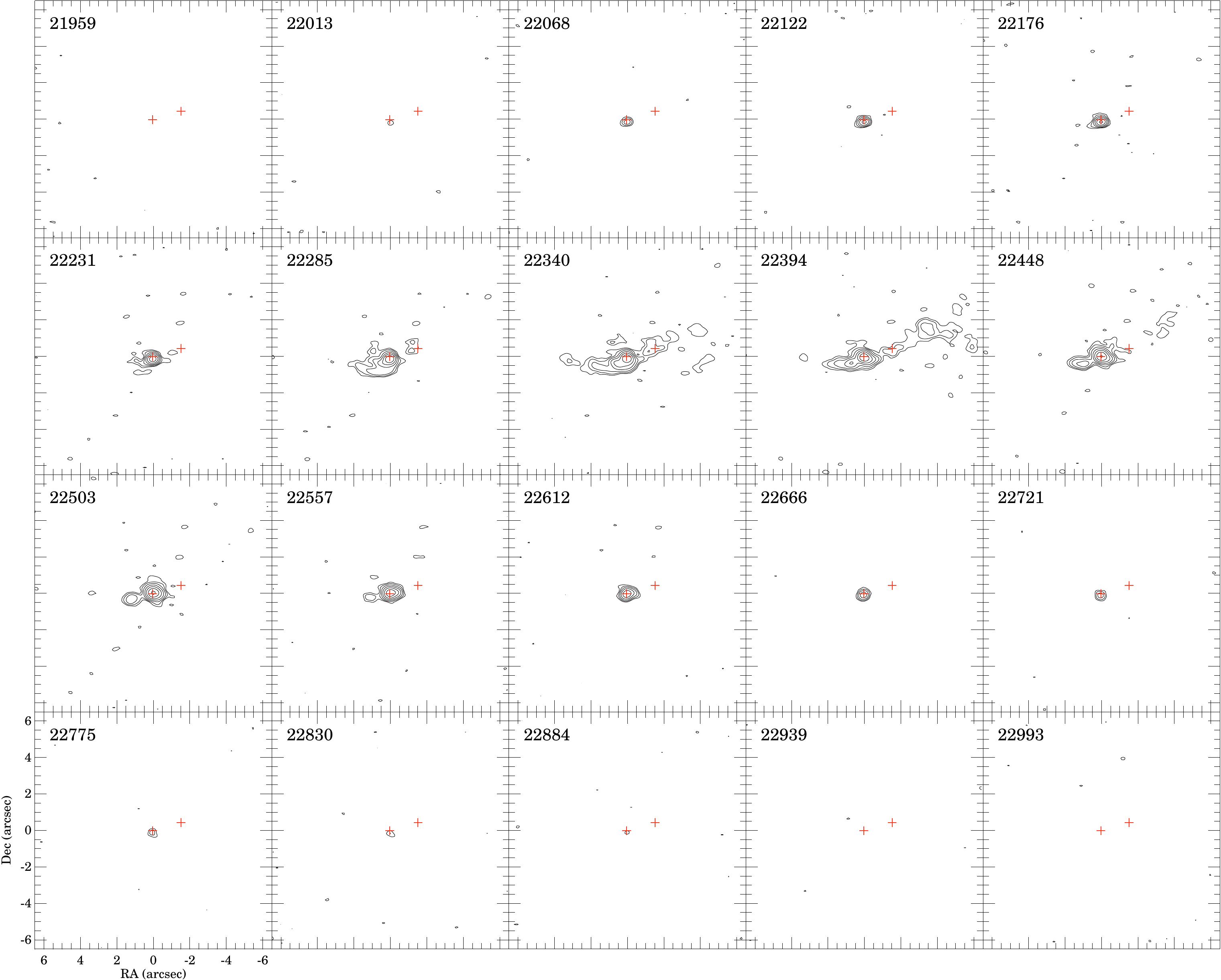}
\caption{Same as Figure~\ref{fig_channels_i12112} but for {IRAS}~22491$-$1808. For this system $\sigma=250-400$\,$\mu$Jy\,beam$^{-1}$ depending on the channel.}
\end{figure}

\end{document}